\documentclass[apj]{emulateapj}
\usepackage{amsmath, amsthm, amssymb}

\usepackage{color}

\slugcomment{Accepted to the Astrophysical Journal Supplement Series}

\shorttitle{Skynet Algorithm I}
\shortauthors{Martin et al.}

\begin{document}


\title{Skynet Algorithm for Single-Dish Radio Mapping I:  Contaminant-Cleaning, Mapping, and Photometering Small-Scale Structures}


\author{J.~R.~Martin\altaffilmark{1}, D.~E.~Reichart\altaffilmark{1}, D.~A.~Dutton\altaffilmark{1}, M.~P.~Maples\altaffilmark{1}, T.~A.~Berger\altaffilmark{1,2}, F.~D.~Ghigo\altaffilmark{3}, J.~B.~Haislip\altaffilmark{1}, {\color{black}O.~H.~Shaban\altaffilmark{1},} A.~S.~Trotter\altaffilmark{1}, L.~M.~Barnes\altaffilmark{1}, M.~L.~Paggen\altaffilmark{1}, R.~L.~Gao\altaffilmark{4}, C.~P.~Salemi\altaffilmark{1}, G.~I.~Langston\altaffilmark{3,5} S.~Bussa\altaffilmark{3}, J.~A.~Duncan\altaffilmark{3}, S.~White\altaffilmark{3}, S.~A.~Heatherly\altaffilmark{3}, J.~B.~Karlik\altaffilmark{1}, E.~M.~Johnson\altaffilmark{1}, J.~E.~Reichart\altaffilmark{6}, A.~C.~Foster\altaffilmark{1}, V.~V.~Kouprianov\altaffilmark{1}, S.~Mazlin\altaffilmark{1}, J.~Harvey\altaffilmark{1}}
\email{reichart@unc.edu}


\altaffiltext{1}{Department of Physics and Astronomy, University of North Carolina at Chapel Hill, Chapel Hill, NC 27599}
\altaffiltext{2}{Institute for Astronomy, University of Hawaii, Honolulu, HI 96822}
\altaffiltext{3}{Green Bank Observatory, Green Bank, WV, 24944}
\altaffiltext{4}{North Carolina School for Science and Mathematics, Durham, NC 27705}
\altaffiltext{5}{Division of Astronomical Sciences, National Science Foundation, Arlington, VA 22230}
\altaffiltext{6}{Jordan High School, Durham, NC 27707}


\begin{abstract}
We present a single-dish mapping algorithm with a number of advantages over traditional techniques.  (1)~Our algorithm makes use of weighted modeling, instead of weighted averaging, to interpolate between signal measurements.  This smooths the data, but without blurring the data beyond instrumental resolution.  Techniques that rely on weighted averaging blur point sources sometimes as much as 40\%.  (2)~Our algorithm makes use of local, instead of global, modeling to separate astronomical signal from instrumental and/or environmental signal drift along the telescope's scans.  Other techniques, such as basket weaving, model this drift with simple functional forms (linear, quadratic, etc.\@) across the entirety of scans, limiting their ability to remove such contaminants.  (3)~Our algorithm makes use of a similar, local modeling technique to separate astronomical signal from radio-frequency interference (RFI), even if only continuum data are available.  (4)~Unlike other techniques, our algorithm does not require data to be collected on a rectangular grid or regridded before processing.  (5)~Data from any number of observations, overlapping or not, may be appended and processed together.  (6)~Any pixel density may be selected for the final image.  We present our algorithm, and evaluate it using both simulated and real data.  We are integrating it into the image-processing library of the Skynet Robotic Telescope Network, which includes optical telescopes spanning four continents, and now also Green Bank Observatory's 20-meter diameter radio telescope in West Virginia.  Skynet serves hundreds of professional users, and additionally tens of thousands of students, of all ages.  Default data products are generated on the fly, but will soon be customizable after the fact.
\end{abstract}





\section{Introduction}

\subsection{Skynet}

Founded in 2005, Skynet is a global network of fully automated, or robotic, volunteer telescopes, scheduled through a common web interface.\footnote{https://skynet.unc.edu}  Currently, our optical telescopes range in size from 14 to 40 inches, and span four continents.  Originally envisioned for gamma-ray burst follow-up (Reichart et al.\@ 2005, Haislip et al.\@ 2006, Dai et al.\@ 2007, Updike et al.\@ 2008, Nysewander et al.\@ 2009, Cenko et al.\@ 2011, Cano et al.\@ 2011, Bufano et al.\@ 2012, Jin et al.\@ 2013, Morgan et al.\@ 2014, Martin-Carrillo et al.\@ 2014, Friis et al.\@ 2015, De Pasquale et al.\@ 2016, Bardho et al.\@ 2016, Melandri et al.\@ 2017), Skynet has also been used to study gravitational-wave sources (Abbott et al.\@ 2017a, 2017b, Valenti et al.\@ 2017, Yang et al.\@ 2017), blazars (Osterman Meyer et al.\@ 2008, Valtonen et al.\@ 2016, Zola et al.\@ 2016, Liu et al.\@ 2017, Goyal et al.\@ 2017), supernovae (Foley et al.\@ 2010, 2012, 2013, Pignata et al.\@ 2011, Valenti et al.\@ 2011, 2014, Pastorello et al.\@ 2013, Milisavljevic et al.\@ 2013, Maund et al.\@ 2013, Fraser et al.\@ 2013, Stritzinger et al.\@ 2014, Inserra et al.\@ 2014, Takats et al.\@ 2014, 2015, 2016, Dall'Ora et al.\@ 2014, Folatelli et al.\@ 2014, Barbarino et al.\@ 2015, de Jaeger et al.\@ 2016, Gutierrez et al.\@ 2016, {\color{black}2018,} Tartaglia et al.\@ 2017, 2018, Prentice et al.\@ {\color{black}2018}), supernova remnants (Trotter et al.\@ 2017), novae (Schaefer et al.\@ 2011), pulsating white dwarfs and hot subdwarfs (Thompson et al.\@ 2010, Barlow et al.\@ 2010, 2011, 2013, 2017, Reed et al.\@ 2012, Bourdreaux et al.\@ 2017, Hutchens et al.\@ 2017), a wide variety of variable stars (Layden et al.\@ 2010, Gvaramadze et al.\@ 2012, Wehrung et al.\@ 2013, Miroshnichenko et al.\@ 2014, Abbas et al.\@ 2015, Khokhlov et al.\@ 2017, 2018), a wide variety of binary stars (Reed et al.\@ 2010, Sarty et al.\@ 2011, Helminiak et al.\@ 2011, 2012, 2015, Strader et al.\@ 2015, Tovmassian et al.\@ 2016, 2017, Fuchs et al.\@ 2016, Pala et al.\@ 2017, Neustroev et al.\@ 2017, Lubin et al.\@ 2017, Swihart et al.\@ 2017, Zola et al.\@ 2017, Kriwattanawong et al.\@ 2018), exoplanetary systems (Fischer et al.\@ 2006, Czesla et al.\@ 2012, Meng et al.\@ 2014, 2015, Kenworthy et al.\@ 2015, Kuhn et al.\@ 2016, Awiphan et al.\@ 2016, Blank et al.\@ 2018), trans-Neptunian objects and Centaurs (Braga-Ribas et al.\@ 2013, 2014, Dias-Oliveira et al.\@ 2015), asteroids (Descamps et al.\@ 2009, Pravec et al.\@ 2010, 2012, 2016, Marchis et al.\@ 2012, Savanevych et al.\@ 2018), and near-Earth objects (NEOs; Brozovic et al.\@ 2011, Pravec et al.\@ 2014).  Skynet is also the leading tracker of NEOs in the southern hemisphere (R. Holmes, private communication).

\begin{figure}
\plotone{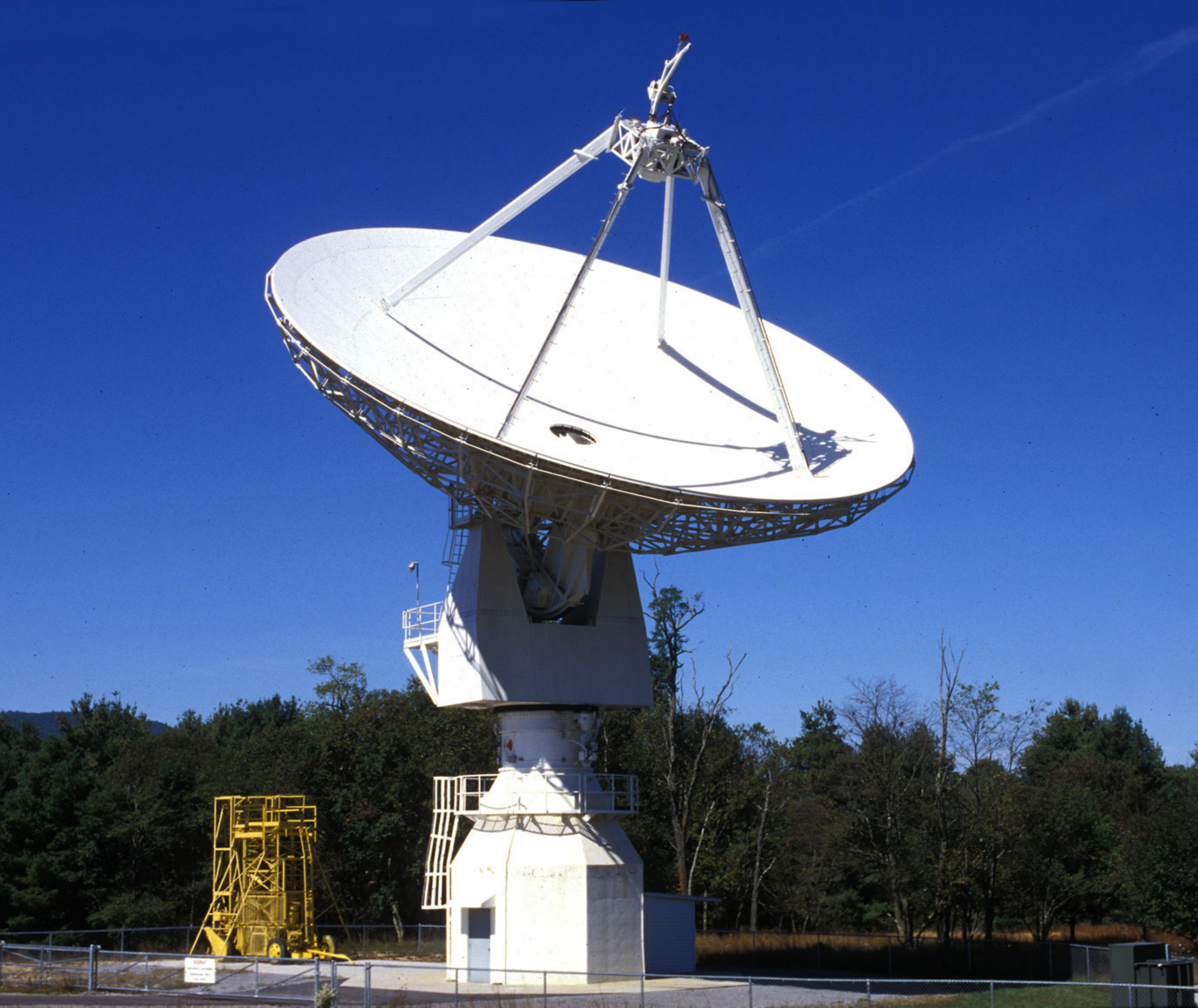}
\caption{Green Bank Observatory 20-meter diameter radio telescope.  (Photo credit:  GBO)}
\end{figure}

\begin{figure*}
\plotone{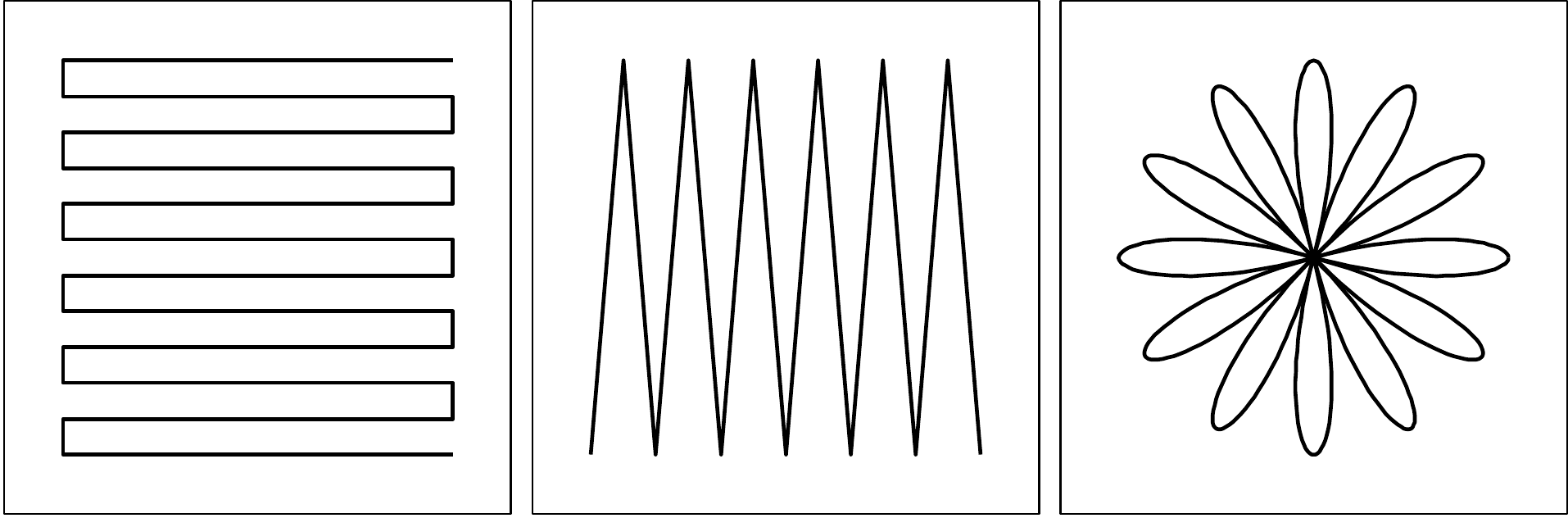}
\caption{OTF mapping patterns.  \textbf{Left:}  Raster (horizontal).  \textbf{Middle:}  Nodding.  \textbf{Right:}  Daisy.}
\end{figure*}

Skynet's mission is split evenly between supporting professional astronomers and supporting students and the public.  Although most of our observations have been for professionals, most of our users are students.  We have developed/are continuing to develop Skynet-based curricula for undergraduates, high-school students (in partnership with Morehead Planetarium and Science Center (MPSC)), and middle school-aged students (in partnership with the University of Chicago/Yerkes Observatory, Green Bank Observatory (GBO), the Astronomical Society of the Pacific, and 4-H), as well as for blind and visually-impaired students (in partnership with Associated Universities, Inc., the University of Chicago/Yerkes Observatory, the Technical Education Research Centers, and the University of Nevada at Las Vegas).  These efforts have reached over 20,000 students, and our public-engagement efforts (also in partnership with MPSC) have also reached over 20,000, mostly elementary and middle school-aged students.  Curriculum-based student users queue observations through the same web interface that the professionals use.  Altogether, over 15 million images have been taken to date.

In partnership with GBO, and funded by the American Recovery and Reinvestment Act, Skynet has added its first radio telescope, GBO's 20-meter in West Virginia (see Figure~1).  We describe the 20-meter, which has been refurbished, in \textsection2.  As with Skynet's optical telescopes, the 20-meter serves both professionals and students.  Professional use consists primarily of timing {\color{black}observations (e.g., pulsar timing; fast radio burst searches in conjunction with \textit{Swift})}, but also some mapping observations{\color{black}, for photometry} (e.g., {\color{black}fading of Cas~A and improved flux-density calibration of the radio sky (}Trotter et al.\@ 2017{\color{black}); intra-day variable blazar campaigns in conjunction with other radio, and optical, telescopes}).  Student use consists of {\color{black}timing, spectroscopic (e.g., Williamson et al.\@ 2018), and mapping observations}, but with an emphasis on mapping, at least for beginners.

Regarding student, as well as public, use, the 20-meter represents a significant opportunity for radio astronomy.  Small optical telescopes can be found on many, if not most, college campuses.  But small radio telescopes are significantly more expensive to build, operate, and maintain, and consequently are generally found only in the remote locations that make the most sense for professional use.  Consequently, most people -- including most students of astronomy -- never experience radio telescopes, let alone use them.  However, under the control of Skynet, the 20-meter is not only more accessible to more professionals, it is already being used by thousands of students per year, of all ages, as well as by the public.

\subsection{Single-Dish Mapping}

In this paper, we present the single-dish mapping algorithm that we developed for Skynet, both for professional use and for student use.  Here we outline the design requirements that we set for ourselves, and outline approaches that we adopted/developed to meet these requirements.

\subsubsection{Mapping Pattern}

Many single-dish mapping algorithms (e.g., Sofue \& Reich 1979, Emerson \& Grave 1988) require the signal to be sampled on a rectangular grid.  Generally, this requires what is called ``step-and-integrate'' or ``point-and-shoot'' mapping.

However, this is an inefficient way to observe, with greater telescope acceleration, deceleration, and settling overheads, greater wear and tear on the telescope, and greater sensitivity to time-variable systematics (see \textsection1.2.3).  ``On-the-fly'' (OTF) mapping, in which signal is integrated while the telescope moves, minimizes these concerns, as long as integrations span no more than $\approx$0.2 beamwidths (FWHM) along the telescope's direction of motion (along the telescope's “scan”), to avoid blurring point sources by more than $\approx$1\% (Mangum, Emerson \& Greison 2007).

A wide variety of OTF mapping patterns might be employed (e.g., see Figure~2):

\begin{itemize}
\item A ``raster'' pattern approximates a rectangular grid, though integrations might not line up from scan to scan.

\item A ``nodding'' pattern, in which the telescope moves up and down, typically in elevation, as Earth's rotation carries the telescope's beam across the sky in the perpendicular direction, is typical of meridian-transit telescopes, which cannot move east-west (e.g., see \textsection2.2).  However, even with full-motion telescopes, a nodding pattern can be used to maintain a constant parallactic angle, if it is desired that the telescope's beam pattern not rotate across the image, or from image to image.

\item ``Spiral'' and ``hypocycloid'' patterns are efficient ways of mapping sources and extended structures, respectively, without abrupt changes to the telescope's motion (Mangum, Emerson \& Greison 2007).

\item A ``daisy'' pattern can also be used to map sources, with the advantage of crossing the source's peak, and sampling the background level, an arbitrarily large number of times.
\end{itemize}

As long as the gaps between scans do not exceed Nyquist sampling, or $\approx$0.4 beamwidths, in theory all information between scans can be recovered (modulo noise, interference, etc.\@)  Given this, and not wishing to limit our users to a particular mapping pattern, or set of patterns, we require that our algorithm work independently of mapping pattern.  Furthermore, since student/quick-look maps may be undersampled, we also require that our algorithm work independently of sampling density (within reason).  Instead, undersampled images will be flagged as not for professional use.

\subsubsection{Signal Averaging vs.\@ Signal Modeling}

Some single-dish mapping algorithms get around the problem of not collecting data on a rectangular grid by resampling the data onto a rectangular grid before processing (e.g., Winkel, Floer \& Kraus 2012).  This is known as ``regridding'', and is typically done by taking a weighted average of the data around each grid point.  This is essentially a convolution of the data with the weighting function, sometimes called the convolution kernel.  

However, this blurs the image.  Many adopt a kernel of width equal to the telescope's beamwidth, but this results in $\approx$40\% blurring and $\approx$40\% errors (near the center of the beam pattern) in the reconstructed image (see Figure~3).  Others oversample so a narrower kernel can be used.  For example, Winkel, Floer \& Kraus (2012) collect $\approx$3 times as much data as required by Nyquist and use a 1/2-beamwidth kernel, but this still results in $\approx$12\% blurring and $\approx$18\% errors (near the center of the beam pattern) in the reconstructed image (Figure~3).

Given that single-dish maps already suffer from poor resolution (compared to interferometric maps), we do not want to degrade resolution further, simply due to processing.  To this end, instead of weighted \textit{averaging}, we use weighted \textit{modeling}:  Instead of averaging the data local to each pixel in the final image, using a weighting function, we fit a model to the data local to each pixel in the final image, using a weighting function (see Figure~4).  As long as (1)~the model is sufficiently flexible over the scale of the weighting function, and (2)~sampling is sufficiently dense to constrain (actually, overconstrain) this model, the signal should be recoverable at any location, without blurring.  For example, the weighted modeling approach that we present in \textsection3.7 is able to recover the simulated data in Figure~3 with $<$1\% errors (near the center of the beam pattern).  

Another requirement that we place on our algorithm is that this -- replacing the data with a locally modeled version of itself -- be the last step, not the first step.  Other algorithms regrid the data before processing (e.g., contaminant removal, see \textsection1.2.3), because their processing operations work only on gridded data.  However, this is a poor modeling practice:  Even if weighted averaging could be done accurately, and even if weighted modeling can be done accurately, every operation on data propagates uncertainty in ways that are both difficult to understand and even more difficult to properly account for before/in the next operation.  It is always preferable to operate on real data, at least for as long as possible, than it is to operate on successive approximations thereof.

\subsubsection{Contaminant Removal}

Single-dish mapping algorithms must address signal contaminants, which we separate into three broad categories:  (1)~en-route signal drift, resulting in what is sometimes called ``the scanning effect'', (2)~radio-frequency interference (RFI), and (3)~elevation-dependent signal.  Each of these is demonstrated in Figure~5.

\textbf{En-Route Drift:}  Even with very stable, modern receivers, the detected signal can still drift in time, due to instrumental reasons, such as $1/f$, or pink, noise, and/or due to environmental reasons, such as changing atmospheric emission, or changing ground emission spilling over the edge of the dish, particular as the dish moves.  This results in low-level variations in the signal along the telescope's scans, and is most noticeable from scan to scan (the scanning effect).  Components of these variations can be made to vary over shorter or longer angular scales by moving the telescope slower or faster, but this does not eliminate them.

Sofue \& Reich (1979) used unsharp masking to separate en-route drift (and small-scale structure in the image) from larger-scale structure, and then modeled the en-route drift along each scan with a second-order polynomial, using sigma-clipping to avoid small-scale structure contamination.  Drawbacks of this approach are:  (1)~Unsharp masking uses a blurred version of the data to correct the data, resulting in a blurred image, at least in the lower signal-to-noise parts of the image (something we wish to avoid; \textsection1.2.2); (2)~Low-order polynomials may adequately model en-route drift over small angular scales, but are too simple/increasingly inaccurate over larger angular scales, such as the length of a scan; and (3)~Sigma-clipping is too crude of an outlier rejection criterion (see \textsection1.3).

\begin{figure*}
\epsscale{0.95}
\plotone{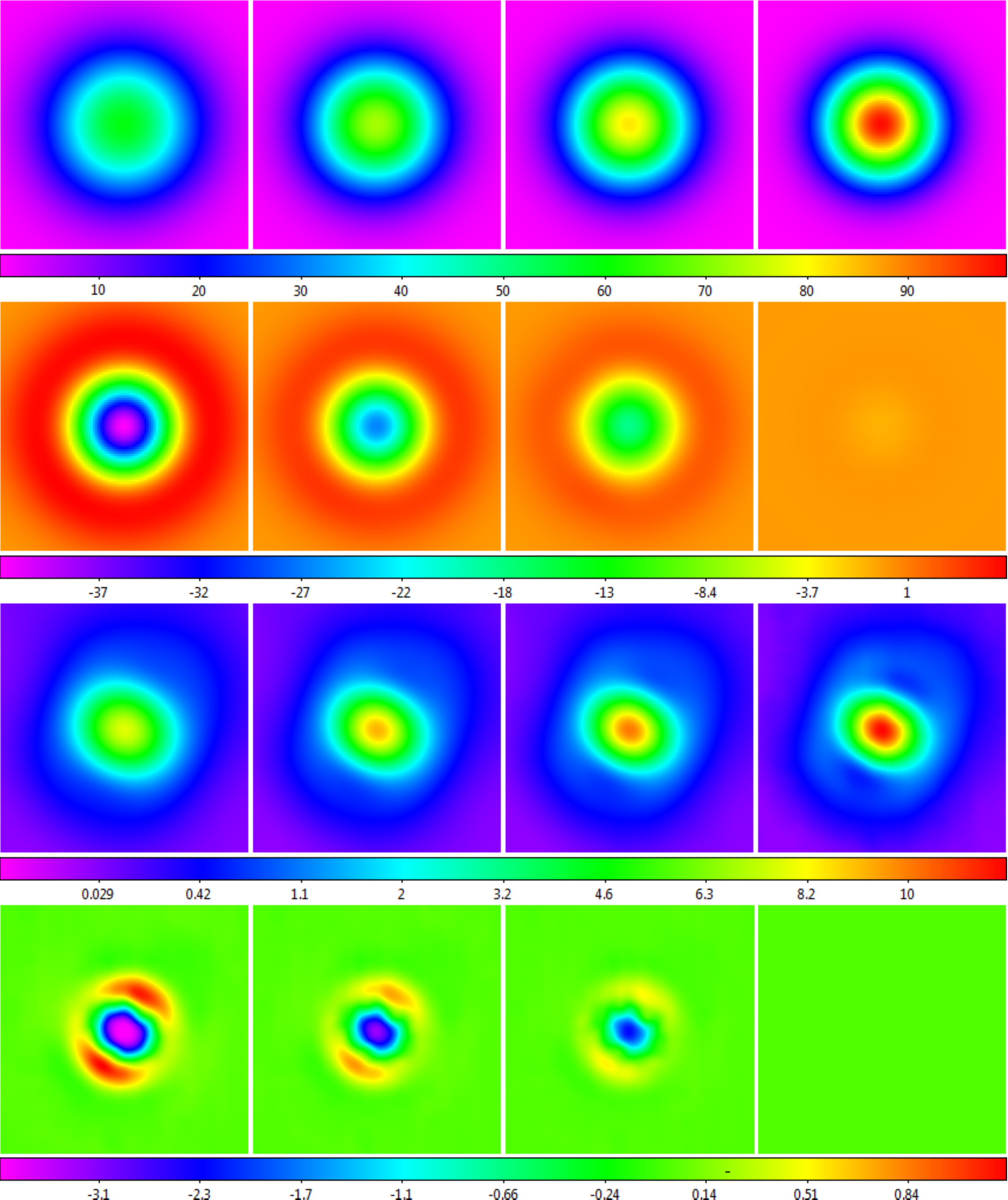}
\caption{\textbf{First Row:}  Simulation of a point source sampled with a Gaussian beam pattern on a 1/5-beamwidth grid, recovered using, from left to right:  1-beamwidth weighted averaging, 2/3-beamwidth weighted averaging, 1/2-beamwidth weighted averaging, and weighted modeling, as described in \textsection3.7.  \textbf{Second Row:}  Residual error associated with each of these techniques.  \textbf{Third Row:}  Cassiopeia~A observed with one of the 20-meter's L-band linear polarization channels using a 1/5-beamwidth raster, and recovered using the above techniques.  Weighted averaging fails to recover the telescope's beam pattern, which is structured.  \textbf{Fourth Row:}  Difference between each of these techniques and weighted modeling.  Square-root and squared scalings are used in the third and fourth rows, respectively, to emphasize fainter beam structure (units are dimensionless, with one corresponding to the noise diode; see \S3.1.)}
\end{figure*}

\begin{figure*}
\plottwo{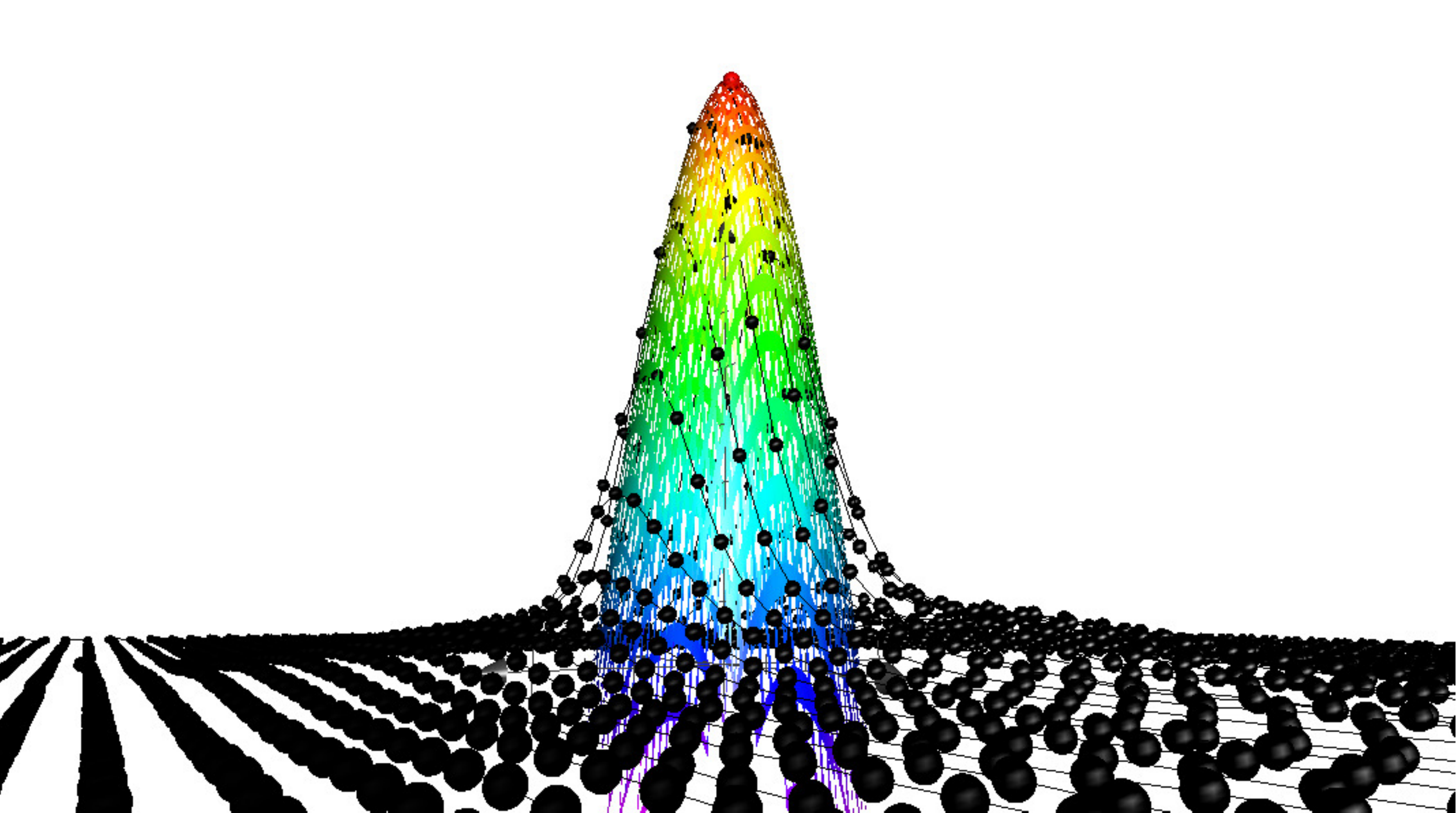}{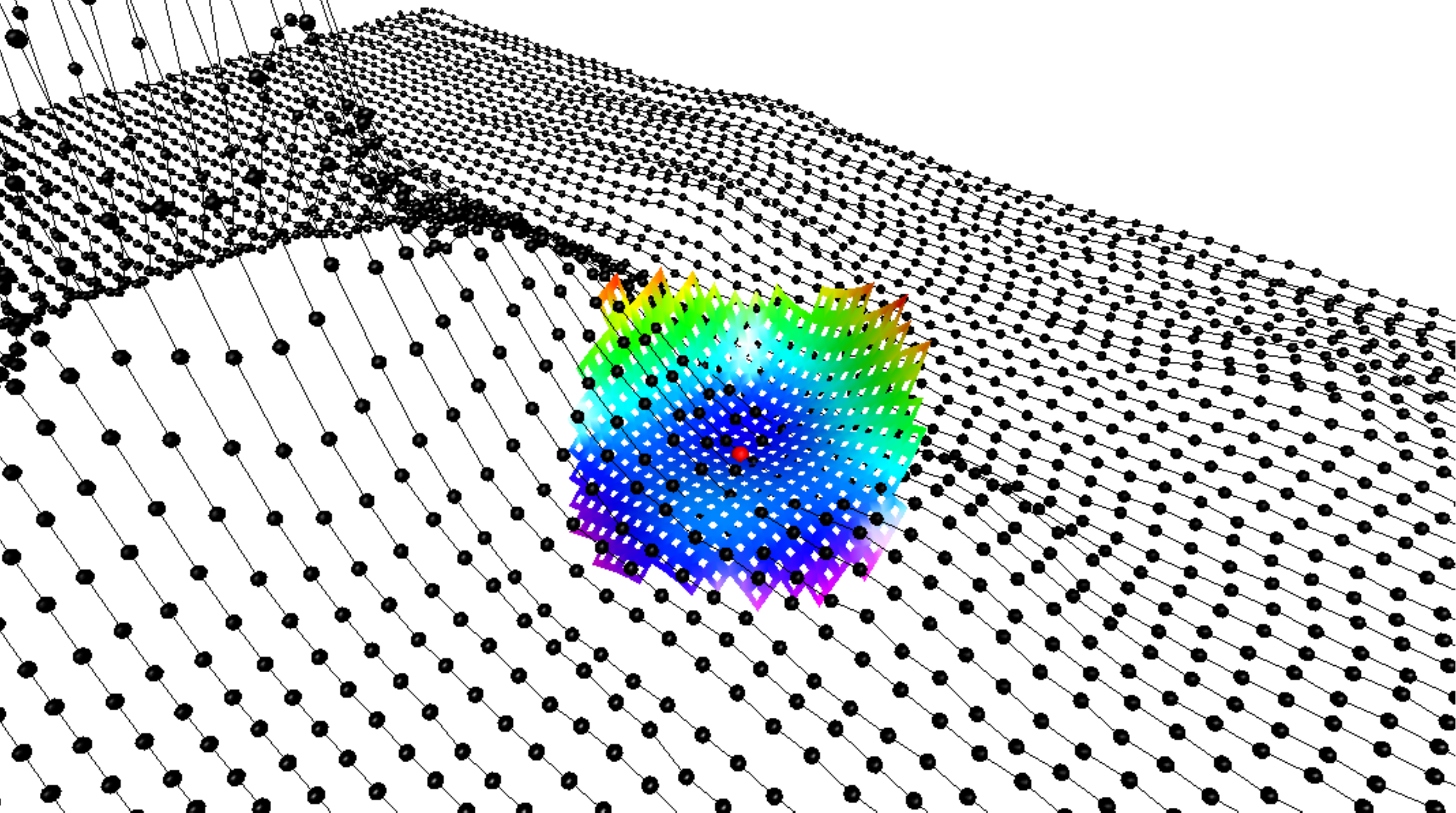}
\caption{Weighted modeling of 20-meter data from the third row of Figure~3, at two representative points.  Modeled surfaces span two beamwidths, but are most strongly weighted to fit the data over only the central, typically, 1/3 -- 2/3 beamwidths, as described in \textsection3.7.  Only the central point (red) is retained.}
\end{figure*}

\begin{figure}
\plotone{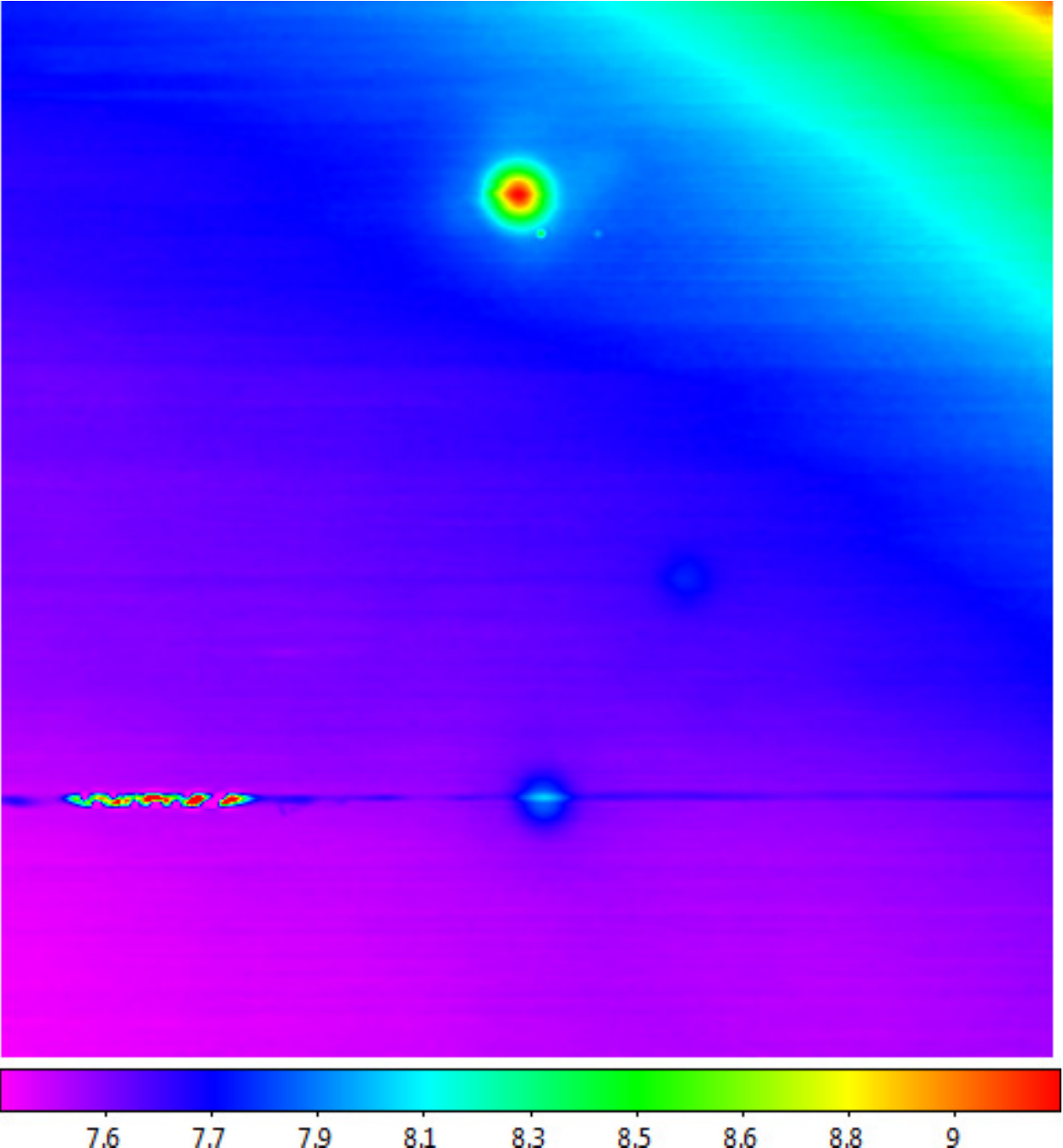}
\caption{Raw map of Virgo~A (top), 3C~270 (center right), and 3C~273 (bottom), acquired with the 20-meter in L band, using a 1/10-beamwidth horizontal raster.  Left and right linear polarization channels have been summed, partially symmetrizing the beam pattern (Figure~3).  Locally modeled surface (\textsection1.2.1, see \textsection3.7) has been applied for visualization only.  All three signal contaminants are demonstrated:  (1)~en-route drift, the low-level variations along the horizontal scans, (2)~RFI, both long-duration, during the scan that passes through 3C~273, and short-duration, near Virgo~A, and (3)~elevation-dependent signal, toward the upper right, which was only $\approx$11$\degr$ above the horizon.}
\end{figure}

\begin{figure}
\plotone{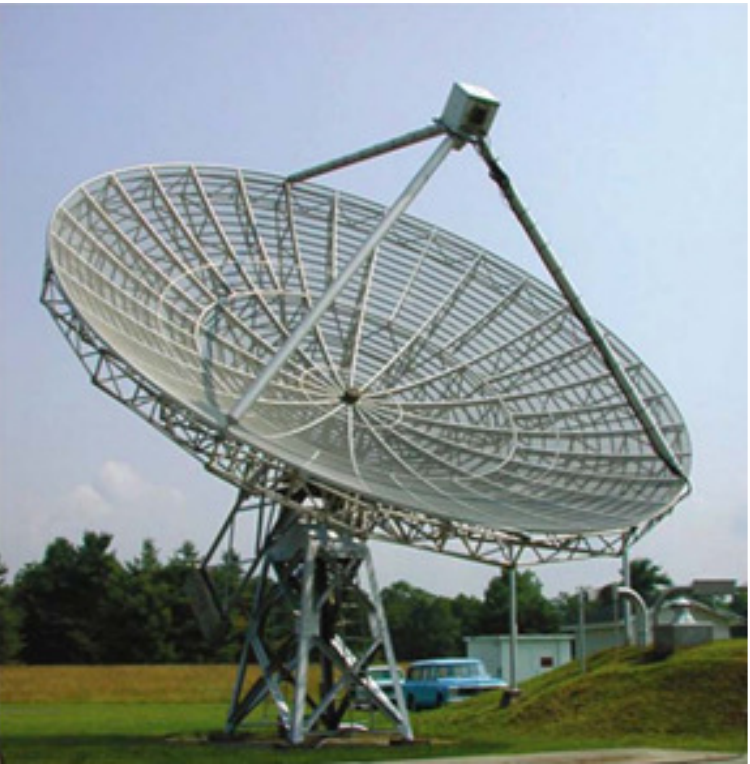}
\caption{Green Bank Observatory 40-foot diameter radio telescope.  (Photo credit:  GBO)}
\end{figure}

Emerson \& Grave (1988) Fourier transform the data, collapsing en-route drift to a single band of near-zero spatial frequencies, perpendicular to the scan direction.  They then mask it in Fourier space and transform back.  Ideally, two orthogonal mappings are combined (in Fourier space), so real spatial frequencies that are masked in one image can be recovered from the other.  The primary advantage of this approach is that it does not assume that en-route drift can be well-modeled by a low-order polynomial over the length of each scan.  However, drawbacks are:  (1)~We will not always have two orthogonal mappings; (2)~Even if we did, OTF rasters do not populate rectangular grids, requiring the data to be regridded before processing (however, \textsection1.2.2); and (3)~This technique would not work well, or at all, with other mapping patterns (e.g., noddings, daisies, etc.; \textsection1.2.1).

Haslam, Quigley \& Salter (1970), Haslam et al.\@ (1974), Seiber, Haslam \& Salter (1979), and Haslam et al.\@ (1981) introduced a technique called basket-weaving, involving two mappings with intersecting scans (not necessarily orthogonal).  Signal differences at the intersections are minimized, but again assuming that en-route drift can be well-modeled by a low-order polynomial over the length of each scan.  The procedure also requires iteration.  Winkel, Floer \& Krauss (2012) introduced an updated version that does not require iteration, but:  (1)~It does require regridding (\textsection1.2.2); (2)~It works only with two, near-orthogonal mappings, so not with single mappings, and not with noddings, daisies, etc.\@ (\textsection1.2.1); and (3)~Although it does permit an arbitrarily high-order parameterization of the en-route drift over the length of each scan, in practice it becomes less effective as the number of parameters times the number of scans approaches the number of regridded pixels (which is typically significantly less than the original number of signal measurements).  This limited them to second-order polynomials in the examples that they presented.  

\textbf{RFI:}  RFI is typically localized to specific frequencies.  If spectral information is available, these frequencies can be identified and masked, or the excess signal at these frequencies can be measured and subtracted off (see \textsection3.1 of Paper II, Dutton et al.\@ 2018).  However, sometimes only continuum data are available (e.g., see \textsection2.2), or spectral data are available but the RFI has a continuous spectrum (e.g., lightning).  In these cases, RFI can be identified only from its temporal signature.  If prolonged, it might be indistinguishable from en-route drift.  If localized in time, it is unlikely to occur at the same position on adjoining scans, appearing as a source that is narrower than the telescope's beamwidth, at least across scans.  In most of the above references, RFI was identified by eye and excised by hand.  However, given the volume, and diversity, of users that we have with the 20-meter, we need to be able to do this automatically.

\textbf{Elevation-Dependent Signal:}  Atmospheric emission increases as elevation decreases.  And if the dish is over-illuminated, terrestrial spillover increases as elevation increases.  If interested only in small-scale structures, these backgrounds can be removed with en-route drift (see \textsection3.3).  However, if large-scale structures are to be retained (or added back in), these backgrounds need to be removed separately (see \textsection2.3 of Paper II, Dutton et al.\@ 2018).

\begin{figure*}
\plotone{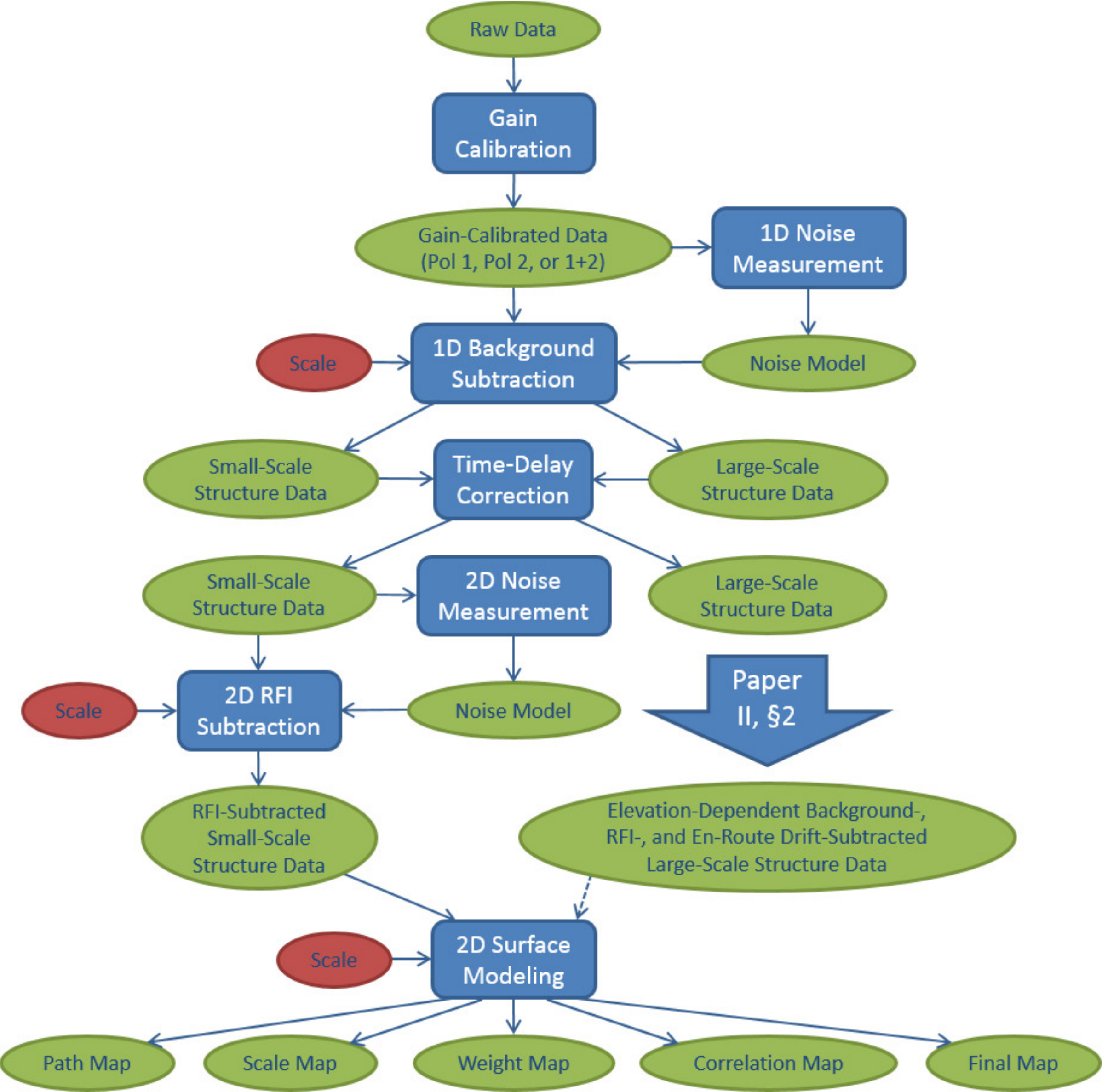}
\caption{Flowchart of our algorithm for contaminant-cleaning and mapping small-scale structures.  Blue are the component algorithms.  Green are the inputs to and outputs of these component algorithms, consisting of data, corresponding noise models, and ultimately maps.  Red are user-selected scales, for separating wanted and unwanted structures, and for modeling the final surface.}
\end{figure*}

\begin{figure}
\plotone{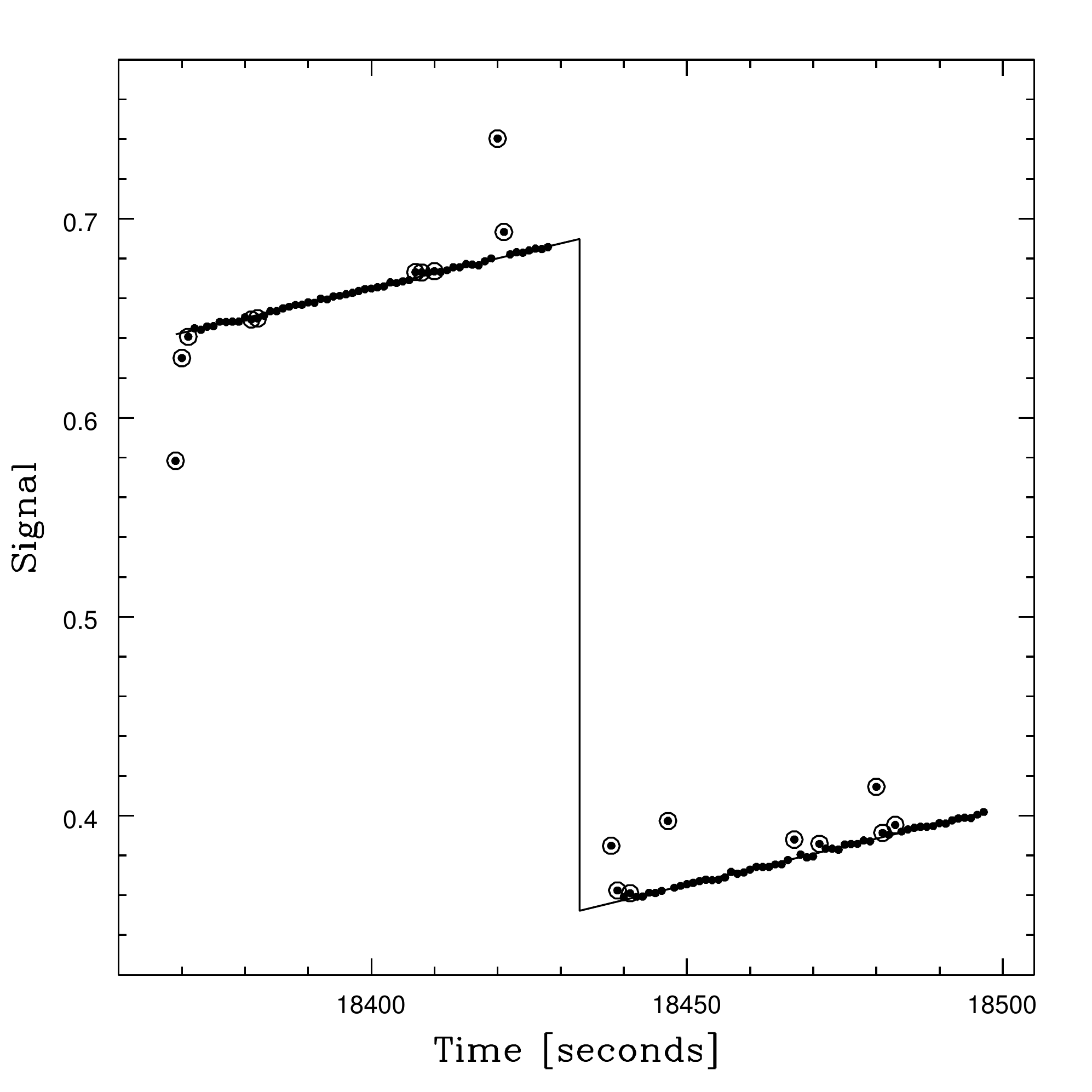}
\caption{40-foot gain calibration data, with the noise diode first on and then off, and best-fit model.  Circled points have been robust-Chauvenet rejected, including data taken during the transitions from off to on and on to off, and RFI-contaminated data.  The background level increased during the calibration, but our model accounts for this:  Simply averaging each level, instead of modeling each level with a line, would have underestimated the result.}
\end{figure}

\begin{figure}
\plotone{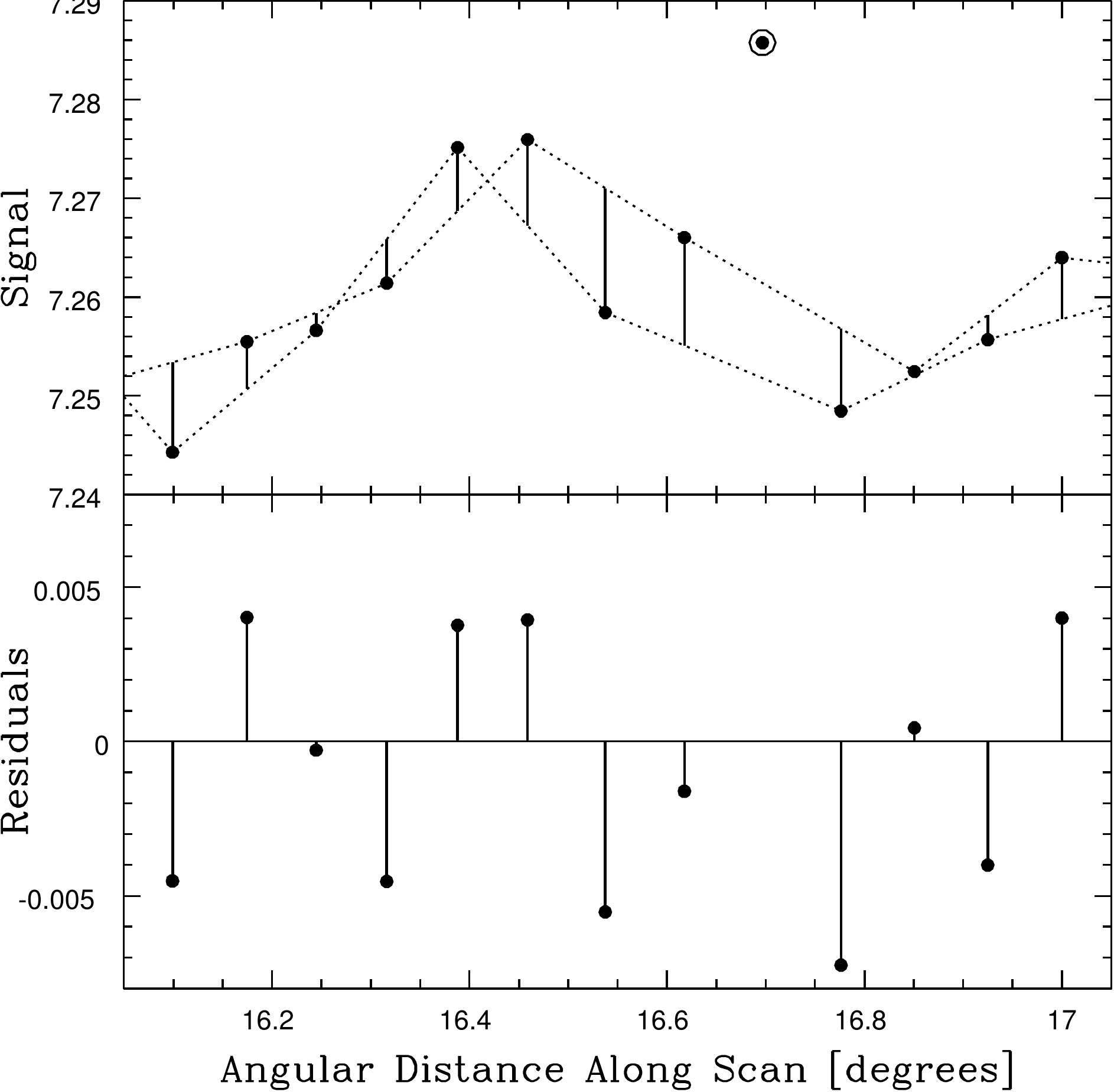}
\caption{Point-to-point noise measurement technique.  \textbf{Top:}  Applied to gain-calibrated 20-meter data.  The circled point has been robust-Chauvenet rejected.  \textbf{Bottom:}  Deviations.  Mean and standard deviations are measured from the non-rejected points, for each scan.}
\end{figure}

\begin{figure}
\plotone{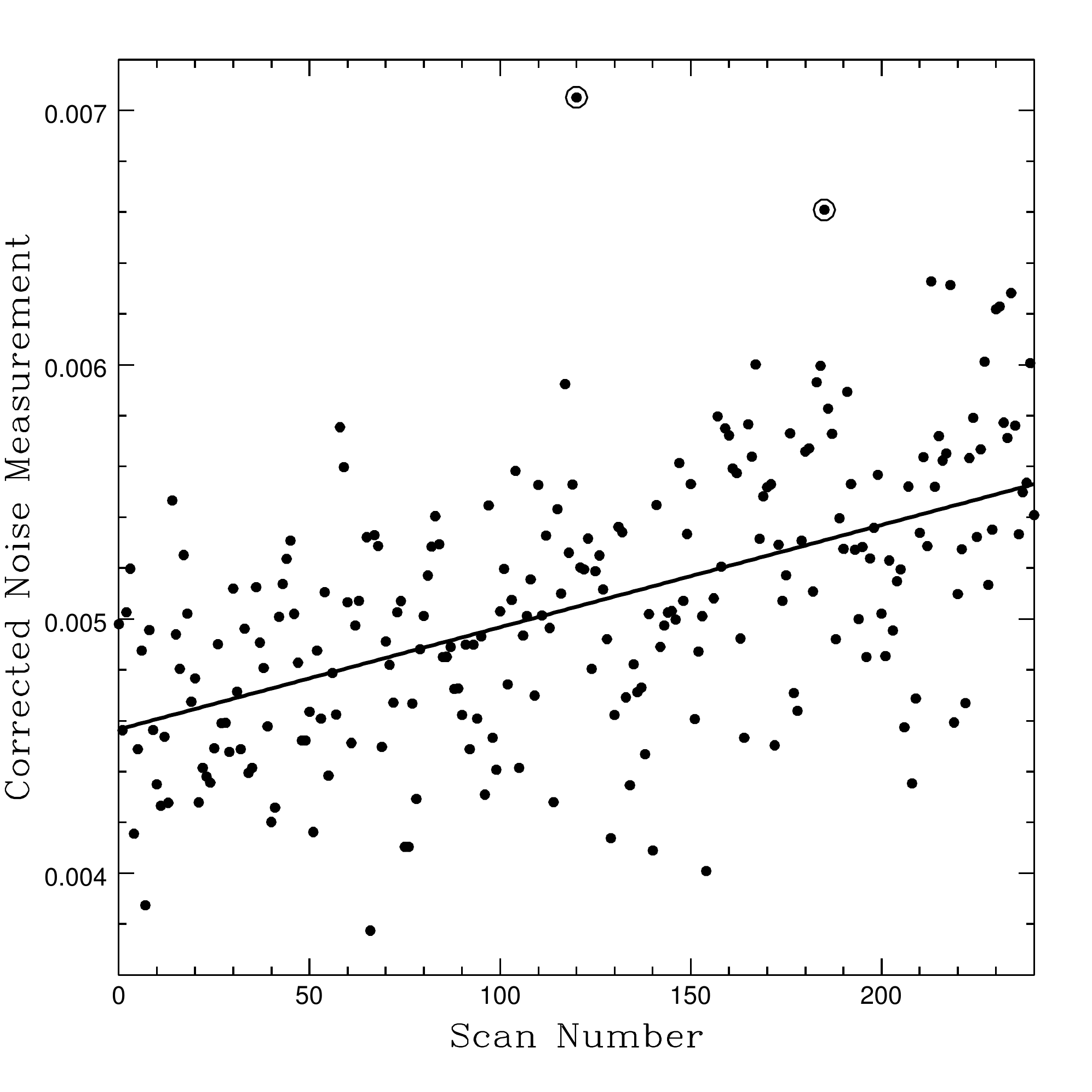}
\caption{Corrected 1D noise measurements vs.\@ scan number for a 20-meter observation, and best-fit model.  Only two points met the robust Chauvenet rejection criterion, and then only barely, which is not unusual given that intra-scan outliers have already been rejected (Figure~9).  The 1D noise level increased by $\approx$20\% over the course of this observation.}
\end{figure}

\begin{deluxetable}{ccc}
\tablewidth{0pt}
\tablecaption{Minimum Recommended 1D Background-Subtraction Scale for the Telescopes and Receivers of \textsection2, in Theoretical Beamwidths}
\tablehead{
\colhead{Telescope}     & \colhead{Receiver}    & \colhead{Scale}}
\startdata
20-meter & L (HI $+$ OH)\tablenotemark{a} & 7\tablenotemark{c} \\
20-meter & L (HI)\tablenotemark{b} & 6\tablenotemark{c} \\
20-meter & L (OH)\tablenotemark{b} & 6\tablenotemark{c} \\
20-meter & X & 6\tablenotemark{c} \\
40-foot & L (HI) & 3 
\enddata
\tablenotetext{a}{Before August 1, 2014}
\tablenotetext{b}{After August 1, 2014}
\tablenotetext{c}{The 20-meter's beam pattern has a low-level, broad component, in both L and X bands, and consequently, we recommend larger background-subtraction scales here.  This component was significant in L band prior to 8/1/14, as can be seen in the third row of Figure~3, as well as in Figure~5.  Post 8/1/14, it was significantly reduced, but not altogether eliminated.  This component corresponds to approximately 2\% -- 3\% and 4\% -- 5\% of the integrated beam pattern in L and X band, respectively.  If this is not a concern, these minimum recommended 1D background-subtraction scales can be lowered to 3 and 4 theoretical beamwidths, respectively.}
\end{deluxetable}

\begin{figure}
\plotone{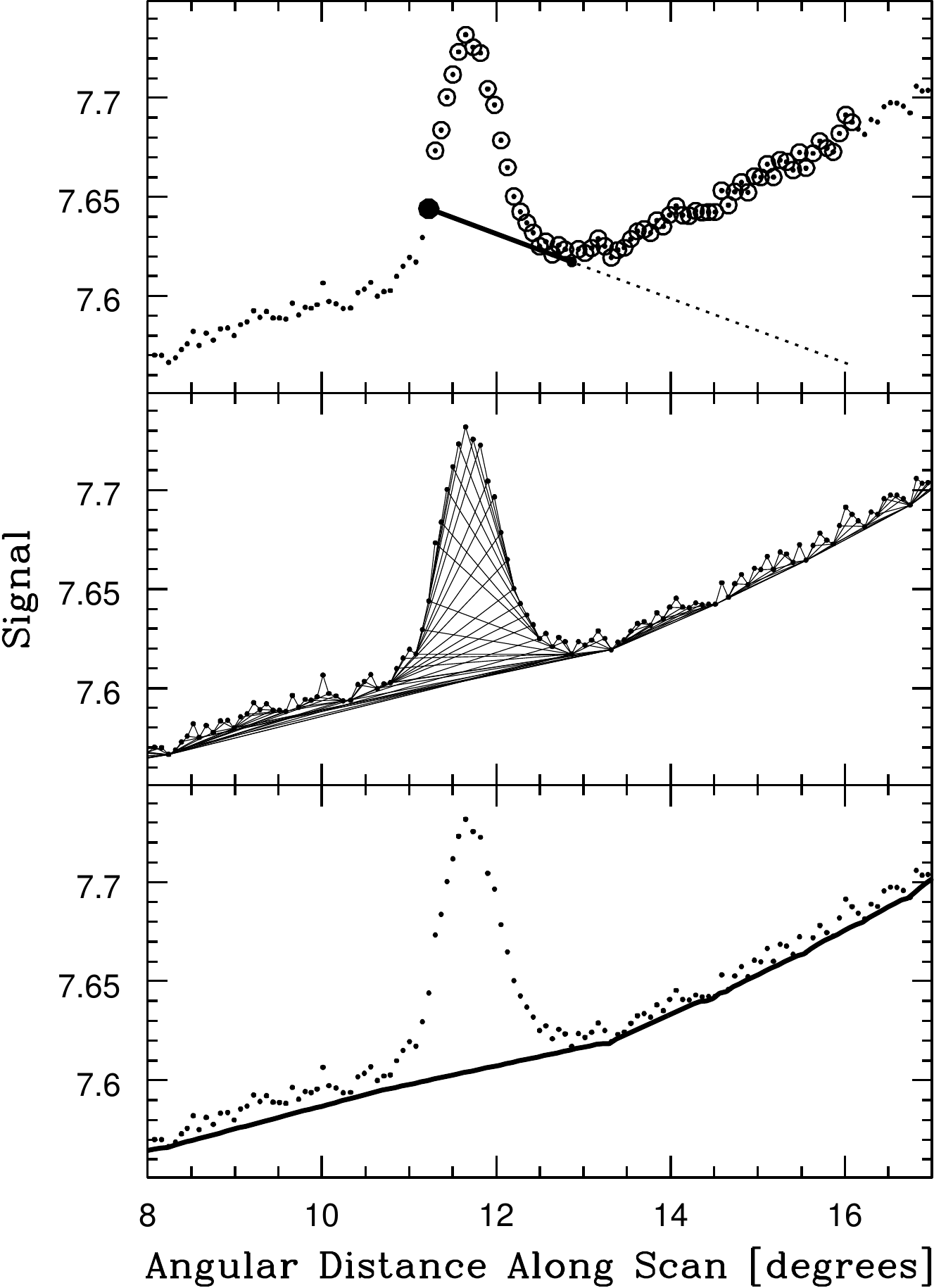}
\caption{\textbf{Top:}  A forward-directed local background model, anchored to an arbitrary point from Figure~5, near 3C~270.  Circled points are within one background-subtraction scale length, but above the model.  \textbf{Middle:}  Forward- and backward-directed local background models, anchored to every point in the scan.  \textbf{Bottom:}  Global background model, constructed from the local background models.}
\end{figure}

\begin{figure}
\plotone{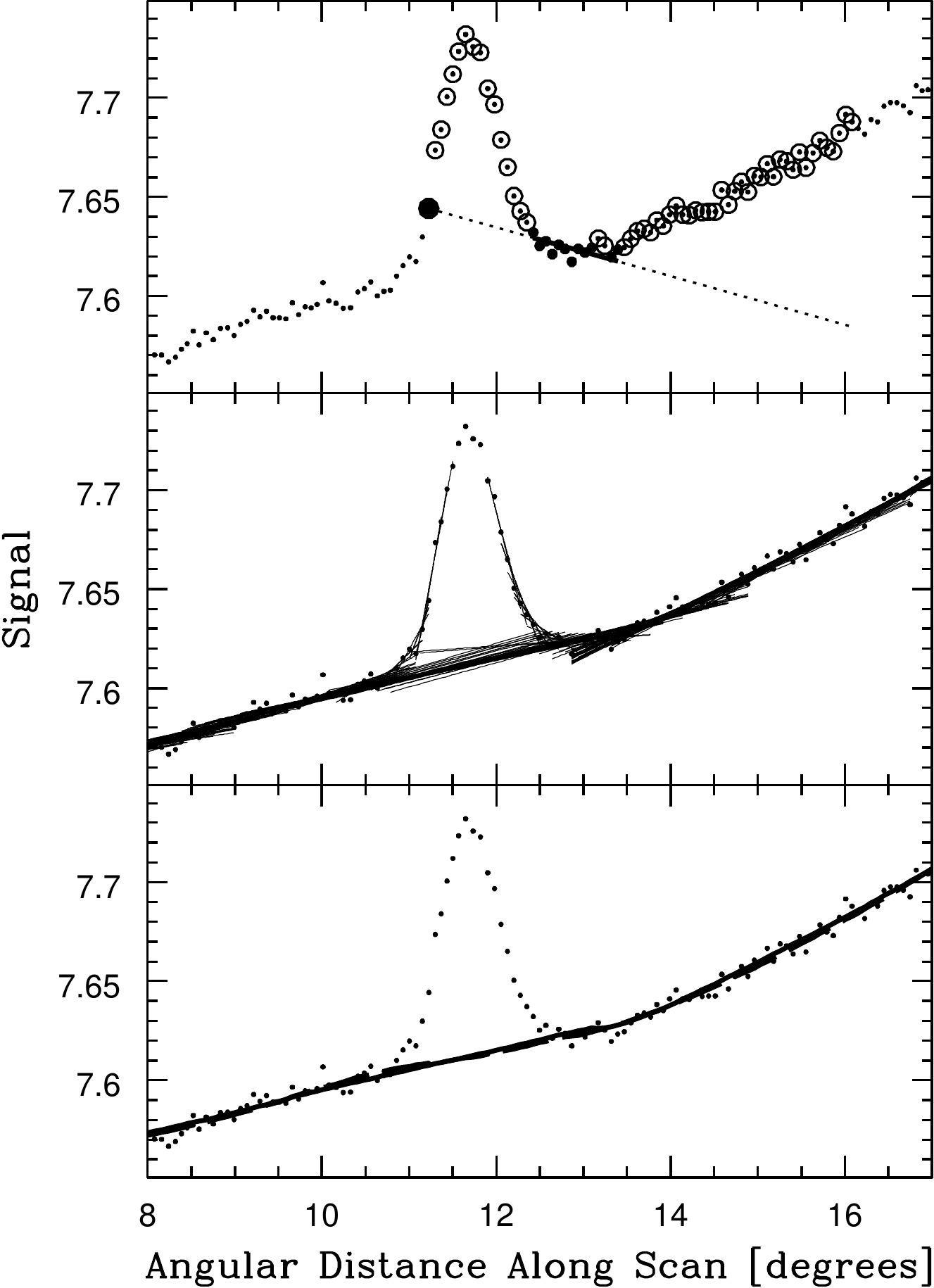}
\caption{\textbf{Top:}  A preliminary local background model, anchored to the same point as in Figure~11.  Circled points have been iteratively rejected as too high, given the modeled noise level (\textsection3.2); the larger points were not rejected.  \textbf{Middle:}  Final local background models, originally, but no longer, anchored to every point in the scan.  \textbf{Bottom:}  Global background model, constructed from the final local background models (solid curve), and from quadratic, instead of linear, local background models (dashed curve).}
\end{figure}

We approach all three types of signal contaminants in the same way.  First, we model them locally, not globally:  We use simple parameterizations, such as first- and second-order polynomials, but we do not expect, nor need, them to hold over large angular scales.  When fitting these models, we use robust, and appropriate, outlier rejection (see \textsection1.3) to separate contaminants from astronomical signal.  Finally, we combine our locally-fitted models into a global model, checking the procedure against simulated data.

\subsection{Robust Chauvenet Outlier Rejection}

Our single-dish mapping algorithm is built upon a new outlier rejection technique, called ``robust Chauvenet rejection'', which we developed for the Skynet Robotic Telescope Network's image-processing library in general, and for this application in particular.  

Sigma clipping (\textsection1.2.3) is one of the simplest outlier rejection techniques, but also one of the crudest.  It suffers from a number of problems, foremost of which is how to set the threshold.  For example, if working with $\approx$100 data points, 2-sigma variations are expected but 4-sigma variations are not.  However, if working with $\approx$10$^4$ data points, 3-sigma variations are expected but 5-sigma variations are not.  Chauvenet rejection is simply sigma clipping plus a reasonable rule for setting the threshold:
\begin{equation}
NP(\rm{>}|z|) = 0.5,
\end{equation}
\noindent where $N$ is the total number of data points and $P(\rm{>}|z|)$ is the cumulative probability of being more than $z$ standard deviations from the mean, assuming a Gaussian distribution (Chauvenet 1863).  

However, sigma clipping, and by extension Chauvenet rejection, also suffers from the following problem:  If the mean and the standard deviation are not known a priori, which is almost always the case, they must be measured from the data, and both of these quantities are sensitive to the very outliers that they are being used to reject.  This limits the applicability of Chauvenet rejection to very small sample-contamination fractions.

Consequently, we developed \textit{robust} Chauvenet rejection, which makes use of the mean and the standard deviation, but also makes use of increasingly robust (but decreasingly precise) alternatives, namely the median and the (half-sample; e.g., Bickel \& Fruhwith 2005) mode, and the 68.3-percentile deviation, measured in three increasingly robust ways.  These quantities approximate the mean and the standard deviation, respectively, and equal them in the case of a Gaussian distribution.  But they are significantly less sensitive to outliers, and are particularly effective, meaning both robust and precise, if applied in proper combination.  

The applicability of this technique is very broad, spanning not only astronomy and science, but all quantitative disciplines.  We have submitted this technique as a companion paper (Maples et al.\@ {\color{black}2018}), and make extensive use of it in this paper, with implementation details offered in the footnotes.  As such, this paper additionally serves the purpose of ``field testing'' this new technique.

\begin{figure}
\plotone{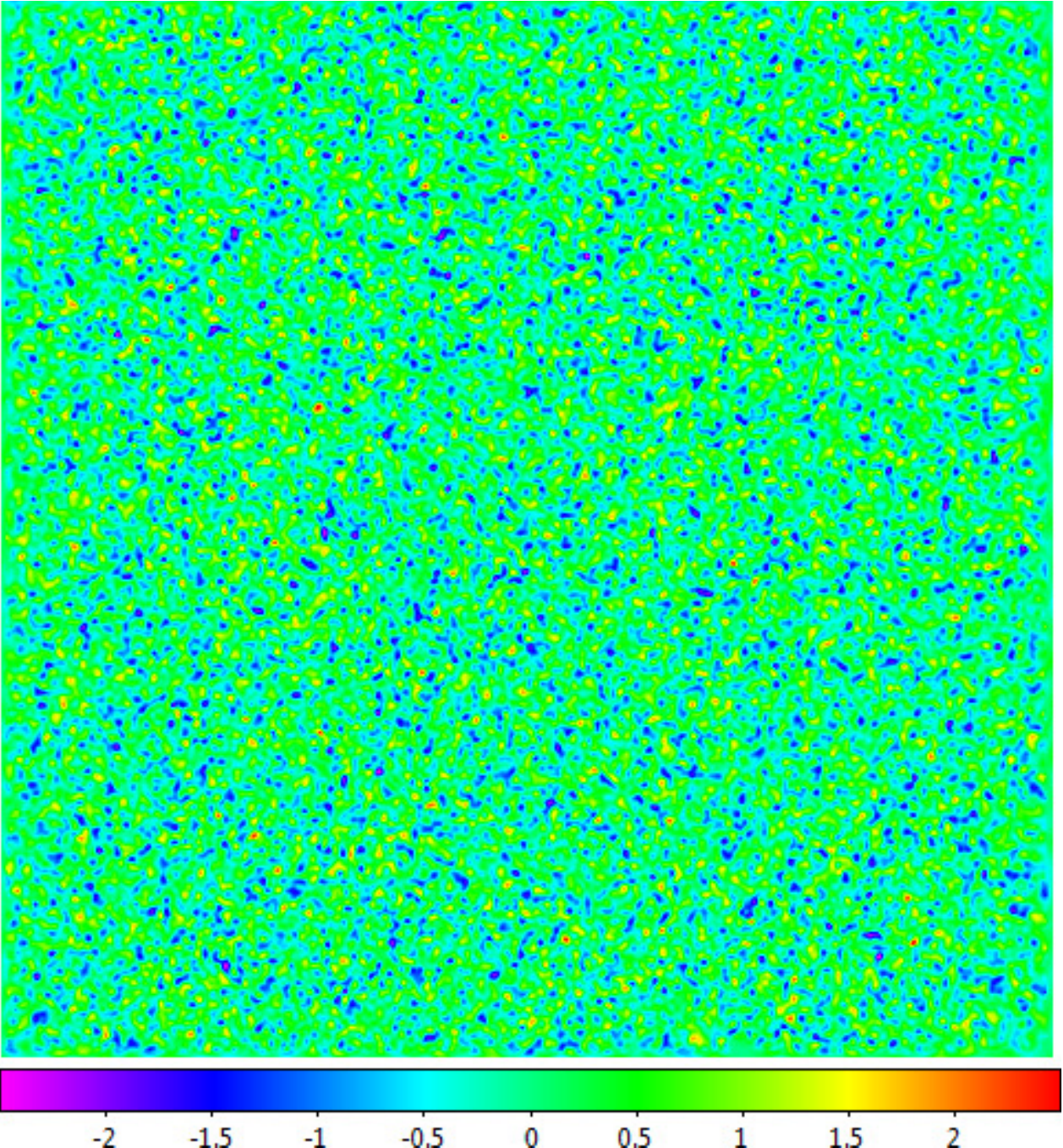}
\caption{20-meter 1/10-beamwidth horizontal raster replaced with Gaussian random noise, of mean zero and standard deviation one.  Locally modeled surface (\textsection1.2.1, see \textsection3.7) has been applied for visualization only.}
\end{figure}

\begin{figure*}
\plotone{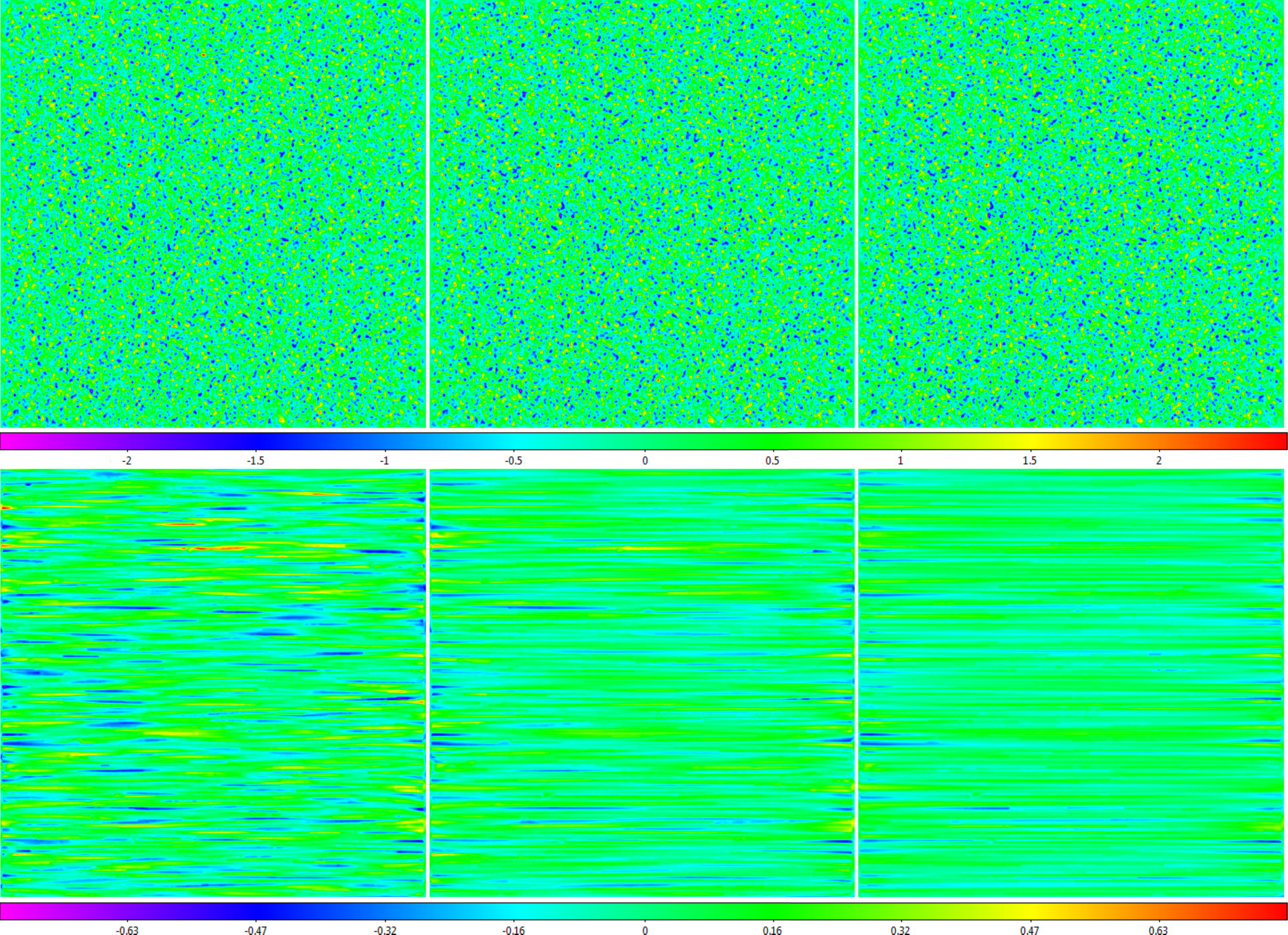}
\caption{\textbf{Top Row:}  Data from Figure~13 background-subtracted, with 6- (left), 12- (middle), and 24- (right) beamwidth scales (the map is 24 beamwidths across).  \textbf{Bottom Row:}  Data from the top row minus the data from Figure~13 (residuals).  Background-subtracted data are not biased high nor low.  To first order, the noise level of the background-subtracted data is $\approx$98.0\% (left), $\approx$98.8\% (middle), and $\approx$99.3\% (right) of that of the original data, and the RMS of the residuals is only $\approx$20.1\% (left), $\approx$15.4\% (middle), and $\approx$12.3\% (right) of the noise level of the original data (see Figure~15 for second-order effect).  Locally modeled surfaces (\textsection1.2.1, see \textsection3.7) have been applied for visualization only.}
\end{figure*}

\begin{figure}
\plotone{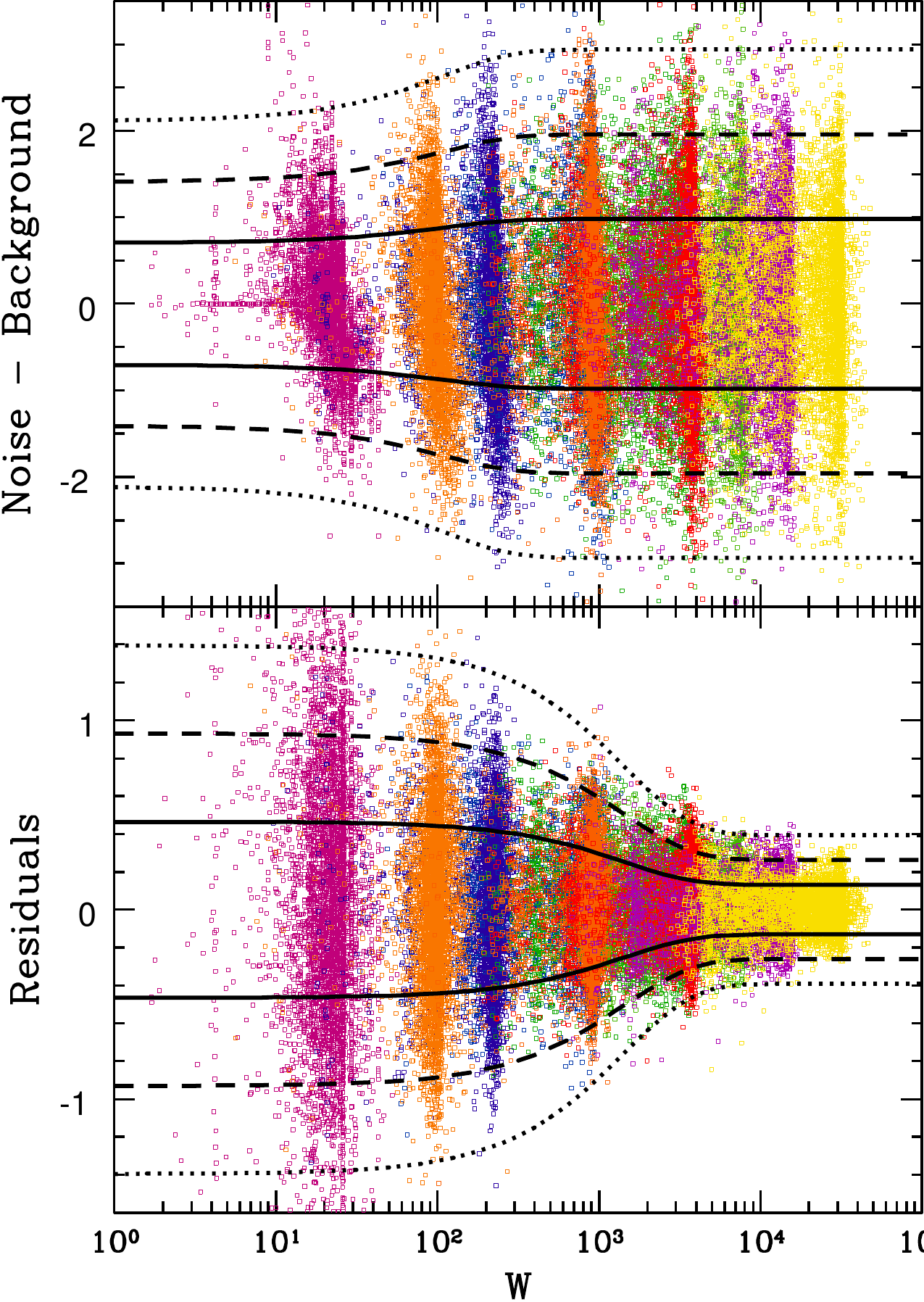}
\caption{Gaussian random noise background-subtracted (top) and residuals (bottom) vs.\@ $W$, the sum of the weights of the non-rejected local background models that determine the global background model at each point, for 1/10- and 1/5-beamwidth rasters, and for 1-, 3-, 6-, 12-, and 24-beamwidth background-subtraction scales (background-subtraction scale times sampling density of the data sets increases from left to right).  The RMS of the data varies with $W$, but not with background-subtraction scale or sampling density independently.  Curves are 1-, 2-, and 3-$\sigma$ model noise envelopes that have been fitted to all of these data simultaneously (Equations 6 and 7).}
\end{figure}

\begin{figure}
\plotone{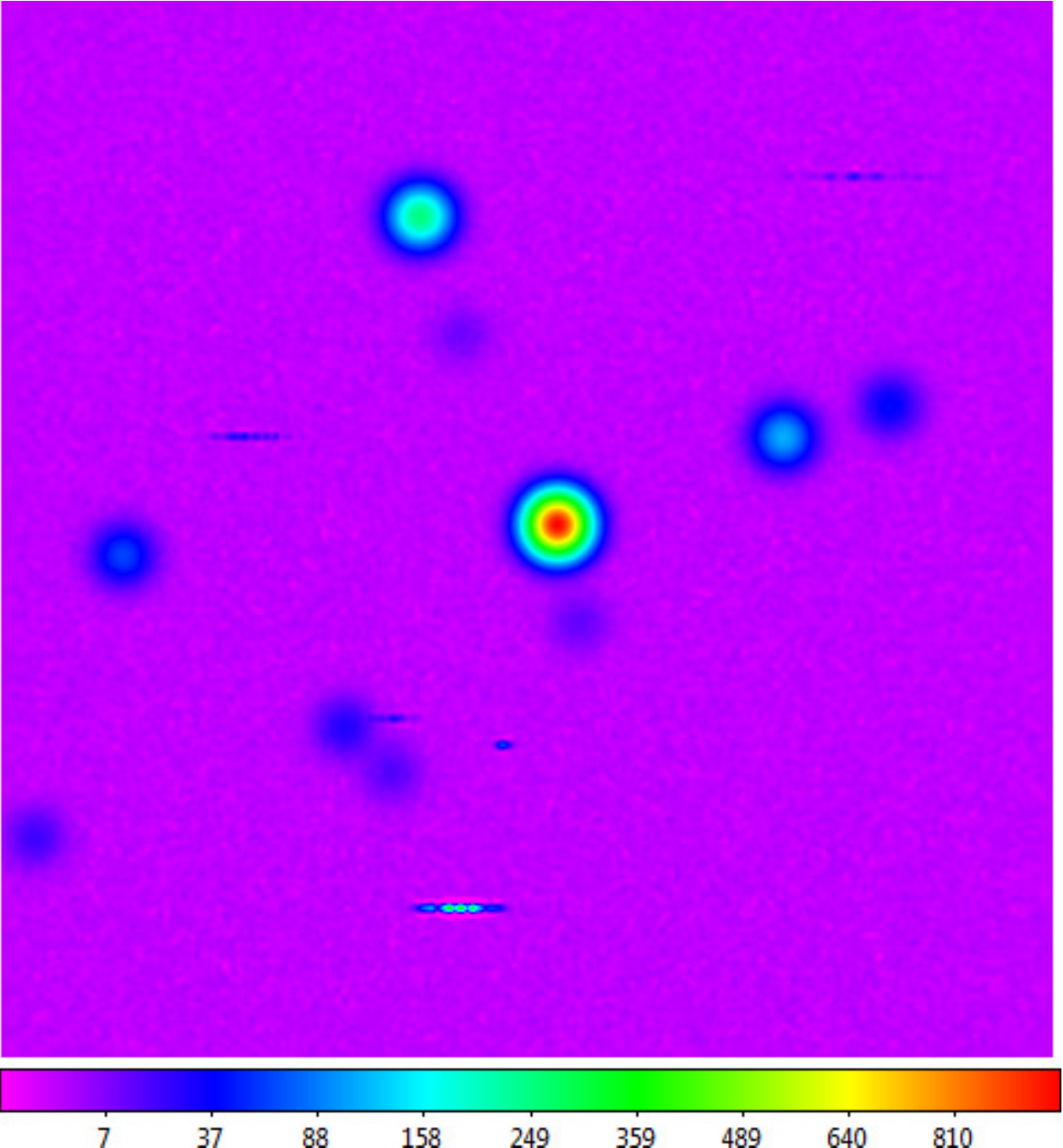}
\caption{Simulated data from Figure~13 to which we have added point sources and short-duration RFI.  For the point sources, we use a Gaussian beam pattern.  For the short-duration RFI, we use the absolute value of a sum of rapidly varying sine functions multiplied by a short-duration Gaussian envelope function.  Locally modeled surface (§\textsection1.2.1, see \textsection3.7) has been applied for visualization only.  Square-root scaling is used to emphasize fainter structures.}
\end{figure}

\begin{figure*}
\epsscale{0.95}
\plotone{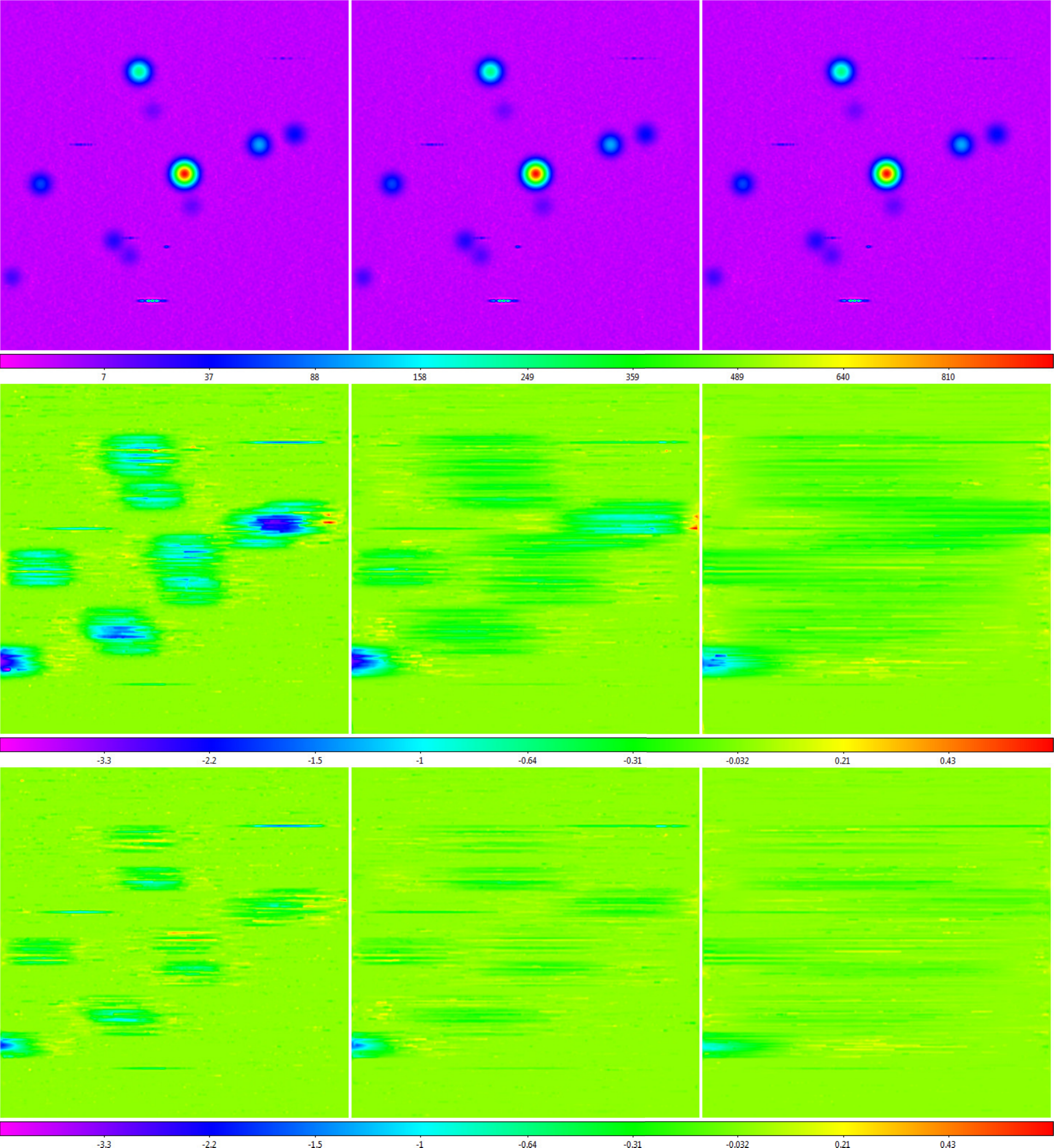}
\caption{\textbf{Top Row:}  Data from Figure~16 background-subtracted, with 6- (left), 12- (middle), and 24- (right) beamwidth scales (the map is 24 beamwidths across).  \textbf{Middle Row:}  Data from the top row (1)~minus the data from Figure~16 (residuals) and (2)~minus the Gaussian random noise residuals from the bottom row of Figure~14 (for greater clarity).  Small-scale structure residuals are biased negative, but typically by at most $\approx$1/2 -- 1 (left), $\approx$1/4 -- 1/2 (middle), and $\approx$1/8 -- 1/4 (right) of the noise level, and independently of the brightness of the proximal small-scale structure (point source or short-duration RFI).  Larger values are possible when small-scale structures blend together into large-scale structures, in the scan direction, where the division between small and large scales is given by the background-subtraction scale.  Larger values are also possible when small-scale structures occur near the ends of scans.  Noise-level biases can be ignored for all but the lowest-S/N sources (see \textsection4), and are further mitigated by our RFI-subtraction algorithm in \textsection3.6.2, and by our large-scale structure algorithm in Paper II.  \textbf{Bottom Row:}  Same as the middle row, but for more-realistic, less-winged sources (given by Equation~9 with $\theta_{RFI} = 1$ beamwidth and $z_0 = 0$; see Figure~31); residuals are $\approx$2 -- 3 times smaller in this case.  Locally modeled surfaces (\textsection1.2.1, see \textsection3.7) have been applied for visualization only.  Square-root and hyperbolic-arcsine scalings are used in the top and bottom two rows, respectively, to emphasize fainter structures.}
\end{figure*}

\begin{figure}
\plotone{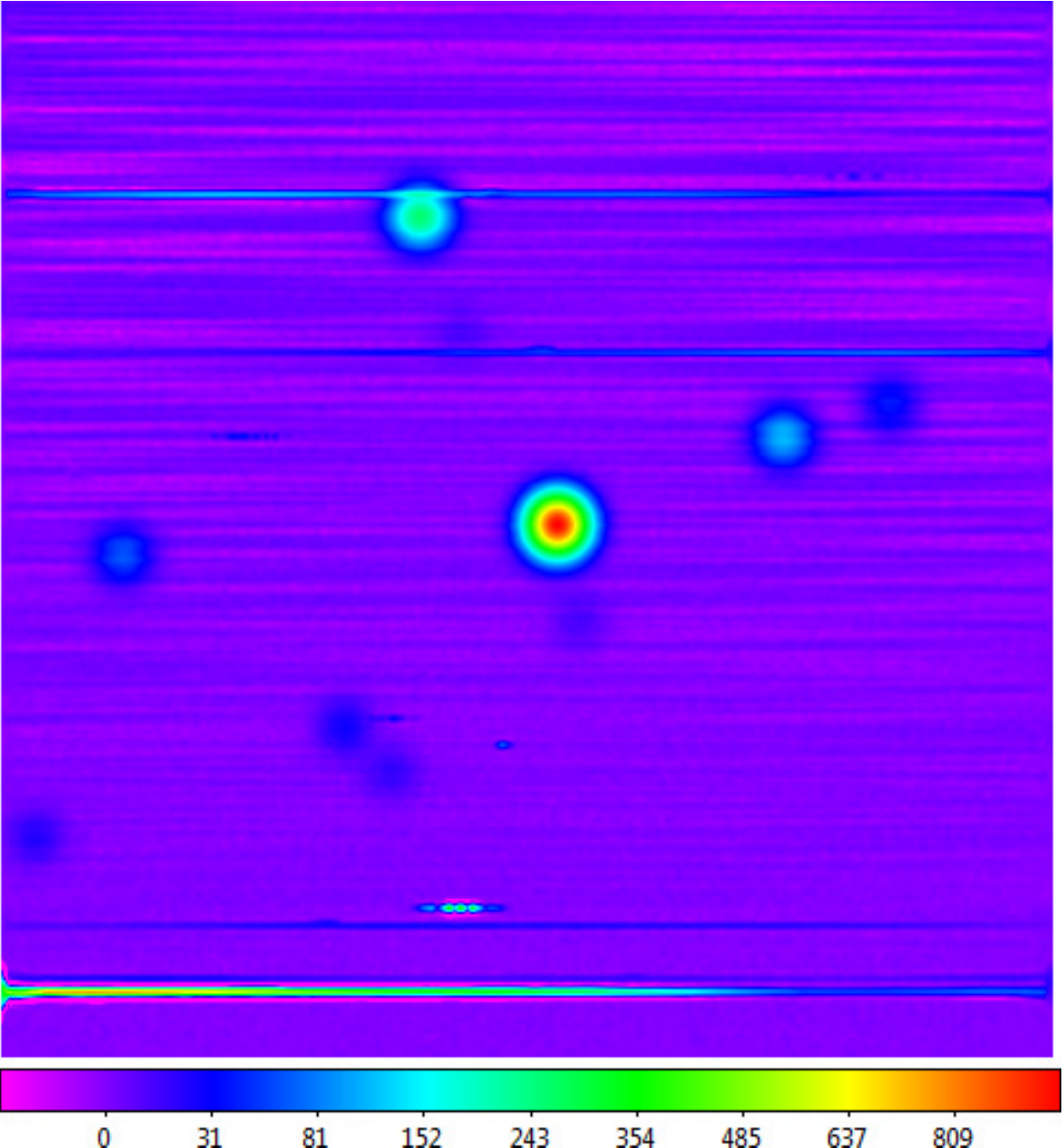}
\caption{Simulated data from Figure~16 to which we have added en-route drift and long-duration RFI.  For the en-route drift, we use a sum of randomly phased sine functions, the shortest period of which is 12 beamwidths.  We linearly increase the maximum amplitude of the en-route drift from zero times the noise level at the bottom of the image to 12 times the noise level at the top of the image.  For the long-duration RFI, we use a similarly constructed sum of sine functions, but plus a constant to ensure that it is always positive, and multiplied by a long-duration Gaussian envelope.  The long-duration RFI is significantly brighter than the en-route drift.  Locally modeled surface (\textsection1.2.1, see \textsection3.7) has been applied for visualization only.  Square-root scaling is used to emphasize fainter structures.}
\end{figure}

\begin{figure*}
\plotone{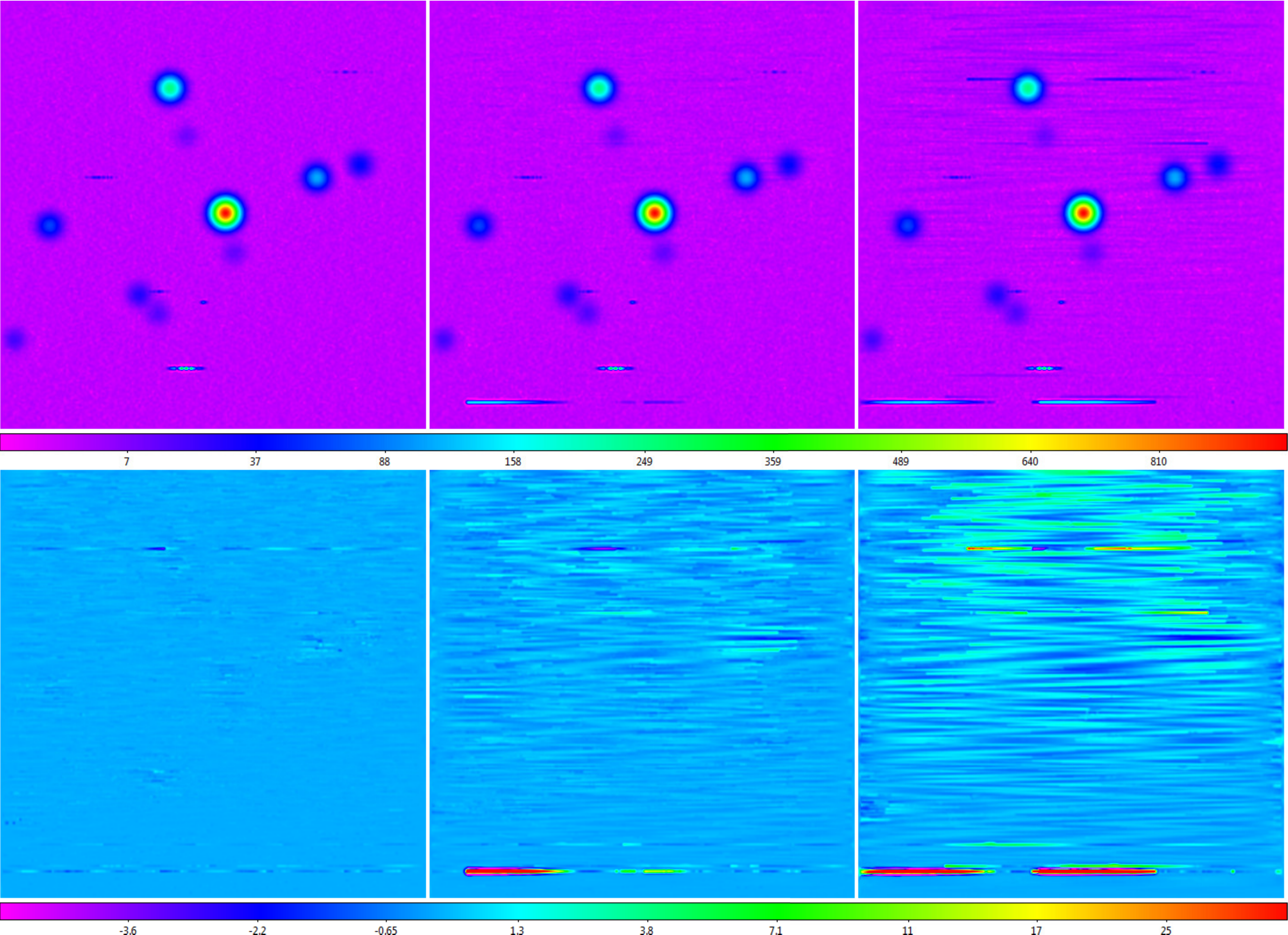}
\caption{\textbf{Top Row:}  Data from Figure~18 background-subtracted, with 6- (left), 12- (middle), and 24- (right) beamwidth scales (the map is 24 beamwidths across).  \textbf{Bottom Row:}  Data from the top row (1)~minus the data from Figure~16 (residuals) and (2)~minus the Gaussian random noise residuals from the bottom row of Figure~14, and the small-scale structure residuals from the middle row of Figure~17 (for greater clarity).  En-route drift and long-duration RFI are not eliminated, but are significantly reduced, especially in the smaller background-subtraction scale maps (see Figure~20).  These gains are furthered, and again significantly, by our RFI-subtraction algorithm in \textsection3.6.3.  Locally modeled surfaces (\textsection1.2.1, see \textsection3.7) have been applied for visualization only.  Square-root and hyperbolic-arcsine scalings are used in the top and bottom rows, respectively, to emphasize fainter structures.}
\end{figure*}

\begin{figure*}
\plotone{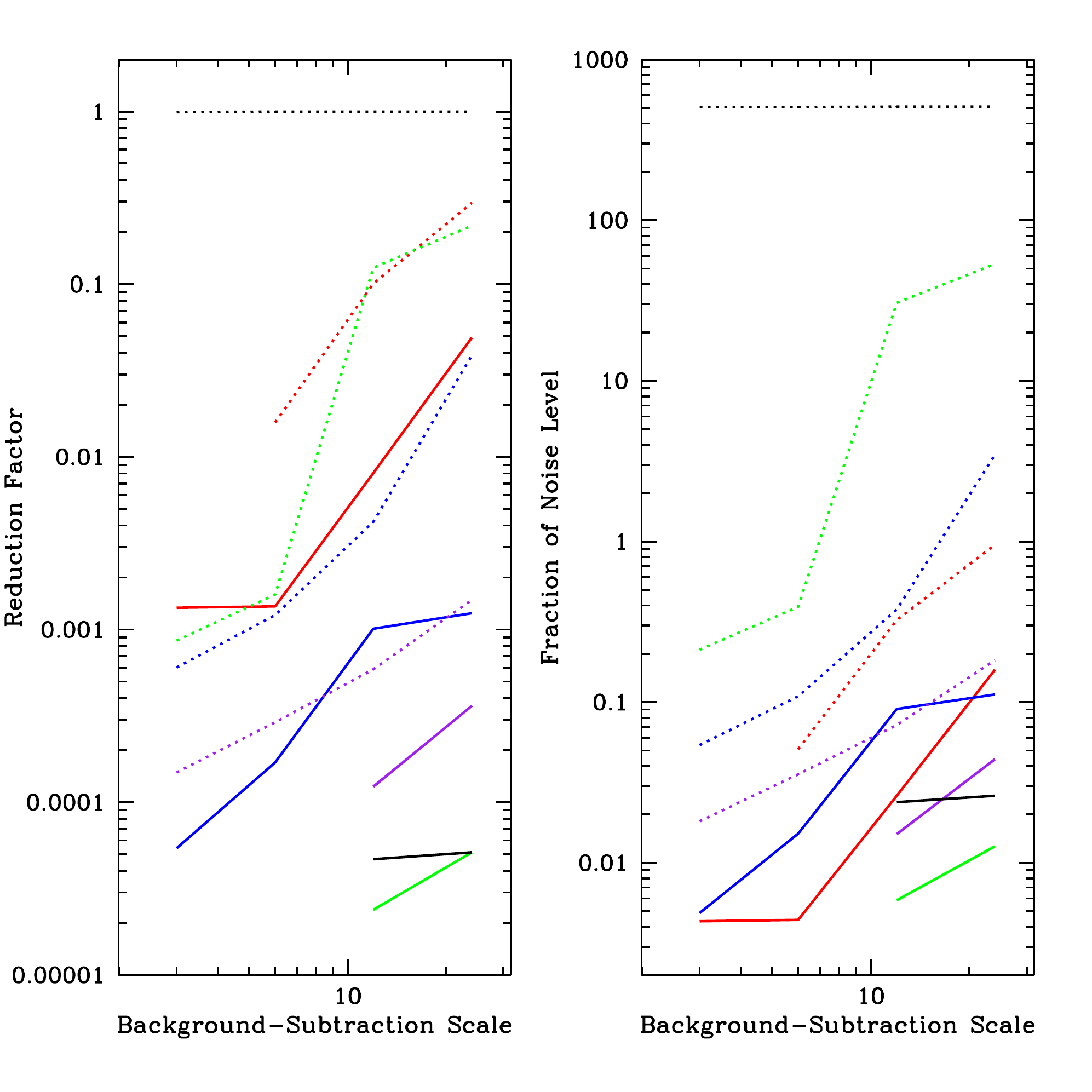}
\caption{\textbf{Left:}  Factor by which background subtraction reduces en-route drift (red), long-duration RFI (green), large-scale astronomical signal (blue; see \textsection3.3.4), elevation-dependent signal (purple; see \textsection3.3.4), and short-duration RFI (black; see \textsection3.6.2) in our simulated data, for background subtraction scales of 3, 6, 12, and 24 beamwidths (dashed curves).  Factor by which background and RFI subtraction (see \textsection3.6) reduce these contaminants (solid curves).  \textbf{Right:}  Fraction of the noise level to which these contaminants are reduced.  If nothing is plotted, the contaminant is completely eliminated on this scale.  For these measurements, each contaminant was simulated separately, and in the absence of sources.}
\end{figure*}

\subsection{Overview}

In \textsection2, we provide a technical description of the refurbished 20-meter, including its L- and X-band receivers.  We also provide a technical description of Green Bank Observatory's 40-foot telescope, used primarily for education and public engagement, but on which we developed many components of our algorithm before the 20-meter was ready.  In \textsection3, we introduce our algorithm for contaminant-cleaning and mapping small-scale structures (e.g., sources) from continuum observations, and test it on both simulated and real data.  In \textsection4, we demonstrate that optical-style, aperture photometry can be carried out on these maps, and introduce an algorithm for calculating photometric error bars, given the different (and more complicated, correlated) noise characteristics of these maps.  We summarize our results in \textsection5.  

In Paper II (Dutton et al.\@ 2018):  (1)~We expand on our small-scale structure algorithm to additionally contaminant-clean and map large-scale structures; (2)~We do the same for spectral (as opposed to just continuum) observations; and (3)~We carry out an X-band survey of the Galactic plane, from $-5\degr < l < 95\degr$, to showcase, and further test, techniques developed in both papers.\\
\\
\\

\section{Telescopes}

\subsection{Green Bank Observatory 20-Meter}

The 20-meter diameter telescope (Figure~1) was constructed by Radiation Systems, Inc. at the Green Bank site, completed in 1994, and outfitted with a dual-frequency 2.3/8.4-GHz feed and receiver (Norrod et al.\@ 1994).  Funded by the U.S.\@ Naval Observatory (USNO), it was part of a network of Very-Long Baseline Interferometry (VLBI) stations that monitored changes in Earth's rotation axis and rate, as well as measured continental drift (Ma et al.\@ 1998).

The antenna has an altitude/azimuth mount, and is able to observe down to 1 degree above the horizon.  Its maximum slew speed is 2 degrees per second in each axis, which met the needs of the geodetic VLBI program, to quickly observe many radio sources, despite being widely distributed across the sky.  Its pointing accuracy is $\approx$35 arcsec rms.

The antenna's surface is a paraboloid with a focal ratio of 0.43 at prime focus.  Its surface accuracy was originally 0.8 mm rms, but this has degraded over the years to $\approx$1.4 mm rms.  Its aperture efficiency is $\approx$50\% at 8 GHz.

Due to budget cutbacks, USNO ended its Green Bank operations in 2000.  From 2000 to 2010, the telescope was used only occasionally, primarily for testing experimental receivers.

In 2010, the University of North Carolina at Chapel Hill (UNC-CH) received a grant from the National Science Foundation to expand its Skynet Robotic Telescope Network, and this included restoring and automating the 20-meter.  

Two receivers were prepared.  The primary receiver is for 1.3 -- 1.8 GHz (L band), and had previously been used on Green Bank's 140-foot diameter telescope.  We originally operated this receiver with three interchangeable feeds, each for a different part of the band.  However, changing these feeds was a many-hour, manual operation.  In 2014, we built and installed a new feed, for the entire 1.3 -- 1.8 GHz band.

The secondary receiver is for 8 -- 10 GHz (X band).  It is a stripped-down version of the original, dual-frequency receiver, outfitted with a new 8 -- 10 GHz feed.  This receiver is used primarily when the L-band receiver is off the telescope for maintenance, which has only been for two extended (many-month) periods since 2010.

A new backend was also prepared.  The spectrometer uses a Field-Programmable Gate Array (FPGA) design that was adapted from the Green Bank Ultimate Pulsar Processing Instrument (GUPPI).  In its low resolution mode, it produces 1024-channel spectra over a 500-MHz band.  In 2013, a high-resolution mode was added; in this mode, it produces 1024-, 2048-, 4096-, 8192-, or 16286-channel spectra in each of two 15.625-MHz bands (see \textsection3 of Paper II).

New software was also written for low-level control of (1) the antenna's pointing, and (2) data acquisition from the spectrometer, as well as for high-level control of the entire system through Skynet's web-based interface (Footnote~7).  For both its optical telescopes and its radio telescope, Skynet features a dynamically updated queue, with options for assigning different users, groups of users, and collaborations of groups different levels of priority access, including target-of-opportunity access.  

Beyond the initial grant, continued operation of the 20-meter has been made possible by a combination of education grants and research grants (primarily to search for fast radio bursts), as well as by operations funds and personnel of the Green Bank Observatory.  Since the 20-meter went online on Skynet in 2012, over 23,000 observations have been carried out, mostly by students, for both education and research objectives.

\subsection{Green Bank Observatory 40-Foot}

The 40-foot diameter telescope (see Figure~6) was obtained in 1961 to determine if several bright radio sources varied on daily timescales.

Purchased from Antenna Systems, Inc., this relatively inexpensive (\$40K) meridian-transit telescope was pre-fabricated in Hingham, Massachusetts, and assembled at the Green Bank site in two days.  

The reflecting surface is aluminum mesh, similar to that of Green Bank's former 300-foot diameter telescope, but with smaller openings.  The reflector backup structure is also aluminum, with a supporting pedestal of galvanized steel.  The telescope's maximum slew speed is 20 degrees per minute along the local meridian, and it can point far enough south to observe the Galactic center, and as far north as a few degrees beyond the celestial pole.

The original receivers and feeds were built onsite, and operated at 750 and 1400 MHz.  Backend electronics are housed in an adjacent, underground bunker, to prevent radio-frequency noise from the electronics from interfering with celestial signals.  

The original electronics automatically repointed the telescope to the declination of each source sufficiently before it transited the meridian each day.  Data were digitized and written out to punched paper tape.  Put into operation in 1962, the 40-foot is thought to have been the world's first fully automated telescope (Bowyer \& Meredith 1963; Lockman, Ghigo, \& Balser 2007).

This project ran for several years, but was abandoned because none of the sources varied on daily timescales.  Afterward, they were instead observed less frequently, but with higher precision, using Green Bank's 300- and 140-foot diameter telescopes.  

The 40-foot fell into disuse for about twenty years.  But in 1987, it was refurbished as an educational resource, for students from 5th grade through graduate school to learn basic observational radio astronomy.  The feed was replaced with one that had been used on Green Bank's 85-foot diameter Tatel telescope for Project OZMA, sensitive to 18 {\color{black}--} 21 cm.  A new receiver was built at the National Radio Astronomy Observatory's Central Development Laboratory in Charlottesville, Virginia.  Spectral-line capability was added using a synthesizer (which has since been upgraded, see below) and narrowband filters that were inherited from Green Bank's 300-foot diameter telescope after its collapse in 1988.

Over the past thirty years, several generations of students, numbering in the high thousands, have used the 40-foot as part of their science education.  For pedagogical reasons, observing with the 40-foot is now almost completely manual.  Students set analog dials and switches, and annotate and measure their data on a chart recorder.  When taking spectra, of the spin-flip transition of neutral hydrogen, students step through each frequency by hand.

That said, a few users have developed their own digitization and recording systems for the 40-foot, which they connect in parallel with the chart recorder.  Most notably, a group that is now based at UNC-CH, and now affiliated with the Skynet Robotic Telescope Network, did this in 1991, and has been collecting data, with large groups of students, for one week each summer since then.  We make use of these data throughout this paper, alongside higher-quality data that we have more recently collected with the 20-meter:  The 40-foot data are useful here specifically because they are lower quality, and consequently more rigorously test our algorithms.  

That said, a subset of these data have proven to be of sufficiently high quality to help constrain the fading history of the Cassiopeia~A supernova remnant (Reichart \& Stephens 2000; Trotter et al.\@ 2017).  And over the past 2 -- 3 years, coaxial cables from the front end have been replaced with fiber optics, and newer synthesizers, mixers, and amplifiers have been installed, making the system more reliable and robust, and eliminating some RFI.

\section{Mapping Small-Scale Structures with Continuum Observations}

In this section, we present our algorithm for contaminant-cleaning and mapping small-scale structures (e.g., sources) from continuum observations.  In \textsection3.1, we calibrate each of a telescope's polarization channels against, typically very small, variations in gain.  In \textsection3.2, we measure point-to-point variations in signal, which can be related to the noise level of the data along each scan, at least on the sampling timescale.  In \textsection3.3, we make use of this noise measurement to separate small-scale structures, both astronomical and short-duration RFI, from large-scale structures (astronomical, en-route drift, long-duration RFI, and elevation-dependent signal) along each scan.  In \textsection3.4, we cross-correlate consecutive scans to measure and correct for any time delay between coordinate and signal measurements.  In \textsection3.5, we measure scan-to-scan variations in signal, which can be related to the noise level of the data across each scan.  In \textsection3.6, we make use of this noise measurement to separate astronomical small-scale structures from RFI, both along and across scans.  In \textsection3.7, we present our surface-modeling algorithm, which interpolates between the contaminant-cleaned data, without blurring these data beyond instrumental resolution (\textsection1.2.2).  A flowchart of the entire algorithm can be found in Figure~7.

\subsection{Gain Calibration}

Signal measurements can be calibrated against variations in gain using a noise diode in the receiver.  By switching the diode on and off, preferably with the telescope tracking (not scanning) the sky, so astronomical signal does not also change, contemporaneous measurements can be calibrated by dividing by the difference, $\Delta$, between the on and off levels.  

This common practice, however, can be contaminated by outlying measurements, due to catching the diode in transition, due to RFI, etc.  Consequently, we apply robust Chauvenet rejection (\textsection1.3) when measuring each level.\footnote{We model each level with a line, in case the telescope is not tracking, or cannot track (as in the case of the 40-foot; see Figure~8).  We fit this model to the data, simultaneously rejecting outliers, as described in {\color{black}\textsection8} of Maples et al.\@ {\color{black}2018}, using iterative bulk rejection followed by iterative individual rejection (using the generalized mode $+$ broken-line deviation technique, followed by the generalized median $+$ 68.3\%-value deviation technique, followed by the generalized mean $+$ standard deviation technique), using the smaller of the low and high one-sided deviation measurements.  Data are weighted by the number of dumps that compose each measurement.  Each fitted line is evaluated at the dump-weighted mean time of the non-rejected diode-on and off measurements; $\Delta$ is their difference at this time.}  We demonstrate this with lower-quality 40-foot data in Figure~8.

In practice, both the 20-meter's and the 40-foot's gains vary negligibly over the timescale of an observation.  Consequently, we calibrate only at the beginning and end of observations, instead of more frequently, say, between scans.  Users may then select the first calibration, $\Delta_1$, the second calibration, $\Delta_2$, or a linear interpolation between them:
\begin{equation}
\Delta(t) = \Delta_1 + (\Delta_2-\Delta_1)\frac{t-t_1}{t_2-t_1}.
\end{equation}

We calibrate each polarization channel separately, but then process three maps, one for each polarization channel and one in which we {\color{black}average} the two channels' calibrated data first (after multiplying each by {\color{black}appropriate flux-density conversion factors}, if this information is available\footnote{{\color{black}To convert from gain-calibration units to Janskys per pixel, we will soon be implementing automatic, dense mappings of primary calibration sources Cyg~A, Tau~A, and Vir~A, multiple times per day.  These will be automatically processed (\textsection3) and photometered (see \textsection4), and used to determine time-stamped conversion factors, from gain-calibration units to Janskys per beam, using the flux-density calibrations of Trotter et al.\@ 2017.  For any observation, the most recently acquired conversion factors, one for each polarization channel, will be written into the raw data file's header, and will be converted to Janskys per pixel, and applied, when these data are processed into maps (since a map's pixel density is user-configurable; see \textsection3.7).  If for whatever reason the necessary conversion factor is, or one or both of the necessary conversion factors are, not available, we will leave the maps} in gain-calibration units, where one corresponds to the noise diode.  All maps in this paper are in gain-calibration units.}).

\subsection{1D Noise Measurement}

Key to the next section will be a measurement of the standard deviation of the now-calibrated data on the smallest scales available, along each scan.  We refer to this as the 1D ``noise'' level of the data, and we measure it directly from the data, using the following technique.

For each non-rejected point, we draw a line between the immediately preceding and proceeding non-rejected points, and measure the central point's deviation from this line (see Figure~9).  For each scan, we robust-Chauvenet reject the point with the most discrepant deviation, update the scan's deviations, and repeat until all outliers are rejected.\footnote{We reject outliers as described in {\color{black}\textsection4} and {\color{black}\textsection6} of Maples et al.\@ {\color{black}2018}, using iterative individual rejection (using the median $+$ broken-line deviation technique, followed by the median $+$ 68.3\%-value deviation technique, followed by the mean $+$ standard deviation technique), using two-sided deviation measurements.  Unlike in \textsection3.1, we do not precede this with iterative bulk rejection, because deviations change, and must be updated, after each rejection.  Deviations are weighted by $\left[N_n^{-1}+N_{line}^{-1}\left(x_n\right)\right]^{-1}$, where $N_n$ is the number of dumps that compose the central point and $N_{line}\left(x_n\right)$ is the corresponding weight of the line at $x_n$, the angular distance of the central point along the scan.  Here, $N_{line}^{-1}\left(x_n\right)=\frac{\left(x_{n+1}-\bar{x}\right)^2N_{n-1}^{-1}+\left(x_{n-1}-\bar{x}\right)^2N_{n+1}^{-1}}{\left(x_{n+1}-x_{n-1}\right)^2}+\frac{\left(N_{n+1}^{-1}+N_{n-1}^{-1}\right)}{\left(x_{n+1}-x_{n-1}\right)^2}\left(x_n-\bar{x}\right)^2$ ({\color{black}from} {\color{black}Equations 29 and 30} of Maples et al.\@ {\color{black}2018, and standard propagation of uncertainties}) and $\bar{x}=\frac{N_{n-1}x_{n-1}+N_{n+1}x_{n+1}}{N_{n-1}+N_{n+1}}$ ({\color{black}the weighted average; see \textsection8.3.5} of Maples et al.\@ {\color{black}2018}).}   Outliers are typically contaminated by residual signal from bright astronomical sources or RFI.  The post-rejection mean deviation is approximately zero and the post-rejection standard deviation is what we call the ``point-to-point'' noise measurement.

We calibrate this technique by applying it to simulated, Gaussian random noise, of known standard deviation.  We find that the point-to-point technique overestimates the noise's true standard deviation by 22.0\%.  We correct each scan's noise measurement accordingly.

Finally, we combine all of the scans' noise measurements into a single model for the entire observation, additionally allowing for a gradual change in the noise level over the course of the observation:  We fit a line to these data, again robust-Chauvenet rejecting outliers (if even necessary; Figure~10).\footnote{We fit this model to the data, simultaneously rejecting outliers, as described in {\color{black}\textsection8} of Maples et al.\@ {\color{black}2018}, using iterative bulk rejection followed by iterative individual rejection (using the generalized mode $+$ 68.3\%-value deviation technique, followed by the generalized median $+$ 68.3\%-value deviation technique, followed by the generalized mean $+$ standard deviation technique), using the smaller of the low and high one-sided deviation measurements.  Data are weighted by the number of non-rejected dumps that compose each scan.}  The final product is a noise model for each signal measurement in the observation.

\subsection{1D Background Subtraction}

Now that we have a model for the noise level at each point, we make use of it to separate small-scale structures from large-scale structures.  Specifically, we separate (1)~small-scale astronomical structures (e.g., sources) and (2)~short-duration RFI from (1)~large-scale astronomical structures, (2)~long-duration RFI, (3)~en-route drift, and (4)~elevation-dependent signal.  Since RFI and en-route drift are 1D structures, varying along the scans, we begin by modeling and subtracting off structures, larger than a user-defined scale, along the scans only.  We refer to this as 1D background subtraction.

This user-defined scale should be larger than the scale over which the telescope's beam pattern distributes signal for bright point sources.  We have determined minimum recommended values for the telescopes and receivers of \textsection2, empirically, by increasing this scale until the measured brightness (see \textsection4) of bright sources, observed at multiple parallactic angles, plateaued.  We list these values in Table~1.

Given this scale, a simple approach to 1D background subtraction, which we do not use here, but improve upon here, is to draw a line from each point to another point within one scale length, say, in the forward direction, such that all other points within this scale length are above the line (see Figure~11).  This is then repeated for each point, but in the backward direction.  The minimum of all of the resulting linear, local background models is a non-linear, non-parameterized, global background model that fits snuggly beneath the data.

This algorithm, however, is sensitive to negative noise fluctuations, and really is appropriate only in the limit of noiseless data.  Consequently, we now modify this algorithm to take into account the noise level of the data such that the final, global model passes through the middle of the background-level data, instead of below it.

\begin{figure}
\plotone{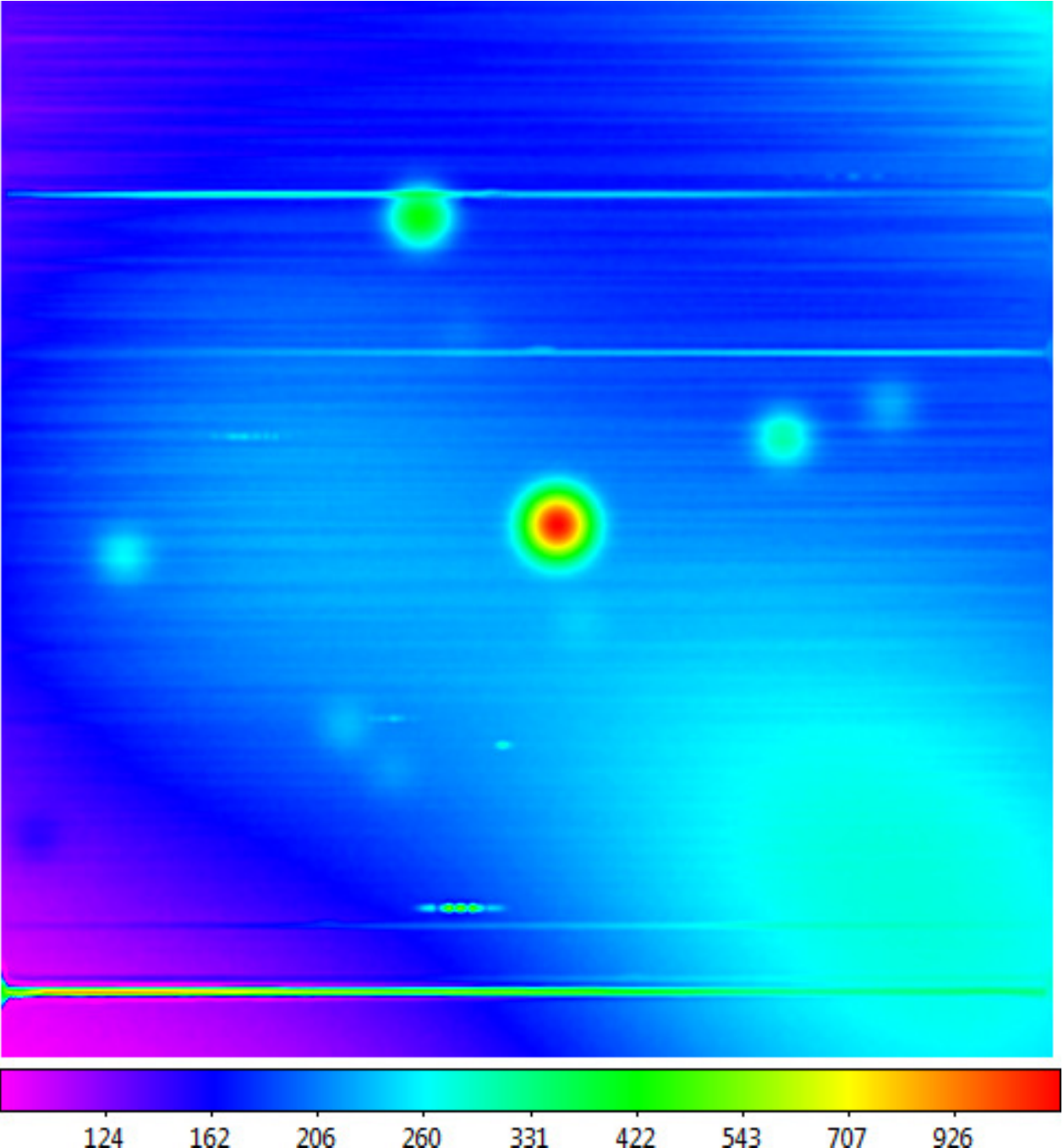}
\caption{Simulated data from Figure~18 to which we have added large-scale astronomical and elevation-dependent signal.  For the large-scale astronomical signal, we use a sum of 2D Gaussian distributions, each with a FWHM of $\approx$12 beamwidths.  For the elevation-dependent signal, we use a cosecant function.  As in Figure~5, elevation is decreasing toward the upper right.  Locally modeled surface (\textsection1.2.1, see \textsection3.7) has been applied for visualization only.  Hyberbolic arcsine scaling is used to emphasize fainter structures.}
\end{figure}

\begin{figure*}
\plotone{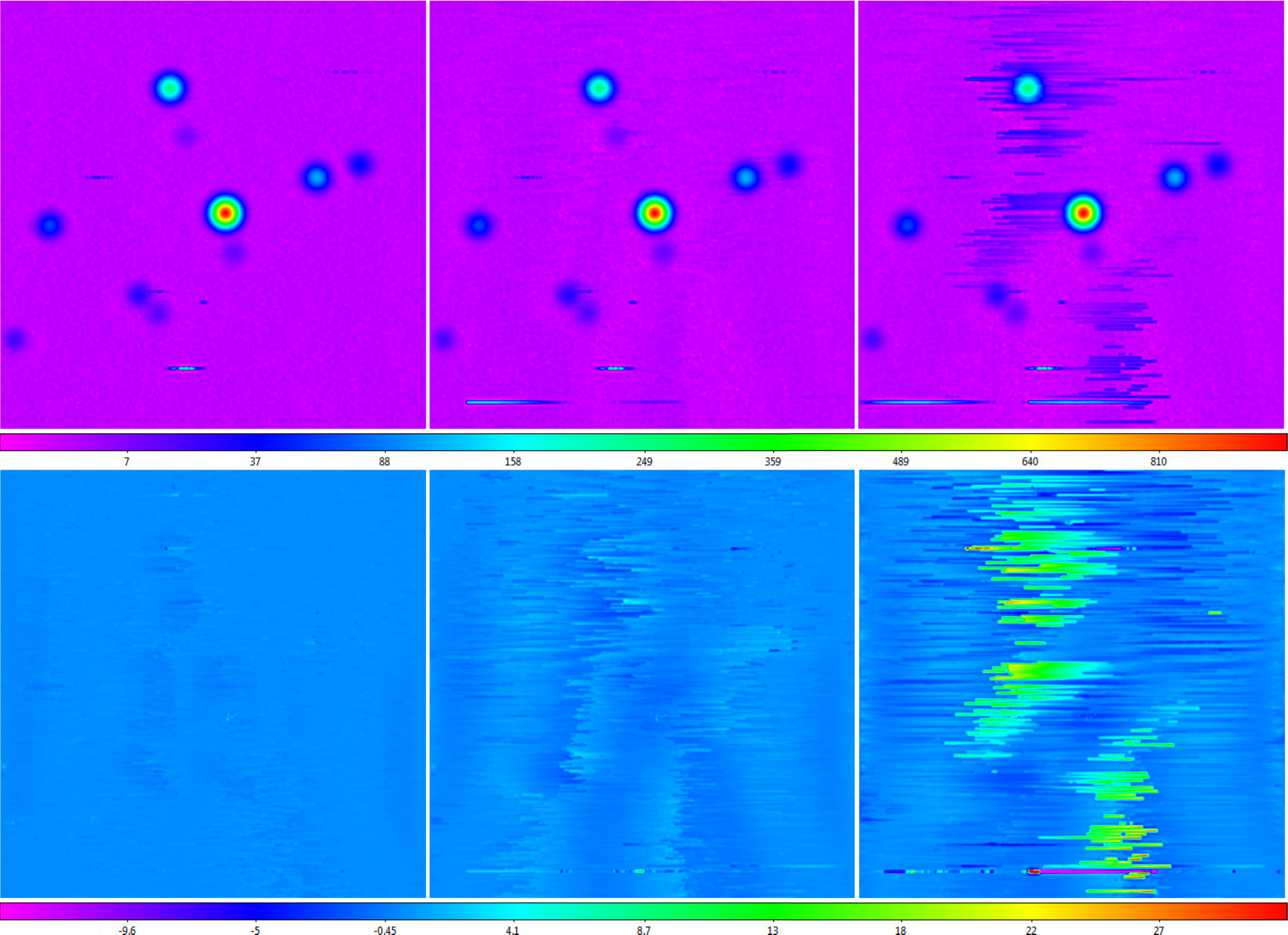}
\caption{\textbf{Top Row:}  Data from Figure~21 background-subtracted, with 6- (left), 12- (middle), and 24- (right) beamwidth scales (the map is 24 beamwidths across).  \textbf{Bottom Row:}  Data from the top row (1)~minus the data from Figure~16 (residuals) and (2)~minus the Gaussian random noise residuals from the bottom row of Figure~14, the small-scale structure residuals from the middle row of Figure~17, and the 1D large-scale structure residuals from the bottom row of Figure~19  (for greater clarity).  Elevation-dependent signal is effectively eliminated.  Large-scale astronomical signal is not eliminated, but is significantly reduced, especially in the smaller background-subtraction scale maps (Figure~20).  These gains are furthered by our RFI-subtraction algorithm in \textsection3.6.4.  Locally modeled surfaces (\textsection1.2.1, see \textsection3.7) have been applied for visualization only.  Square-root scaling is used in the top row to emphasize fainter structures.}
\end{figure*}

We continue to anchor the local background model, in the simplest case a straight line (but see below), to a point at the beginning or end of its scale-length domain.  We then fit this model to all of the points within this domain and calculate the standard deviation of these points about the model.\footnote{Data are again weighted by the number of dumps that compose each signal measurement.}  If this standard deviation is greater than the noise level, as measured by the noise model from \textsection3.2, we reject the greatest positive outlier and refit.  We repeat this process until the standard deviation of the non-rejected points is consistent with the noise model (see Figure~12, top panel).\footnote{Here, we are making an assumption, that the noise is Gaussian.  This is of course true on any single timescale, such as the sampling timescale of the data.  However, this may not be true across all timescales, if in the $1/f$ regime (i.e., if the noise is pink, instead of white).  In this case, our point-to-point noise model is an underestimate, and longer-timescale noise will manifest itself as signal, in the form of en-route drift (\textsection1.2.3).  That said, en-route drift is very successfully subtracted out by our background-subtraction algorithm (see \textsection3.2.3), and what is not removed in background subtraction is removed by our RFI-subtraction algorithm in \textsection3.6.}  

We then reject the anchor point and refit to the non-rejected points.  This results in a slightly lower standard deviation, so, selected from one point below the domain of the non-rejected points to one point above, not to exceed the original domain of the points, we iteratively add the least outlying rejected point back in and refit until the standard deviation of the non-rejected points is again consistent with the noise model.  This results in a final local background model, spanning the domain of the non-rejected points.  We repeat this process for every point in the scan, in both directions, resulting in a collection of final local background models, each of which goes through the middle of the non-rejected data in its domain, instead of below it (Figure~12, middle panel).

Finally, we construct the global background model from the local background models, not by taking the minimum at each point, but by taking the mean, after robust Chauvenet rejection of outliers (Figure~12, bottom panel).\footnote{We reject outliers as described in {\color{black}\textsection4 -- \textsection6} of Maples et al.\@ {\color{black}2018}, using iterative bulk rejection followed by iterative individual rejection (using the mode $+$ broken-line deviation technique, followed by the median $+$ 68.3\%-value deviation technique, followed by the mean $+$ standard deviation technique), using the smaller of the low and high one-sided deviation measurements.  Data are weighted as described below.}$^,$\footnote{In the occasional event that a point has no local background models associated with it, we complete the global background model by linearly interpolating between the immediately preceding and proceeding points for which the global background model was determined.}  When rejecting outliers, and when computing the post-rejection mean, we weight each point (1)~by the number of non-rejected dumps that contributed to its local background model, and (2)~by its position in its local background model, since fitted models are better constrained near the fitted data's center, vs.\@ its end points:
\begin{equation}
w_{ij}=\frac{\sum_j{N_{ij}}}{1+ \left(\frac{x_{ij}-\mu_i}{\sigma_i}\right)^2+ \delta\left(\frac{x_{ij}-\mu_i}{\kappa_i}\right)^4},
\end{equation}
\noindent where $\sum_j{N_{ij}}$ is the number of non-rejected dumps that contributed to the $i$th local background model, $w_{ij}$ is the weight of the $j$th point from the $i$th local background model, $x_{ij}$ is the angular distance of this point along the scan, $\mu_i$ is the dump-weighted mean angular distance of all of the non-rejected points from the $i$th local background model, $\sigma_i$ is the dump-weighted standard deviation of these values, $\kappa_i$ is analogous to standard deviation, and is related to these values' kurtosis:
\begin{equation}
\kappa_i=\left[\frac{\sum_jN_{ij}(x_{ij}-\mu_i)^4}{\sum_j{N_{ij}}}\right]^{1/4},
\end{equation}
\noindent and $\delta$ is zero for linear local background models and one for quadratic local background models (analogous terms can be added for higher-order local background models).  We justify Equation~3 in Appendix~A.

Higher-order local background models results in more flexible global background models, and sometimes additional flexibility is needed.  However, too high of an order can result in global background models that are too flexible, performing poorly around sources.  Fortunately, one does not have to go to too high of an order to find a good solution:  In this paper, we use quadratic local background models (e.g., Figure~12, bottom panel), which result in global background models that are sufficiently flexible to subtract off most large-scale signal (see \textsection3.3.3, \textsection3.3.4), but are also sufficiently robust to make at most noise-level errors around sources (see \textsection3.3.2).  {\color{black}(That said, linear local background models perform marginally better when using a background-subtraction scale that is near the minimum recommended value; Table~1.)}

\subsubsection{Simulation:  Gaussian Random Noise}

We test this algorithm by applying it to simulated data of increasing complexity.  We begin by applying it to just Gaussian random noise (see Figure~13), to evaluate its performance in the absence of small- or large-scale structures.  We background-subtract these data on 6-, 12-, and 24-beamwidth scales (see Figure~14, top row).  Residuals are presented in the bottom row of Figure~14.  We find that:  (1)~The background-subtracted data are not biased high nor low; and (2)~The noise level of the background-subtracted data is nearly that of the original data, and the RMS of the residuals is much less than the noise level of the original data (Figure~14).  

How nearly and how much less is a measure of the quality of the background subtraction, which at any point $j$ depends (1)~on the number of local background models that contributed to point $j$'s global background model, and (2)~on the weights, $w_{ij}$, of these local background models at this point.  For example, the RMS of the residuals is greater in the smaller background-subtraction scale maps, and near the ends of scans, where fewer, and lower weight, local background models inform each point's background subtraction.

\begin{figure*}
\plotone{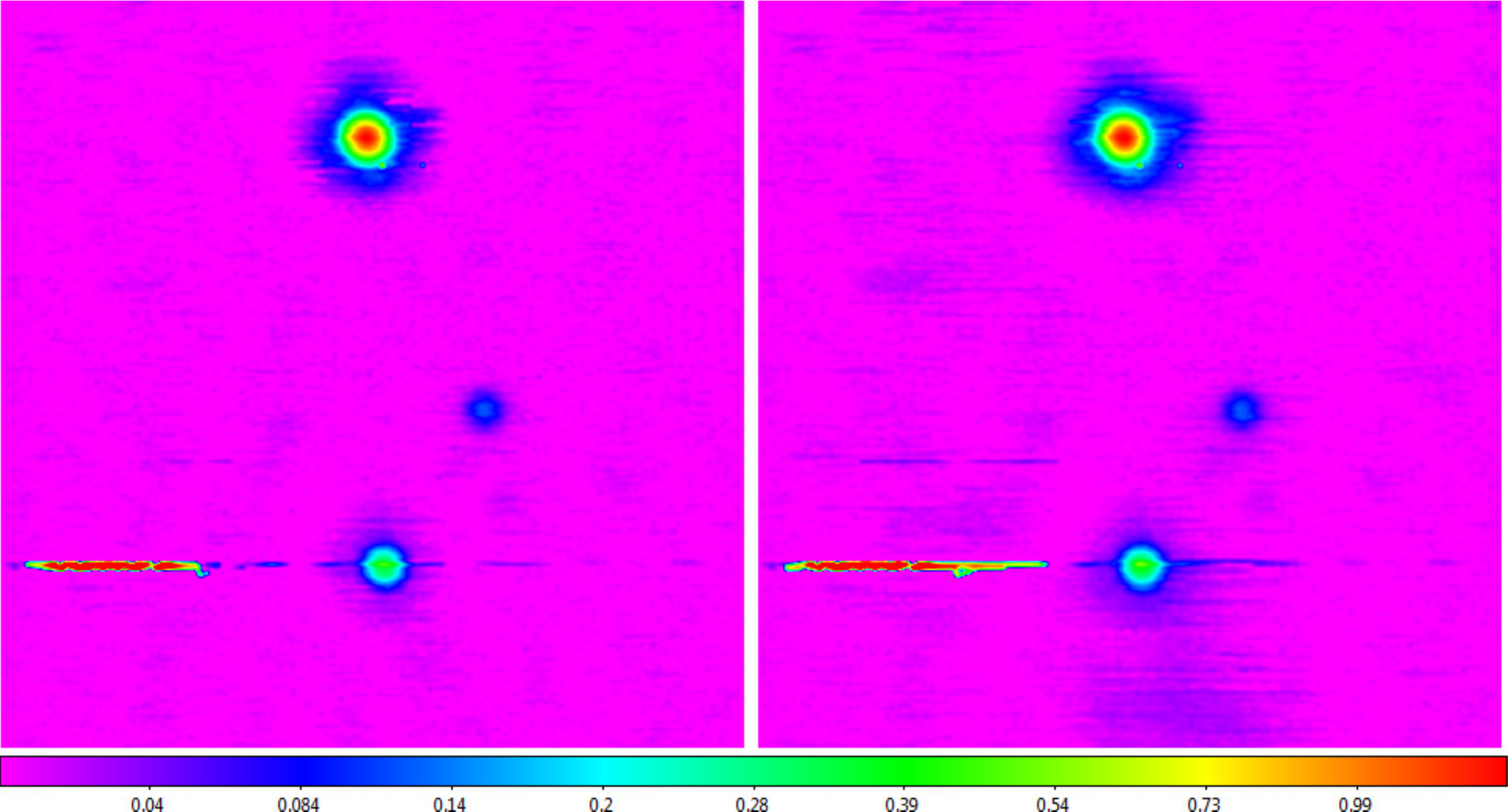}
\caption{20-meter L-band raster from Figure~5 background-subtracted, with 7- (left; Table~1) and 24- (right) beamwidth scales (the map is 24 beamwidths across).  Locally modeled surfaces (\textsection1.2.1, see \textsection3.7) have been applied for visualization only.  Hyperbolic-arcsine scaling is used to emphasize fainter structures.}
\end{figure*}

\begin{figure*}
\plotone{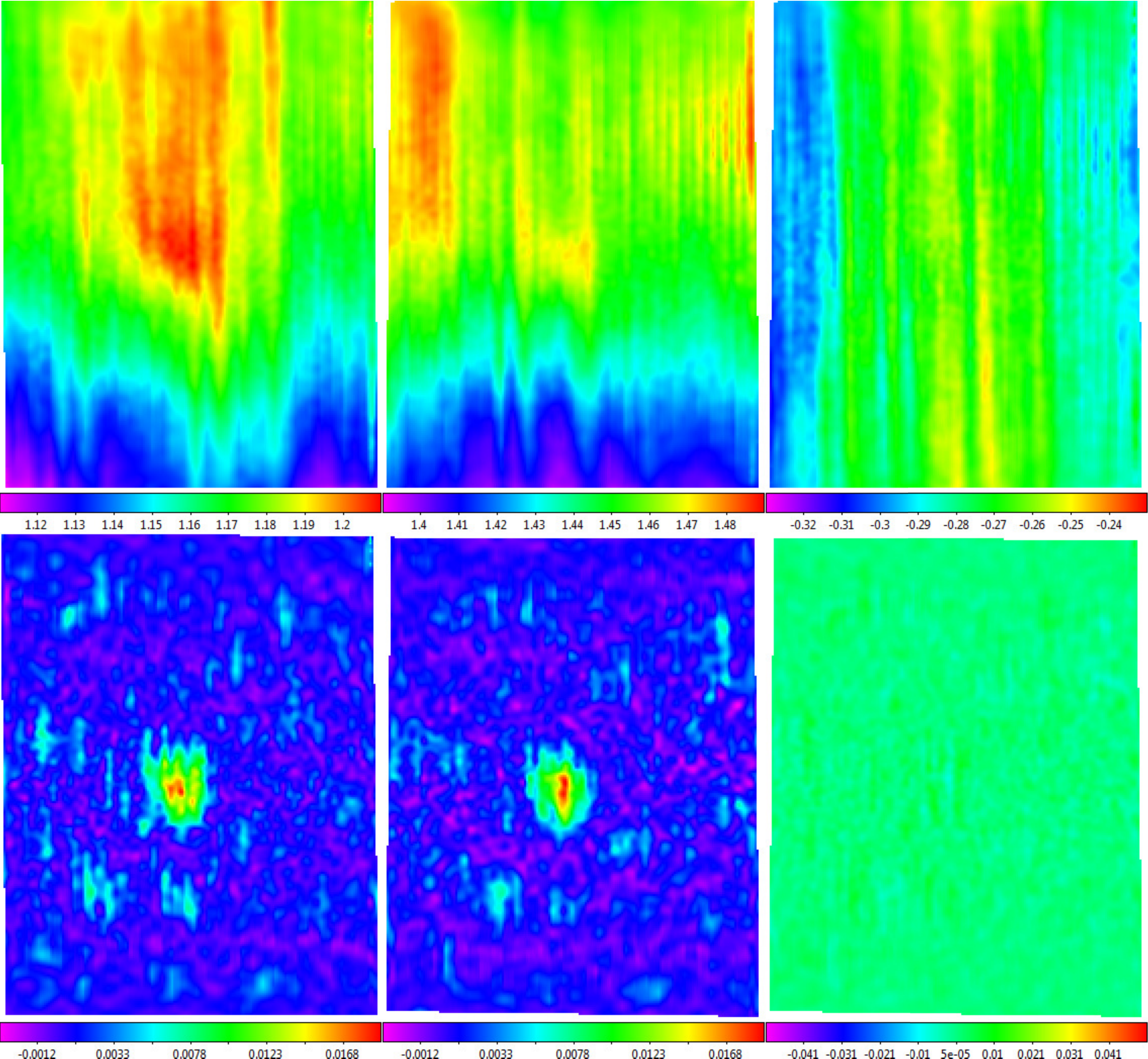}
\caption{\textbf{Top Row:}  (Time-delay corrected, see \textsection3.4) raw maps of Andromeda (left and middle), acquired with the 40-foot in L band, using a maximum slew speed nodding pattern, and their difference (right).  Instrumental signal drift dominates each map.  \textbf{Bottom Row:}  Data from the top row background-subtracted (left and middle), with a 5-beamwidth scale (larger than the minimum recommended scale from Table~1, given the size of the source), and their difference (right, spanning the same scale range as above).  Similar maps are extracted, despite the large systematics.  Locally modeled surfaces (\textsection1.2.1, see \textsection3.7) have been applied for visualization only.}
\end{figure*}
  
\begin{figure*}
\plotone{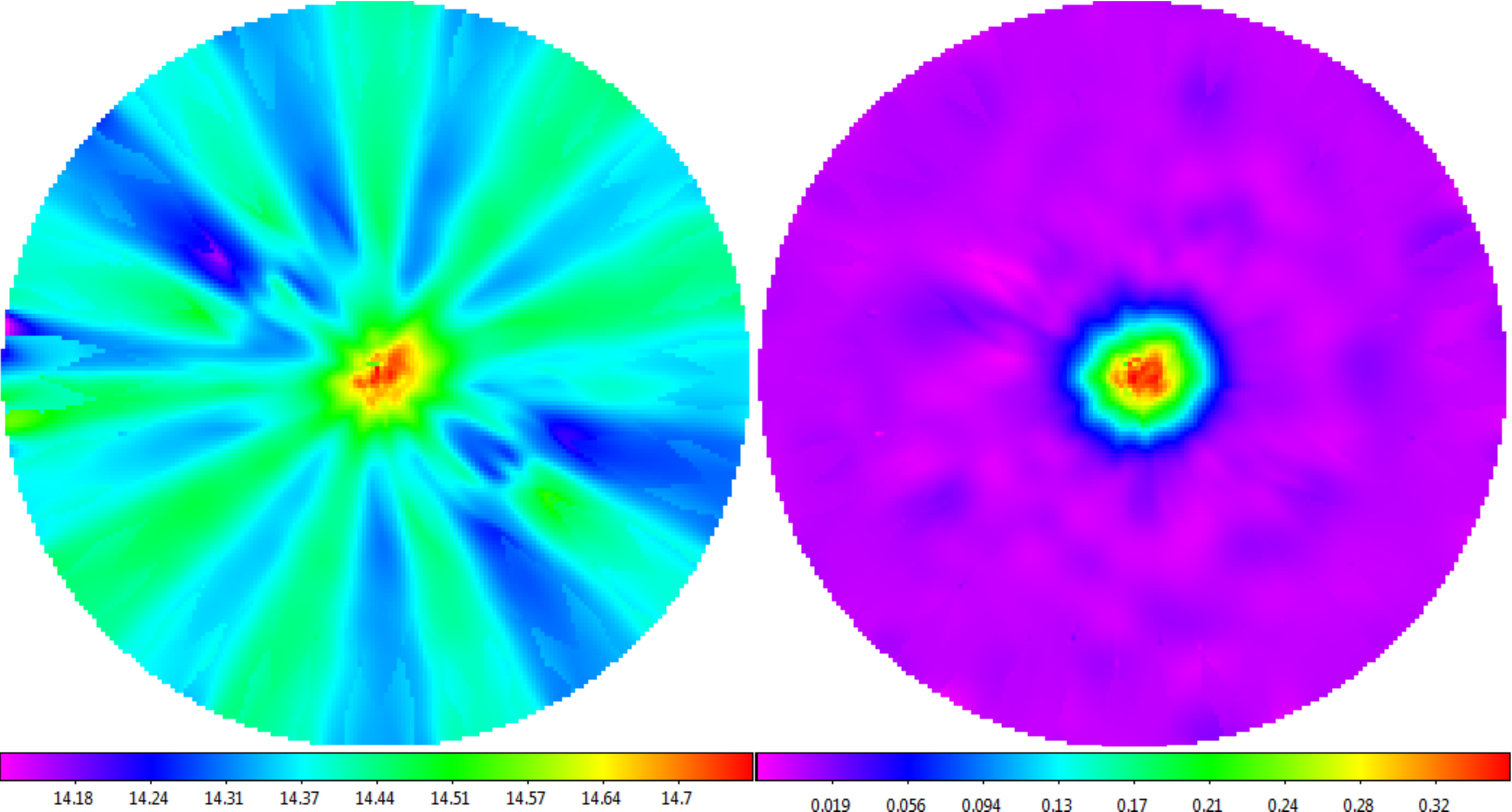}
\caption{\textbf{Left:}  Raw map of 3C~84, acquired with the 20-meter in X band using a 20-petal daisy pattern.  \textbf{Right:}  Data from the left panel background-subtracted, with a 6-beamwidth scale (Table~1).  Locally modeled surfaces (\textsection1.2.1, see \textsection3.7) have been applied for visualization only.}
\end{figure*}

In Figure~15, we plot these background-subtracted data (top panel) and residuals (bottom panel) vs.\@ the sum of the weights of the local background models that contributed to each point's global background model:
\begin{equation}
W = \sum_iw_{ij}.
\end{equation}
\noindent We supplement these data with 1- and 3-beamwidth background-subtraction scale data, and with analogous data from a lower-density, 1/5-beamwidth raster.  We find that the noise level of the background-subtracted data is well-modeled by:
\begin{equation}
\sigma_1 \approx 0.979 - 0.275e^{-W/109}
\end{equation}
\noindent and that the RMS of the residuals is well-modeled by:
\begin{equation}
\sigma_2 \approx 0.131 + 0.334e^{-W/1415},
\end{equation}
\noindent relatively independently of choice of background-subtraction scale and of choice of sampling density.  The noise level of the background-subtracted data is less than that of the original data, because some of the original data's variability is being incorporated into what is being subtracted off, though it approaches that of the original data for moderate and large values of $W$.  Likewise, the RMS of the residuals decreases as more information is incorporated, but to a limit.  Finally, $\sigma_1^2 + \sigma_2^2 \approx 1$ for all values of $W$, which means that additional variability is not being introduced by the background-subtraction technique itself.  

In summary, background-subtraction errors in the absence of small- or large-scale structures range from only $\approx$47\% of the original noise level for small values of $W$ to only $\approx$13\% of the original noise level for large values of $W$.  Furthermore, these are random errors, biased neither high nor low, reducing the noise level of the background-subtracted data by only $\approx$30\% for very small values of $W$ to only $\approx$2\% for moderate and large values of $W$.  

\subsubsection{Simulation:  Small-Scale Structures}

Next, we add simulated point sources and short-duration RFI to our simulated noise (see Figure~16).  We again background-subtract these data on 6-, 12-, and 24-beamwidth scales (see Figure~17, top row).  We present residuals in the middle row of Figure~17, but here we have additionally subtracted off the residuals from the bottom row of Figure~14, to help distinguish residuals that are due to the new, small-scale structures from those that are due to the Gaussian random noise.  There are three categories of residuals due to the new structures:

1.  Background-subtracted data are biased low in the vicinity of small-scale structures (whether point sources or short-duration RFI), but (with two exceptions, which we address in points 2 and 3, respectively) this bias is at or below the noise level, and furthermore is independent of the brightness of the proximal small-scale structure.  For example, the 1000 peak-S/N source near the center of the image is as biased, and as negligibly biased, as the $\approx$10 peak-S/N sources across the image.

The biased regions are rectangular, centered on the small-scale structures, and the length of the background-subtraction scale in the scan direction.  The bias level is greatest in the center of these regions, with peak values roughly given by:
\begin{equation}
\rm{peak\,bias\,level} \approx -\frac{(0.5-1)\times\rm{noise\,level}}{\left(\frac{\rm{background\,subtraction\,scale}}{6\,\rm{beamwidths}}\right)}.
\end{equation}
\noindent Even though this is a systematic error, vs.\@ a random error, it is at a sufficiently low level that it can be ignored.  Furthermore, in the more realistic case of a less-winged beam function (e.g., see Figure~31), this bias is smaller, by a factor of up to $\approx$2 -- 3 (Figure~17, bottom row).  Background-subtraction bias is further mitigated by our RFI-subtraction algorithm in \textsection3.6.2, at least in the regions around sources, and by our large-scale structure algorithm in Paper II, which adds over-subtracted signal back in.

2.  This bias can be larger when small-scale structures are blended together, resulting in a structure that is larger than the background-subtraction scale in the scan direction.  For example, the two sources toward the right of the image are marginally blended in the scan direction (the low point between them, in Figure~16, is $\approx$1.5 times the noise level).  The 6-beamwidth background-subtraction scale is smaller than the blended structure, resulting in residuals that are, in this case, $\approx$3 times the noise level (in the middle row, but much less in the bottom row).  However, the 12- and 24-beamwidth background-subtraction scales are larger than the blended structure, resulting in typical, sub-noise level residuals.  

3.  Finally, this bias can also be larger when small-scale structures occur within $\approx$1 -- 2 beamwidths of the ends of scans, since there is less/no data on the other side of the structure to inform background models.  For example, the integrated brightness of the source in the lower left of Figure~17 is underestimated by $\approx$15\% (in the middle row, but, again, by less in the bottom row).  This is a known deficiency of such approaches, but one that affects only the edges of maps where the telescope stops and reverses direction.  As such, we give the user the option to clip these data before surface-modeling (\textsection1.2.1, see \textsection3.7), if desired.

\subsubsection{Simulation:  1D Large-Scale Structures}

Next, we add simulated en-route drift and long-duration RFI (see Figure~18).  We again background-subtract these data on 6-, 12-, and 24-beamwidth scales (see Figure~19, top row).  We present residuals in the bottom row of Figure~19, but here we have additionally subtracted off the residuals from the bottom and middle rows of Figures 14 and 17, to help distinguish residuals that are due to the new, 1D large-scale structures from those that are due to the Gaussian random noise and the small-scale structures, respectively.  

Since we background subtract in the same 1D in which en-route drift and long-duration RFI occur, along the scans, we are effective at reducing these structures, especially when the background-subtraction scale is smaller than the scale over which these structures are varying (roughly 12 beamwidths in this simulation).  We find that en-route drift and long-duration RFI are reduced by factors of $\approx$3 and $\approx$5, to $\approx$96\% of and $\approx$53 times the noise level, respectively, when background-subtracted on double this scale (24 beamwidths); by factors of $\approx$63 and $\approx$630, to $\approx$5\% and $\approx$39\% of the noise level, respectively, when background-subtracted on half of this scale (6 beamwidths); and by even greater factors when background-subtracted on even smaller scales (see Figure~20).  These gains are furthered, and significantly, by our RFI-subtraction algorithm in \textsection3.6.3. 

\subsubsection{Simulation:  2D Large-Scale Structures}

Next, we add simulated large-scale structures, including elevation-dependent signal (see Figure~21).  We again background-subtract these data on 6-, 12-, and 24-beamwidth scales (see Figure~22, top row).  We present residuals in the bottom row of Figure~22, but here we have additionally subtracted off the residuals from the bottom, middle, and bottom rows of Figures 14, 17, and 19, to help distinguish residuals that are due to the new, 2D large-scale structures from those that are due to the Gaussian random noise, the small-scale structures, and the 1D large-scale structures, respectively.  

We are also effective at reducing these structures, especially when the background-subtraction scale is smaller than the scale over which these structures are varying (in this case, roughly 12 and 24 beamwidths, respectively).  We find that large-scale astronomical and elevation-dependent signal is reduced by factors of $\approx$26 and $\approx$670, to $\approx$3 times and $\approx$18\% of the noise level, respectively, when background-subtracted on the scale of the map (24 beamwidths); by factors of $\approx$830 and $\approx$3400, to $\approx$11\% and $\approx$4\% of the noise level, respectively, when background-subtracted on the 6-beamwidth scale; and by even greater factors when background-subtracted on even smaller scales (Figure~20).  These gains are furthered by our RFI-subtraction algorithm in \textsection3.6.4. 

\begin{figure*}
\plotone{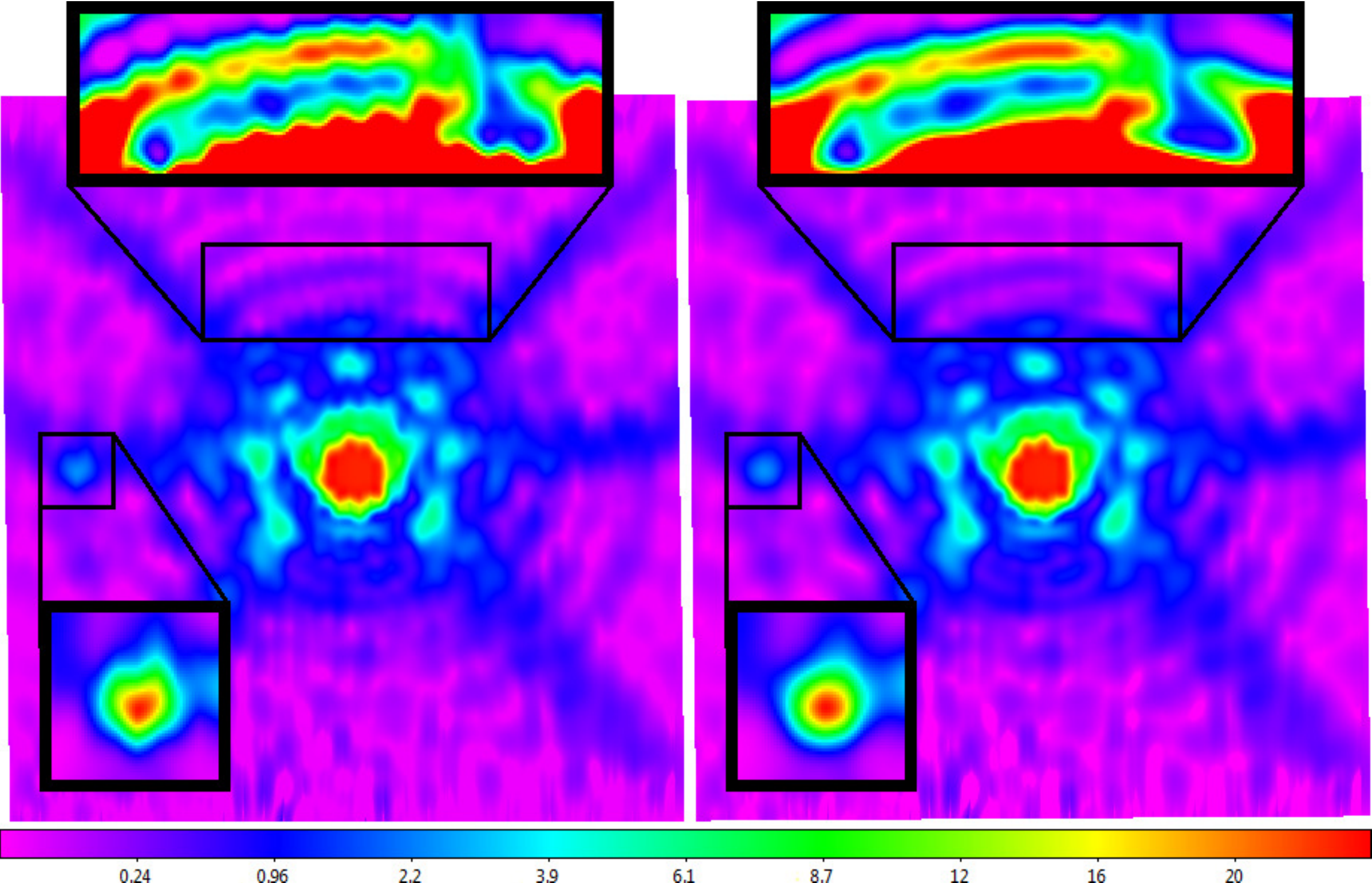}
\caption{Background-subtracted map of the sun in L band, highlighting the 40-foot's diffraction pattern, before (left) and after (right) time-delay correction.  The center is saturated.  Taurus~A is to the left.  {\color{black}The time-delay correction is most noticeable in the Airy rings (e.g., top zoom), and in (unsaturated) sources (e.g., left zoom).}  Locally modeled surfaces (\textsection1.2.1, see \textsection3.7) have been applied for visualization only.  Square-root scaling is used to emphasize fainter structures.}
\end{figure*}

\begin{figure*}
\plotone{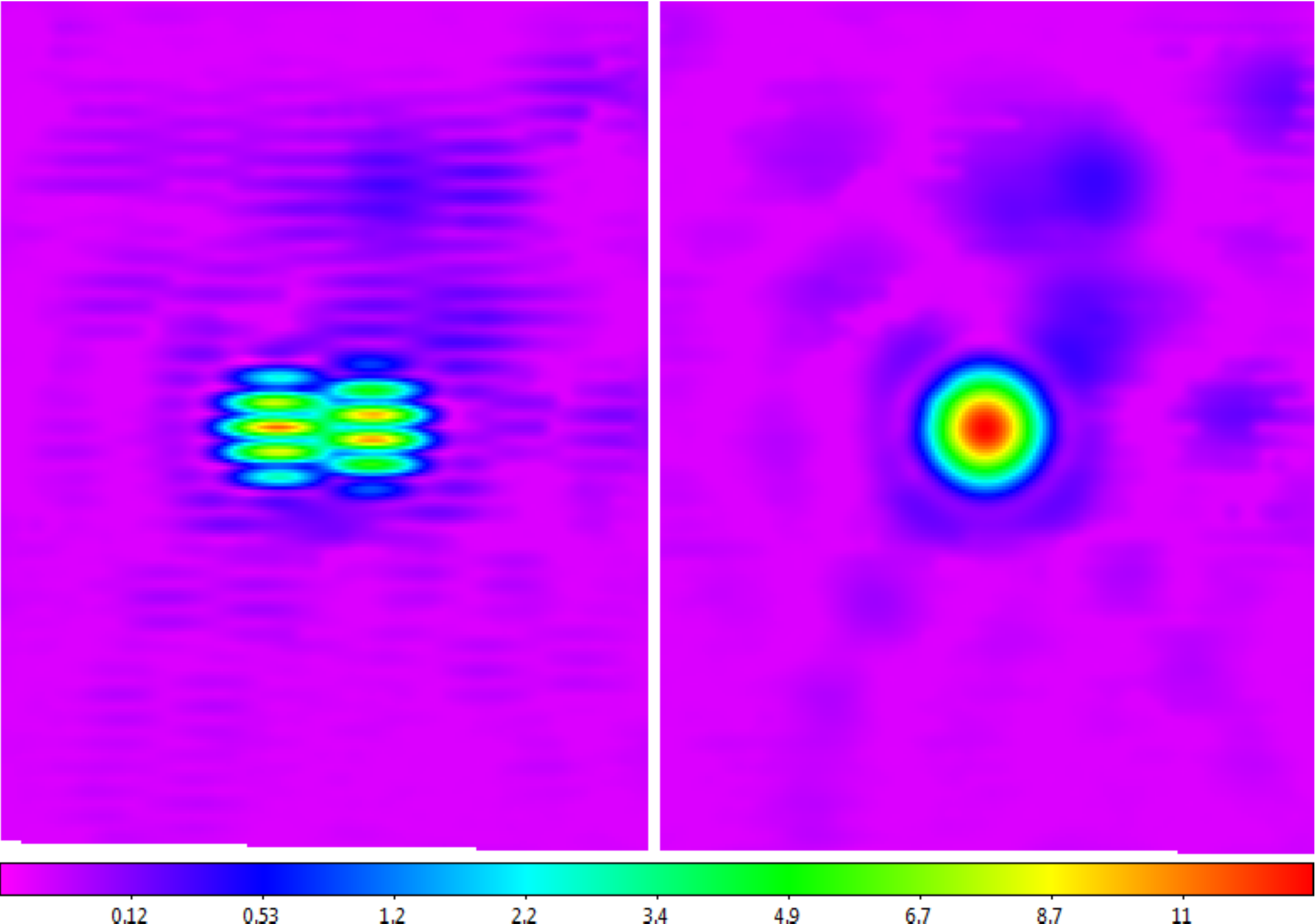}
\caption{Background-subtracted map of Cassiopeia~A in L band, acquired with the 20-meter with signal and position computers' clocks unsychronized, before (left) and after (right) time-delay correction.  Locally modeled surfaces (\textsection1.2.1, see \textsection3.7) have been applied for visualization only.  Square-root scaling is used to emphasize fainter structures.}
\end{figure*}

\begin{figure}
\plotone{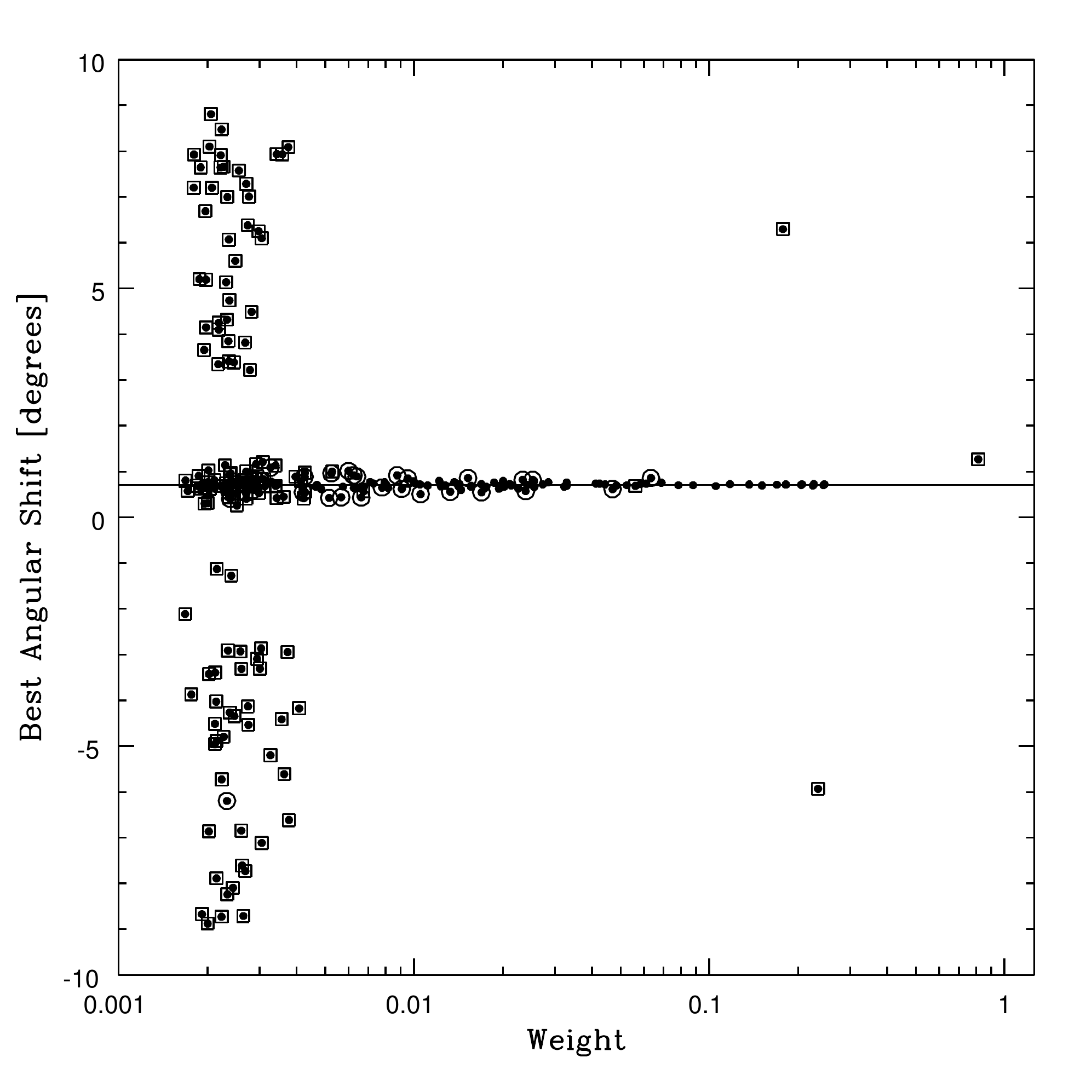}
\caption{Best angular shift between adjacent scans vs.\@ cross-correlation weight for all scans in a 20-meter observation.  Many of the low-weight best angular shifts are not well defined, because their corresponding scans are noise-dominated.  A few of the high-weight best angular shifts are incorrect, due to RFI contamination (consequently, simply taking the weighted mean of these values would yield a poor result).  To eliminate both cases:  For each best angular shift, we calculate the probability that it could, by chance alone, be as close as it is to both its preceding and proceeding values,\footnote{For two, adjacent, best angular shifts, $\delta_i$ and $\delta_{i+1}$, it is not difficult to show that this probability is given by $p_{i,i+1}=\frac{\left|\delta_{i+1}-\delta_i\right|}{\Delta}\left(2-\frac{\left|\delta_{i+1}-\delta_i\right|}{\Delta}\right)$, where $\Delta$ is the angular length of the scans.  For three, this probability is given by $2p_{i-1,i}p_{i,i+1}$.} and eliminate all best angular shifts for which this probability exceeds one half divided by the best angular shift sample size (e.g., Chauvenet 1863, Maples et al.\@ {\color{black}2018}; squared points).  The remaining best angular shifts repeat consistently for at least three consecutive measurements, and consequently are likely due to astronomical signal, not noise or RFI contamination.  We take the unweighted mean of these values (line), robust-Chauvenet rejecting any remaining outliers (circled points).\footnote{We reject outliers as described in {\color{black}\textsection4 -- \textsection6} of Maples et al.\@ {\color{black}2018}, using iterative bulk rejection followed by iterative individual rejection (using the mode $+$ broken-line deviation technique, followed by the median $+$ 68.3\%-value deviation technique, followed by the mean $+$ standard deviation technique), using the smaller of the low and high one-sided deviation measurements.  Data are weighted equally, in case any of the remaining high-weight data are still biased by RFI contamination (or by source saturation, as is the case in Figure~26).}}
\end{figure}

\begin{figure}
\plotone{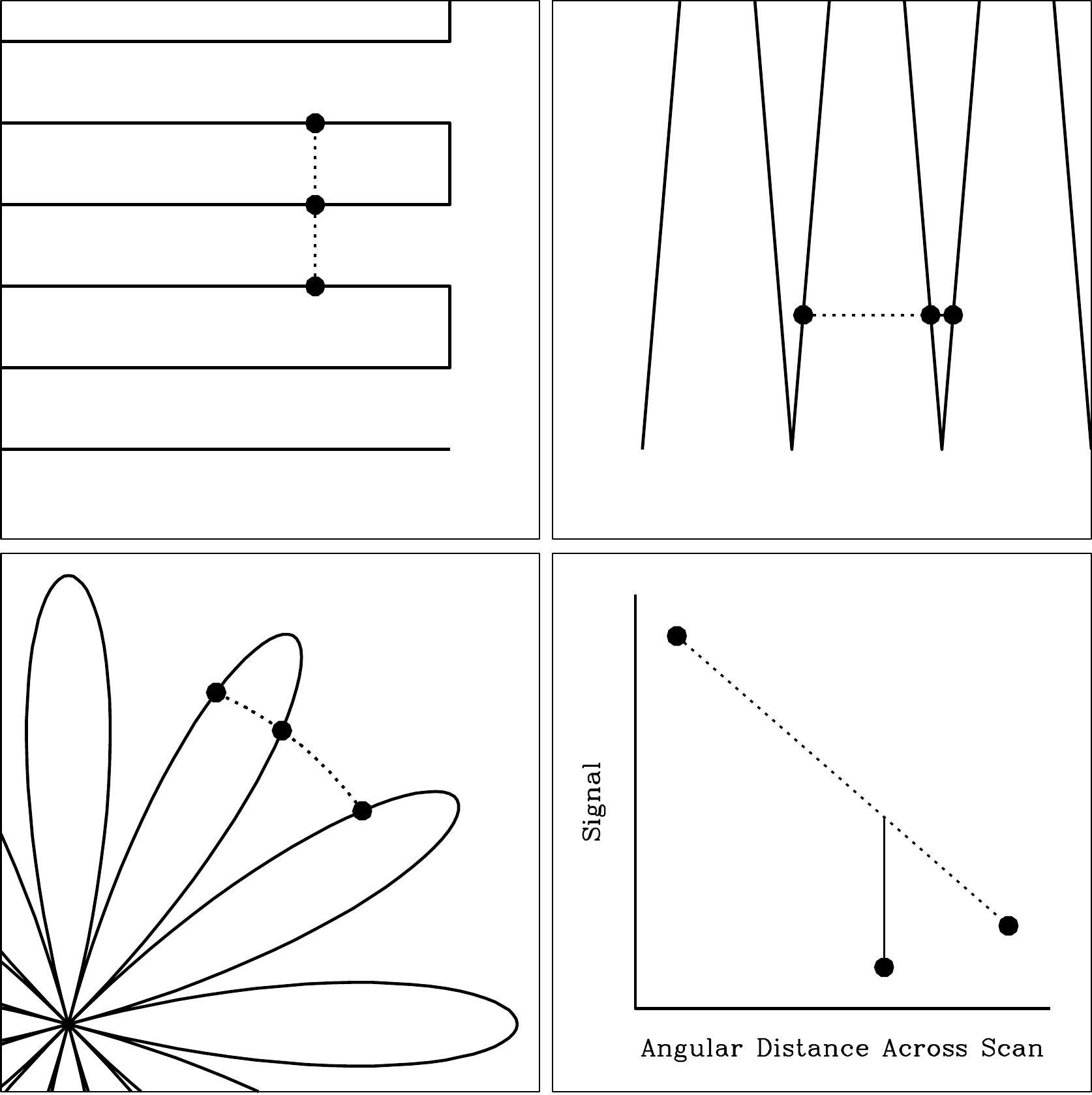}
\caption{Scan-to-scan measurement technique, applied to raster (top left), nodding (top right), and daisy (bottom left) mappings.  Residuals are measured (bottom right), and mean and standard deviations are measured from the non-rejected points, for each scan.}
\end{figure}

\begin{figure*}
\plotone{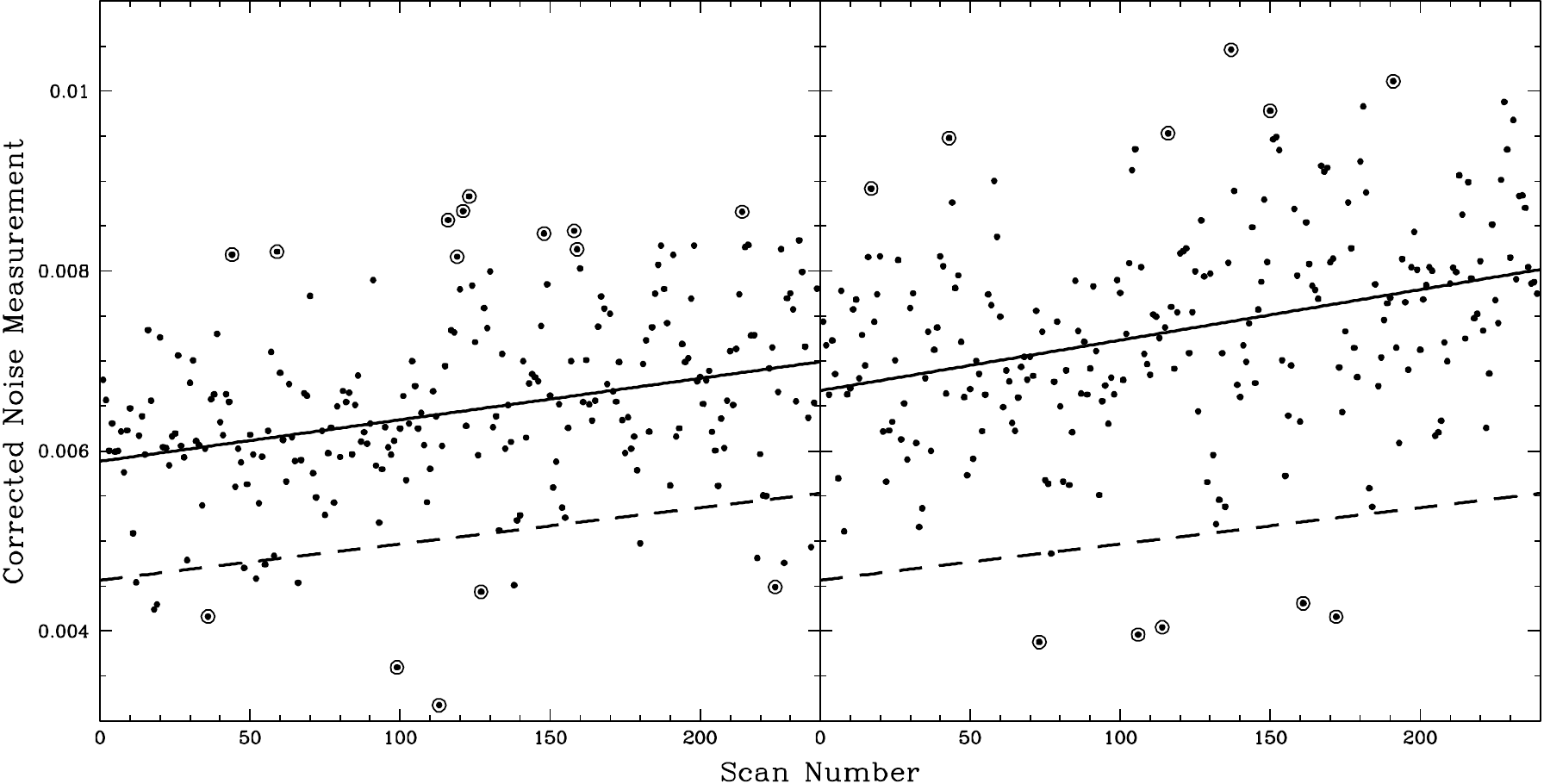}
\caption{\textbf{Left:}  Corrected 2D noise measurements vs.\@ scan number for the 20-meter observation from Figure~10, after background-subtracting on a 6-beamwidth scale, and best-fit model (solid line).  Circled points have been robust-Chauvenet rejected.  1D noise model from Figure~10 is included for comparison (dashed line).  The 2D noise level is $\approx$28\% higher, due to residual 1D structures (e.g., residual en-route drift) post-background subtraction.  \textbf{Right:}  Same, but with a 24-beamwidth background-subtraction scale.  In this case, the 2D noise level is $\approx$46\% higher than the 1D noise level, because contaminants are less completely eliminated on longer background-subtraction scales (e.g., Figure~20).}
\end{figure*}

\begin{figure}
\plotone{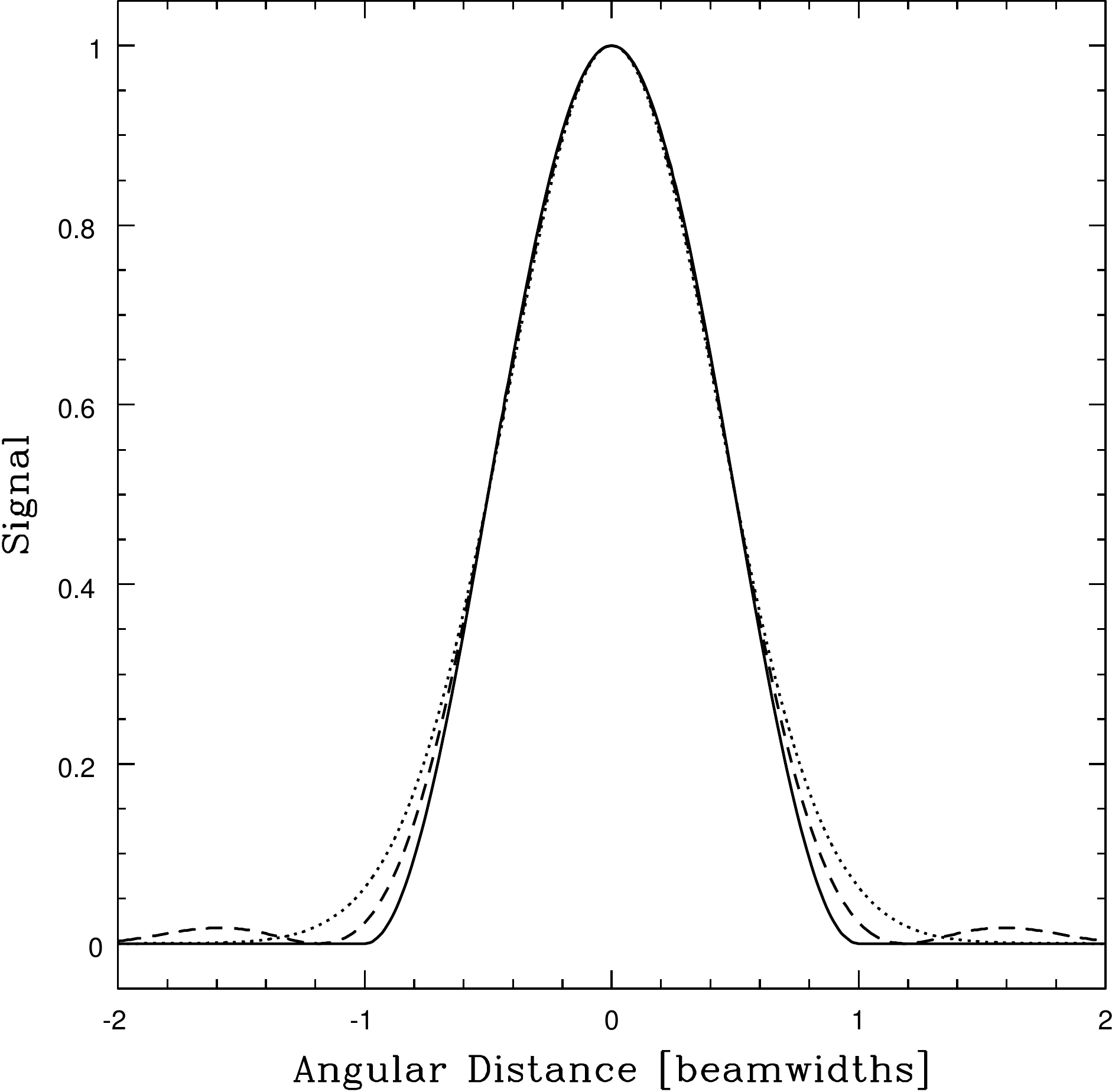}
\caption{Local model (Equation~9) with $\theta_{RFI} = 1$ beamwidth and $z_0 = 0$ (solid curve), and Airy (dashed curve) and Gaussian (dotted curve) functions, each with FWHM $=$ 1 beamwidth.}
\end{figure}

\begin{figure*}
\plotone{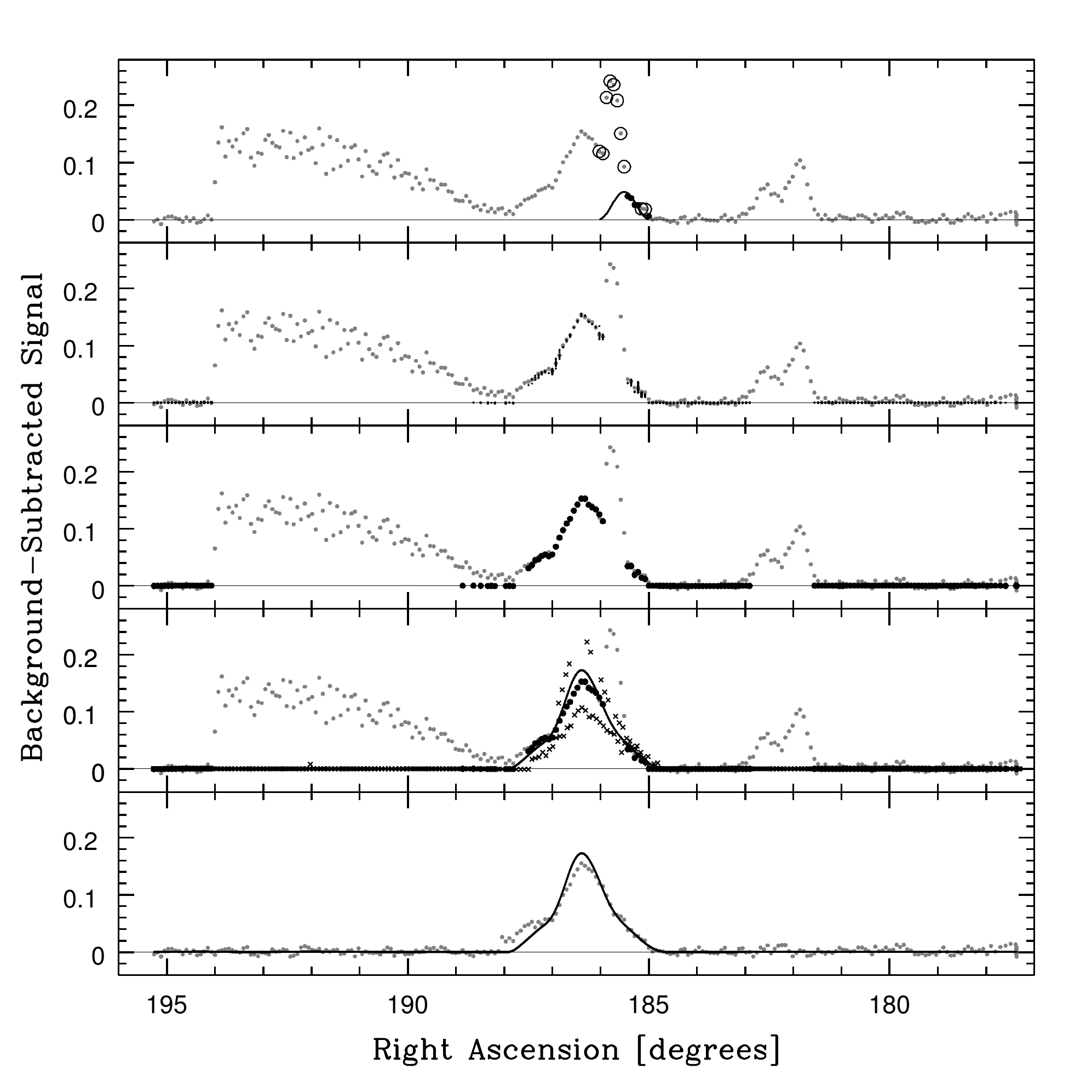}
\caption{\textbf{First:}  1D cross-section, along a scan, of a 2D local model, centered on an arbitrary point from the left panel of Figure~23, near Virgo~A (curve).  We have contaminated this scan with three instances of simulated RFI (the original, uncontaminated data are plotted in the fifth panel).  Circled points have been iteratively rejected as too high, given the modeled noise level (\textsection3.5); the larger, darker points were not rejected.  \textbf{Second:}  Every local model that intersects this scan, evaluated at each local model's non-rejected points (smaller, darker points).  All instances of simulated RFI have been rejected as too narrow, either along this scan or across adjacent scans, compared to the RFI-subtraction scale (in this case $\theta_{RFI} = 0.9$ beamwidths; Table~2).  \textbf{Third:}  Global model, constructed from the local model values (larger, darker points).  \textbf{Fourth:}  1D cross-section of the 2D surface model (\textsection1.2.2, see \textsection3.7), constructed from the 2D global model, three scans of which are shown (this scan $=$ larger, darker points; adjoining scans $=$ crosses).  Differences between data and model are due primarily to residual en-route drift, which differs from scan to scan.  \textbf{Fifth:}  The same, but constructed from the original, uncontaminated data, demonstrating the effectiveness of this approach to RFI subtraction.}
\end{figure*}

\begin{figure*}
\epsscale{0.9} 
\plotone{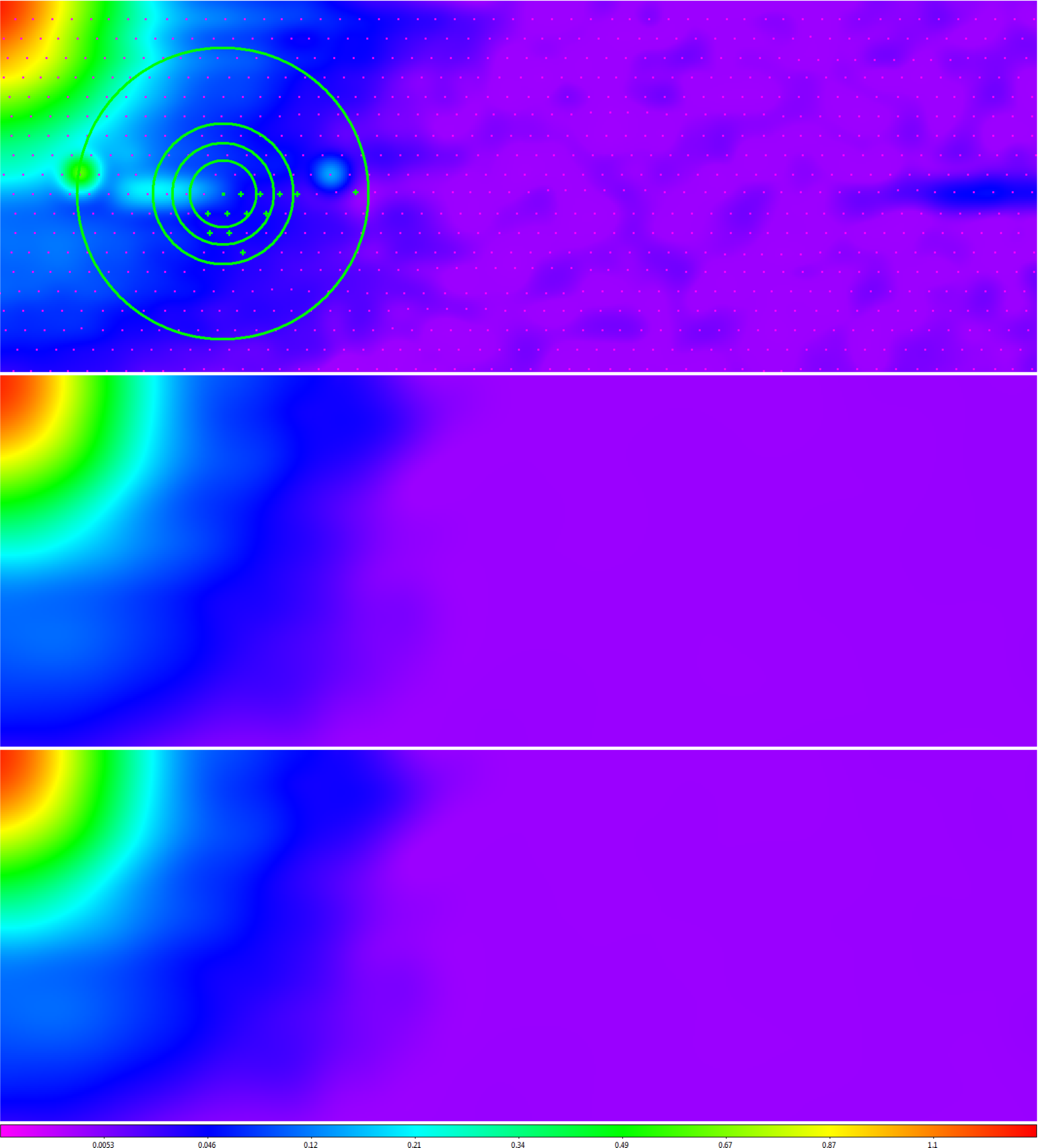}
\caption{\textbf{Top:}  Zoom-in of the first panel of Figure~32, but in 2D.  Purple dots mark observation points.  Green contours mark the 0\%-, 25\%-, 50\%-, 75\%-, and 100\%-of-peak levels of the local model.  Green plusses mark observation points within the domain of the local model that were not rejected, for being too high, given the modeled noise level (\textsection3.5).  Locally modeled surface (\textsection1.2.1, see \textsection3.7) has been applied for visualization only.  \textbf{Middle:}  Zoom-in of the fourth panel of Figure~32, in 2D.  \textbf{Bottom:}  Zoom-in of the fifth panel of Figure~32, in 2D, which is nearly identical, despite not being contaminated with the simulated RFI.}
\end{figure*}

\subsubsection{20-Meter and 40-Foot Data}

We now apply this algorithm to real data.  First, we apply it to the 20-meter L-band raster from Figure~5, using small and large background-subtraction scales (see Figure~23).  The 1D and 2D large-scale structures are significantly reduced in both cases, and will be further reduced by our RFI-subtraction algorithm in \textsection3.6.5 (see Figure~38).

Next, we apply the algorithm to two nodding maps of the same region of the sky, taken with the 40-foot, two days apart.  This is a good test of the algorithm in that the signal stability of the 40-foot is not nearly as good as that of the 20-meter, as can be seen in the raw data in the top row of Figure~24.  Despite this, the two background-subtracted maps are the same, to the noise level (Figure~24, bottom row).

Finally, we apply the algorithm to a daisy map, of 3C~84, taken with the 20-meter in X band, to demonstrate its application to a non-rectangular mapping pattern (see Figure~25).  Each slew of the telescope across the diameter of the observing region is treated as a separate scan (see Figure~49).

\subsection{Time-Delay Correction}

In the case of the 20-meter, signal is integrated over a user-defined time, and coordinate information is recorded at the midpoint of this integration time.  However, with the 40-foot, signal is run through an RC filter, with a user-defined time constant, typically 0.1 seconds, but signal and coordinate values are sampled simultaneously, resulting in an effective delay between the two due to the time constant.  This results in alternating coordinate errors in alternating scans (see Figure~26, left panel).  Even with the 20-meter, the same can happen if the signal and coordinate computers' clocks become unsynchronized (see Figure~27, left panel).

To correct for this, in the case of the 40-foot, and to check for this, and correct if necessary, in the case of the 20-meter, we cross-correlate adjacent scans.  The position of the maximum value of the cross-correlation gives the best angular shift between the scans, to at least 1/30 beamwidths, and the square root of this maximum value gives a weight.  If the scans intersect a source, the best angular shift is well defined, and this weight is correspondingly high.  If they intersect only noise, the best angular shift is not well defined and this weight is low.  For all adjacent pairs of scans in an observation, we measure the best angular shift, reject outliers (in two steps, see Figure~28), and take the mean of the remaining values.  Half of this value is how much each scan is misaligned, on the whole, in alternating directions.  

However, the data are not misaligned in angle, but in time.  Consequently, we divide this angular shift by the mean slew speed of the telescope, measured from the data, again robust-Chauvenet rejecting outliers (corresponding primarily to periods of acceleration near the beginnings of scans).\footnote{We reject outliers as described in {\color{black}\textsection4 -- \textsection6} of Maples et al.\@ {\color{black}2018}, using iterative bulk rejection followed by iterative individual rejection (using the mode $+$ broken-line deviation technique, followed by the median $+$ 68.3\%-value deviation technique, followed by the mean $+$ standard deviation technique), using the smaller of the low and high one-sided deviation measurements.  Data are weighted equally.}$^,$\footnote{In the case of daisies, the telescope's slew speed is variable, and maximum when the telescope passes through the source at, or nearly at, the center of the daisy.  Since this source will dominate the angular shift measurement, instead of dividing by the mean slew speed of the observation, we divide by the mean of each scan's maximum slew speed.}   We then take this time and shift each signal measurement, interpolating the telescope's coordinate values accordingly.  The results can be seen in the right panels of Figures 26 and 27.

If any time-delay correction is required, it is best to do this after background subtraction, so the cross-correlation is dominated by sources, instead of by instrumental effects (e.g., Figure~24).  It is also best to do this before RFI subtraction (\textsection3.5, \textsection3.6), which would treat misaligned sources as regions of at least partial RFI contamination.  

\subsection{2D Noise Measurement}

Key to RFI subtraction in the next section will be a re-measurement of the standard deviation of the now-background subtracted data on the smallest scales available (\textsection3.2), but this time across each scan instead of along each scan.  We refer to this as the 2D noise level of the data, and we measure it directly from the data, using the following technique.

For each point, we draw a line between the point with the most similar position along the preceding scan to that along the proceeding scan, and measure the central point's deviation from this line (see Figure~29).  For each scan, we robust-Chauvenet reject the point with the most discrepant deviation, update the scan's deviations, and repeat until all outliers are rejected.\footnote{We reject outliers as described in {\color{black}\textsection4} and {\color{black}\textsection6} of Maples et al.\@ {\color{black}2018}, using iterative bulk rejection followed by iterative individual rejection (using the mode $+$ broken-line deviation technique, followed by the median $+$ 68.3\%-value deviation technique, followed by the mean $+$ standard deviation technique), using the smaller of the low and high one-sided deviation measurements.  As in \textsection3.2, we do not precede this with iterative bulk rejection, because deviations change, and must be updated, after each rejection.  Deviations are again weighted by $\left[N_n^{-1}+N_{line}^{-1}\left(x_n\right)\right]^{-1}$, where $N_n$ is the number of dumps that compose the central point and $N_{line}\left(x_n\right)$ is the corresponding weight of the line at $x_n$, the angular distance of the central point, in this case, across the scan (see Footnote~10).}  Outliers are typically contaminated by RFI, as well as by residual signal from bright astronomical sources.  The sum in quadrature of the post-rejection mean deviation (the systematic component of the deviation) and the post-rejection standard deviation (the random component of the deviation) is what we call the ``scan-to-scan'' noise measurement.

As in \textsection3.2, we calibrate this technique by applying it to simulated, Gaussian random noise, of known standard deviation.  We find that the scan-to-scan technique overestimates the noise's true standard deviation by 22.9\%.  We correct each scan's noise measurement accordingly.

Finally, we combine all of the scans' noise measurements into a single model for the entire observation, again, allowing for a gradual change in the noise level over the course of the observation:  As in Figure 10, we fit a line to these data, robust-Chauvenet rejecting outliers (see Figure~30).\footnote{We fit this model to the data, simultaneously rejecting outliers, as described in {\color{black}\textsection8} of Maples et al.\@ {\color{black}2018}, using iterative bulk rejection followed by iterative individual rejection (using the generalized mode $+$ 68.3\%-value deviation technique, followed by the generalized median $+$ 68.3\%-value deviation technique, followed by the generalized mean $+$ standard deviation technique), using the smaller of the low and high one-sided deviation measurements.  Data are weighted by the number of non-rejected dumps that compose each scan.}  The final product is a new noise model for each signal measurement in the observation.

\subsection{2D RFI Subtraction}

Our algorithm for subtracting RFI, given continuum observations (or given spectral observations but in the case of broadband RFI), is similar to our algorithm for subtracting unwanted backgrounds.  As in \textsection3.3, we fit a collection of local models to the data, from which we construct a global model, but with three key differences:

1.  In \textsection3.3, we separated sub-background subtraction scale structures from larger structures, along the scans.  In this section, we separate sub-beamwidth structures from larger structures.  RFI can be smaller or larger than this scale along the scans, depending on its duration and the slew speed of the telescope.  But unless it coincidentally repeats at the same position along the scan, scan after scan, it will almost always be smaller than this scale across scans (e.g., Figures 5, 16, and 18), as will any residual en-route drift (e.g., Figures 19 and 23).  Consequently, instead of modeling the data in 1D, as we did in \textsection3.3, we now model the data in 2D (making use of our 2D noise measurement; \textsection3.5).  

2.  We use the following, two-parameter local model (instead of a three-parameter quadratic; \textsection3.3):
\begin{equation}
z(\Delta\theta) = 
\begin{cases}
f\cos^2\left(\frac{\pi\Delta\theta}{2\theta_{RFI}}\right) + z_0 & \text{if $\Delta\theta < \theta_{RFI}$} \\
0 & \text{otherwise}
\end{cases},
\end{equation}
\noindent where $\Delta\theta$ is the 2D angular distance from the model's center, $\theta_{RFI}$ is a user-defined RFI-subtraction scale, and $f$, the first of the two fitted parameters, normalizes the function in the first term, and $z_0$, the other fitted parameter, adds a small, local, background value.  This value should be small because the data are already background subtracted, but is allowed to be non-zero because background subtraction cannot be done perfectly.  We use the function in the first term because, when $\theta_{RFI} \approx 1$ beamwidth, it mimics a point source, but with shorter wings (see Figure~31).  Given this, by setting $\theta_{RFI}$ to just under the true FWHM of a telescope's beam pattern (see below), it can be used to separate astronomical signal from even marginally sub-beamwidth structures, in the following way:

Centered on each point in the observation, and on additional locations around the peaks of high-S/N sources (see below), we fit this model to all points within $\Delta\theta = \theta_{RFI}$ and calculate the standard deviation of these points about the model.  If this standard deviation is greater than the noise level, as measured by the noise model from \textsection3.5, we reject the greatest positive outlier (or the greatest outlier, positive or negative, if $f<0$),\footnote{Although RFI will always be positive, we do this such that noise-dominated regions, which have both positive and negative outliers, albeit only by chance and hence only occasionally, are treated in a relatively unbiased way (see \textsection3.6.1).} and refit.  We repeat this process until the standard deviation of the non-rejected points is consistent with the noise model (see Figure~32, first panel, for 1D cross-section, and Figure~33, top panel, for 2D zoom-in). 

This fit can be done analytically (see Appendix~B).  Furthermore, since $z(\Delta\theta)$ is independent of $f$ and $z_0$ beyond $\Delta\theta = \theta_{RFI}$, only (non-rejected) points within $\Delta\theta = \theta_{RFI}$ need be included in the fit, streamlining the computation.

This results in a single local model, defined only at its non-rejected points (again, Figure~32, first panel, and, Figure~33, top panel).  We repeat this process, centering the model on each point in the observation, and on additional locations around the peaks of high-S/N sources,\footnote{For observation points on the sides of sufficiently high S/N sources, where the slope is steepest, to not be rejected, a denser sampling of local-model center locations is required.  Let $\theta_s$ be whichever is smaller, the mean spacing of points along, or across, scans, measured as a fraction of the telescope's beamwidth.  It is not difficult to show that for an approximately Gaussian local model to be tangent to an approximately Gaussian source function at any observation point along the source function, out to, say, $\Delta\theta=1$ beamwidth, additional local-model center locations must be added within $\Delta\theta\approx1-\theta_{RFI}^2$ beamwidths of the source center, if $\theta_{RFI}\gtrsim\sqrt{\theta_s}$ beamwidths.  It is also not difficult to show that a grid of spacing $\approx$$\theta_s(\theta_{RFI}^{-2}-1)$ results in approximately one center location for every observation point within $\Delta\theta=1$ beamwidth, which we have found to be sufficient.  Note, we have also found that these additional center locations are necessary only for the highest-S/N sources, specifically for sources with a peak S/N greater than about a hundred (we add them wherever peak-S/N $>$ 75, using a centroiding algorithm to determine peak locations).  For 20-meter and 40-foot observations, this corresponds to only a handful of sources, the brightest in the sky.} resulting in a collection of local models, each of which goes through the middle of its non-rejected points (Figure~32, second panel).

\begin{figure*}
\plotone{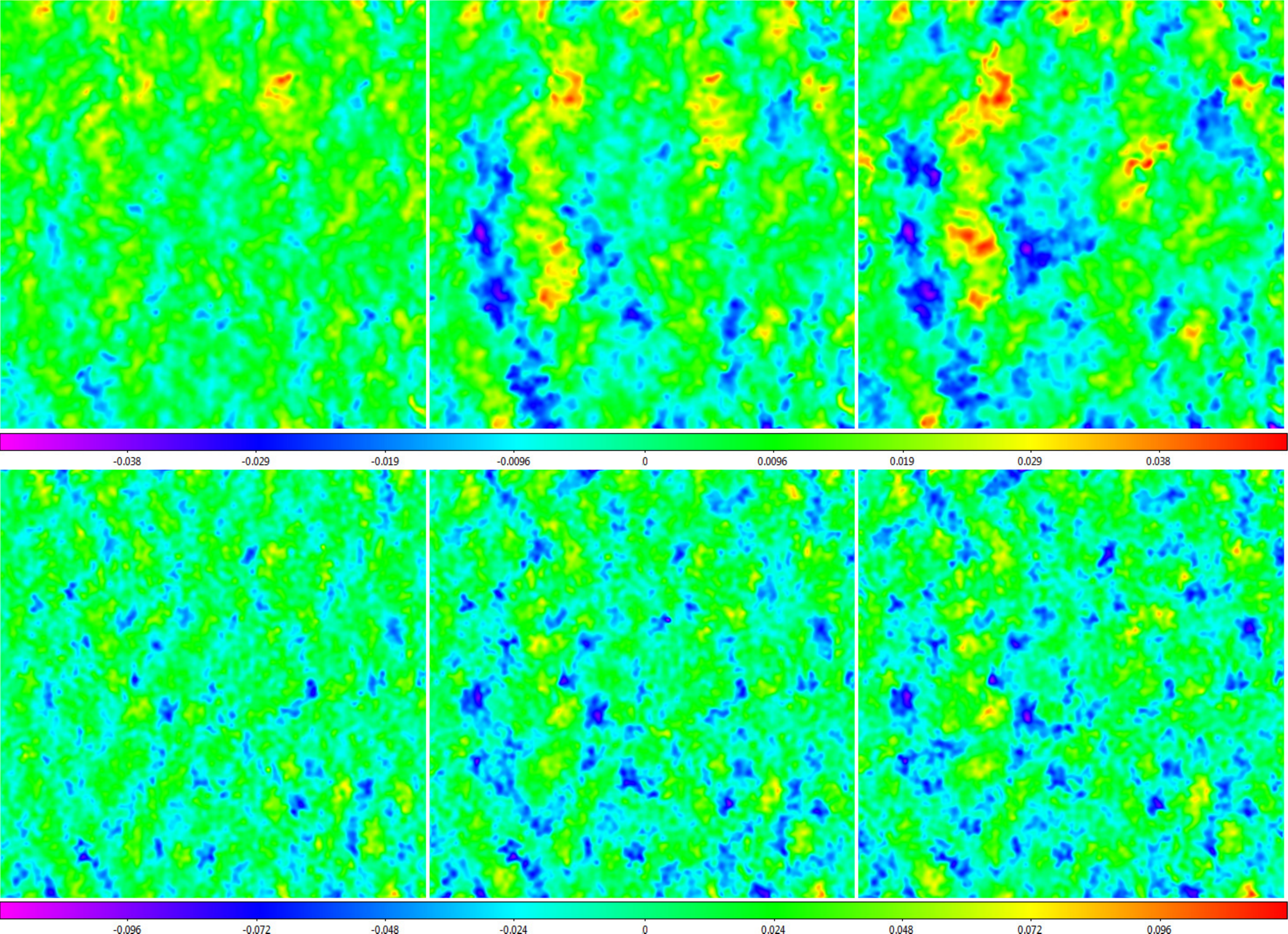}
\caption{Data from the top row of Figure~14, corresponding to 6- (left column), 12- (middle column), and 24- (right column) beamwidth background-subtraction scales, RFI-subtracted, with 0.95- (top row) and 0.5- (bottom row) beamwidth scales.  RFI-subtracted data are not biased high nor low.  On the 0.95-beamwidth scale, the noise level of the RFI-subtracted data is $\approx$1.1\% (left), $\approx$1.3\% (middle), and $\approx$1.4\% (right) of that of the background-subtracted data.  On the smaller, 0.5-beamwidth scale, the noise level of the RFI-subtracted data is roughly twice that:  $\approx$2.5\% (left), $\approx$2.8\% (middle), and $\approx$3.0\% (right) of that of the background-subtracted data.  Locally modeled surfaces (\textsection1.2.1, see \textsection3.7) have been applied for visualization only.}
\end{figure*}

3.  As in \textsection3.3, we construct the global model from the local models by taking the mean of the local models at each point, after robust-Chauvenet rejecting outliers (Figure~32, third panel).\footnote{We reject outliers as described in {\color{black}\textsection4 -- \textsection6} of Maples et al.\@ {\color{black}2018}, using iterative bulk rejection followed by iterative individual rejection (using the mode $+$ broken-line deviation technique, followed by the median $+$ 68.3\%-value deviation technique, followed by the mean $+$ standard deviation technique), using the smaller of the low and high one-sided deviation measurements.  Local-model values are weighted as described in Appendix~C.  The weight of the resulting global-model value is then given by summing the weights of the non-rejected local-model values, but not before dividing each of these weights by a number that, at least approximately, corrects for the non-independence of that particular weight's local-model value, over a scale that is related to the RFI-subtraction scale (see Appendix~D).  This is the weighted number of dumps that contributed to the global-model value, which we make use of again in \textsection3.7.2.}  However, if a point has no local models associated with it, instead of interpolating between surrounding global model values as we do in \textsection3.3, we simply excise the point from the observation and leave it to the surface-modeling algorithm (\textsection1.2.2, see \textsection3.7) to fill the additional gap (Figure~32, fourth and fifth panels, and Figure~33, middle and bottom panels).

\begin{deluxetable}{cccc}[!t]
\tablewidth{0pt}
\tablecaption{Maximum Recommended 2D RFI-Subtraction Scale for the Telescopes and Receivers of \textsection2, in Theoretical Beamwidths}
\tablehead{
\colhead{Telescope}     & \colhead{Receiver}       & \multicolumn{2}{c}{Scale} \\
& & \colhead{Left or Right} & \colhead{Left $+$ Right} \\
& & \colhead{Channel} & \colhead{Channel}}
\startdata
20-meter & L (HI $+$ OH)\tablenotemark{a} & 0.8 & 0.9 \\
20-meter & L (HI)\tablenotemark{b} & 0.7 & 0.8 \\
20-meter & L (OH)\tablenotemark{b} & 0.9 & 1.1 \\
20-meter & X & 0.8 & 0.8 \\
40-foot & L (HI) & 0.7 & 0.7
\enddata
\tablenotetext{a}{Before August 1, 2014}
\tablenotetext{b}{After August 1, 2014}
\end{deluxetable}

Also unlike in \textsection3.3, we do not subtract the resulting global model from the data.  Rather, the global model is the RFI-subtracted result.  Furthermore, given that this is a modeled version of the data, incorporating information from, usually many, nearby points at each point, it is a significantly smoother (but not additionally blurred; \textsection1.2.2) version of the data (see \textsection3.6.1 and \textsection3.6.2).  

\begin{figure*}
\plotone{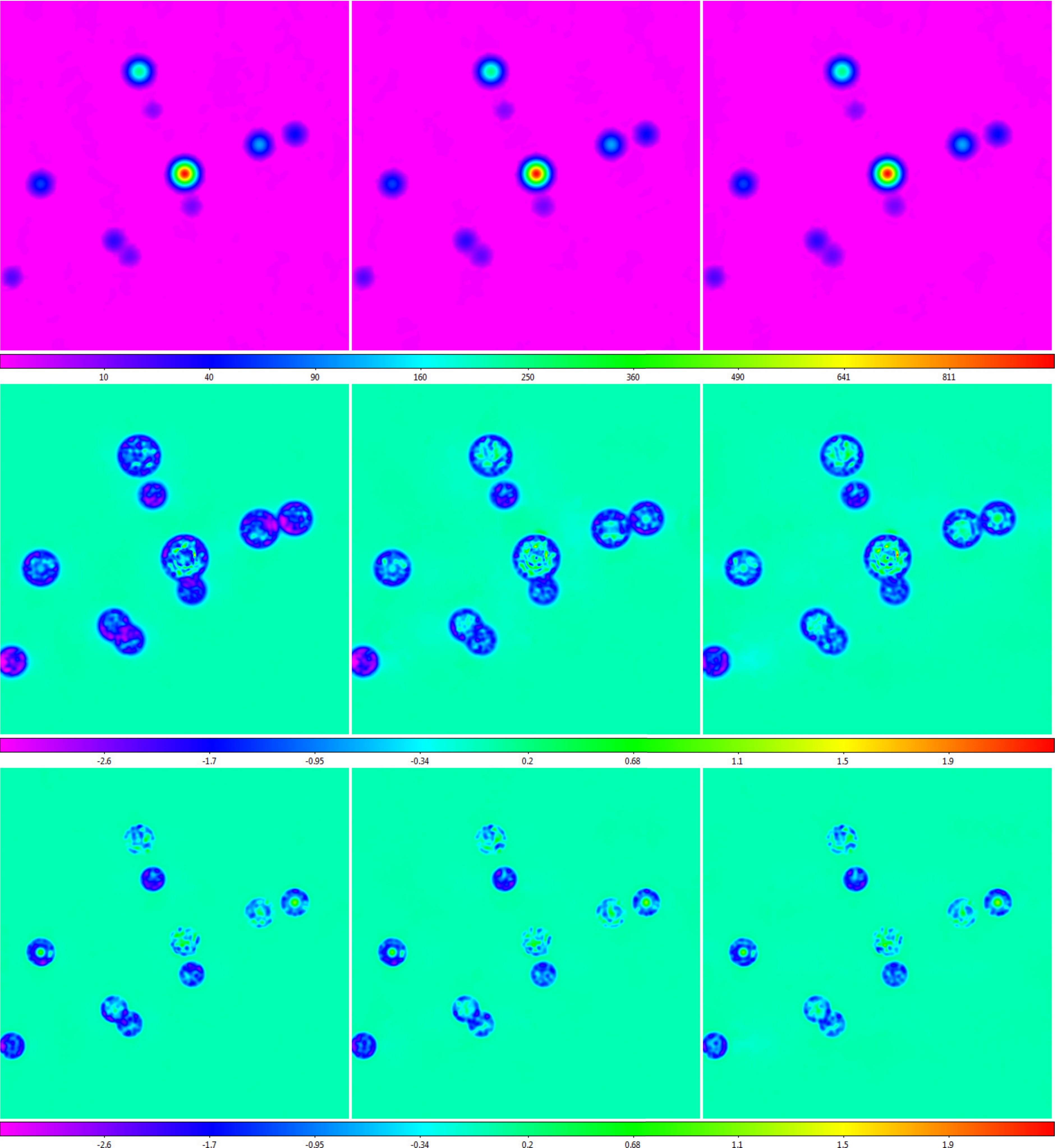}
\caption{\textbf{Top Row:}  Data from the top row of Figure~17, corresponding to 6- (left column), 12- (middle column), and 24- (right column) beamwidth background-subtraction scales, RFI-subtracted, with a 0.95-beamwidth scale.  \textbf{Middle Row:}  Data from the top row (1)~minus the point sources from Figure~16 (residuals) and (2)~minus the Gaussian random noise residuals from the top row of Figure~34 (for greater clarity).  Short-duration RFI is effectively eliminated (Figure~20).  Source residuals, overall, are biased negative, but (1)~significantly less so than in Figure~17, beyond where the sources intersect the noise level, (2)~slightly more so at this boundary, corresponding to post-RFI subtraction sources having slightly clipped wings, and (3)~at a similar level within this boundary, all relatively independently of the brightness of the source.  \textbf{Bottom Row:}  Same as the middle row, but for more-realistic, less-winged sources (given by Equation~9 with $\theta_{RFI} = 1$ beamwidth and $z_0 = 0$; Figure~31); residuals are again $\approx$2 -- 3 times smaller in this case.  Locally modeled surfaces (\textsection1.2.1, see \textsection3.7) have been applied for visualization only.  Square-root scaling is used in the top row and squared scaling is used in the middle and bottom rows to emphasize fainter structures.}
\end{figure*}

\begin{figure*}
\plotone{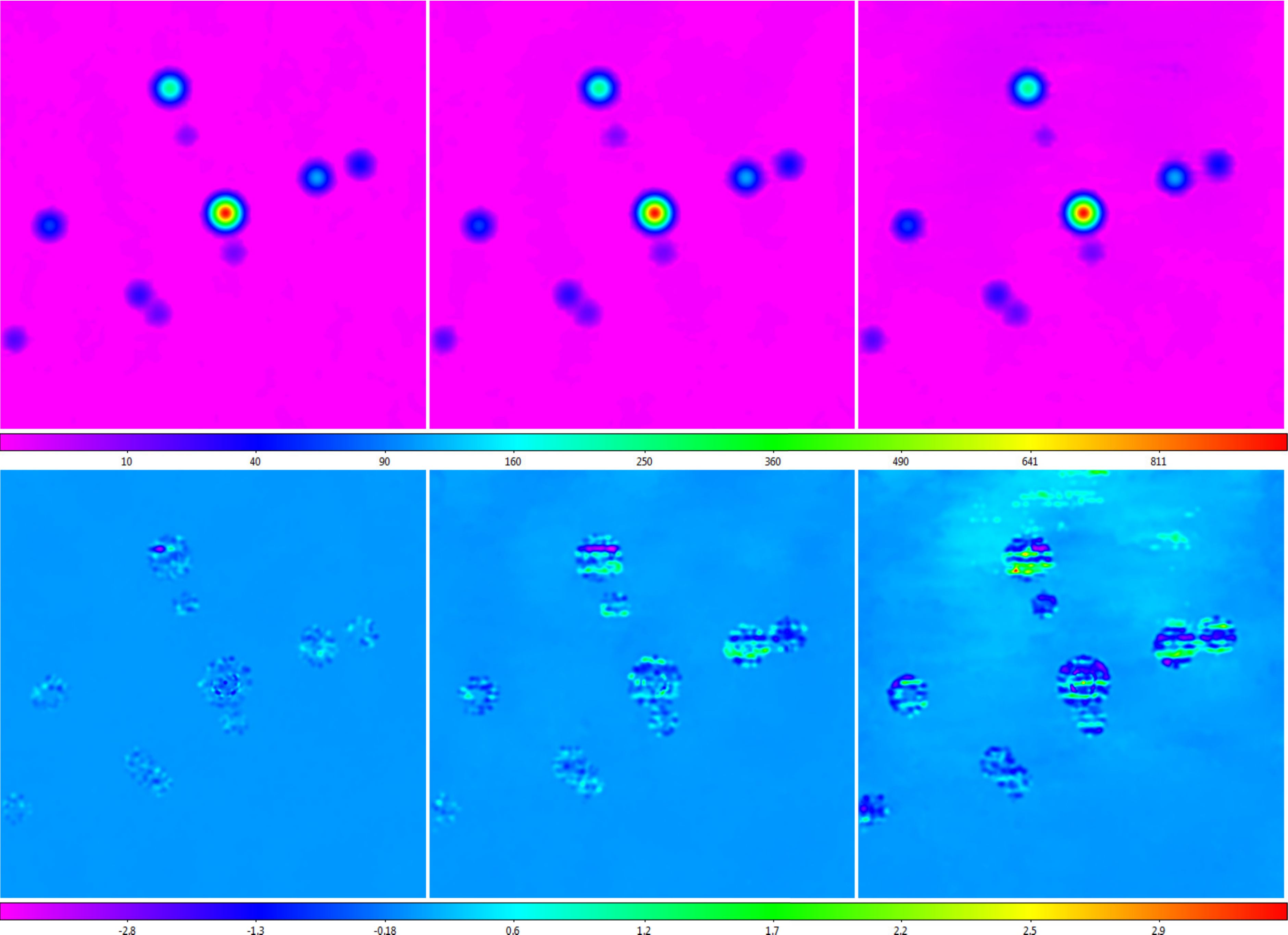}
\caption{\textbf{Top Row:}  Data from the top row of Figure~19, corresponding to 6- (left column), 12- (middle column), and 24- (right column) beamwidth background-subtraction scales, RFI-subtracted, with a 0.95-beamwidth scale.  \textbf{Bottom Row:}  Data from the top row (1)~minus the point sources from Figure~16 (residuals) and (2)~minus the Gaussian random noise residuals from the top row of Figure~34, and the small-scale structure residuals from the middle row of Figure~35 (for greater clarity).  Beyond where the sources intersect the noise level, en-route drift and long-duration RFI are effectively eliminated (Figure~20).  Within these boundaries, the residuals are fairly consistent with the post-background subtraction residuals of Figure~19; these are biased neither high nor low, and are noise-level.  Locally modeled surfaces (\textsection1.2.1, see \textsection3.7) have been applied for visualization only.  Square-root and hyperbolic-arcsine scalings are used in the top and bottom rows, respectively, to emphasize fainter structures.}
\end{figure*}

\begin{figure*}
\plotone{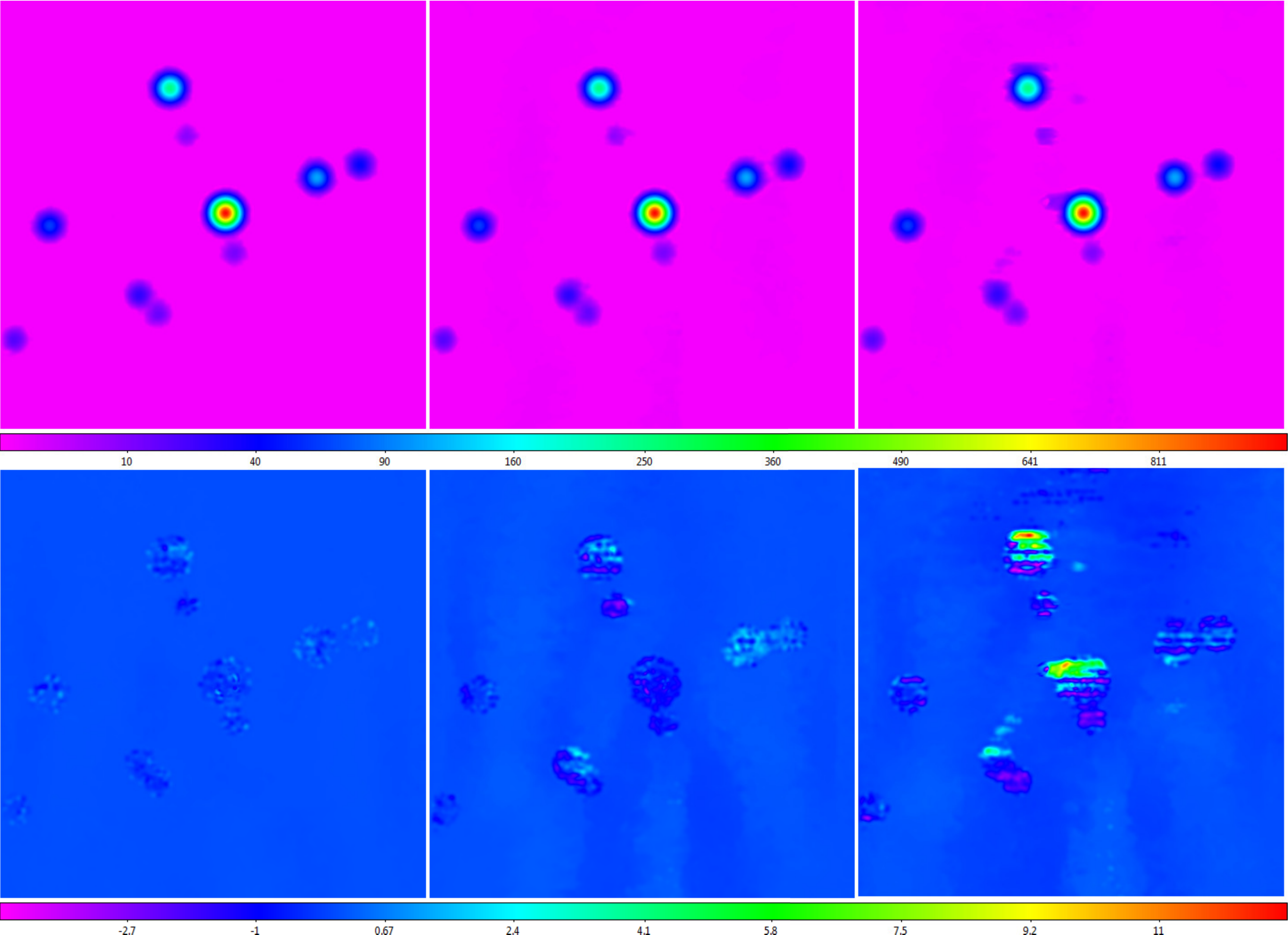}
\caption{\textbf{Top Row:}  Data from the top row of Figure~22, corresponding to 6- (left column), 12- (middle column), and 24- (right column) beamwidth background-subtraction scales, RFI-subtracted, with a 0.95-beamwidth scale.  \textbf{Bottom Row:}  Data from the top row (1)~minus the point sources from Figure~16 (residuals) and (2)~minus the Gaussian random noise residuals from the top row of Figure~34, the small-scale structure residuals from the middle row of Figure~35, and the 1D large-scale structure residuals from the bottom row of Figure~36 (for greater clarity).  Elevation-dependent signal is effectively eliminated (Figure~20).  Large-scale astronomical signal is not eliminated, but is significantly reduced, especially in the smaller background-subtraction scale maps (Figure~20).  Locally modeled surfaces (\textsection1.2.1, see \textsection3.7) have been applied for visualization only.  Square-root and hyperbolic-arcsine scalings are used in the top and bottom rows, respectively, to emphasize fainter structures.}
\end{figure*}

\begin{figure*}
\plotone{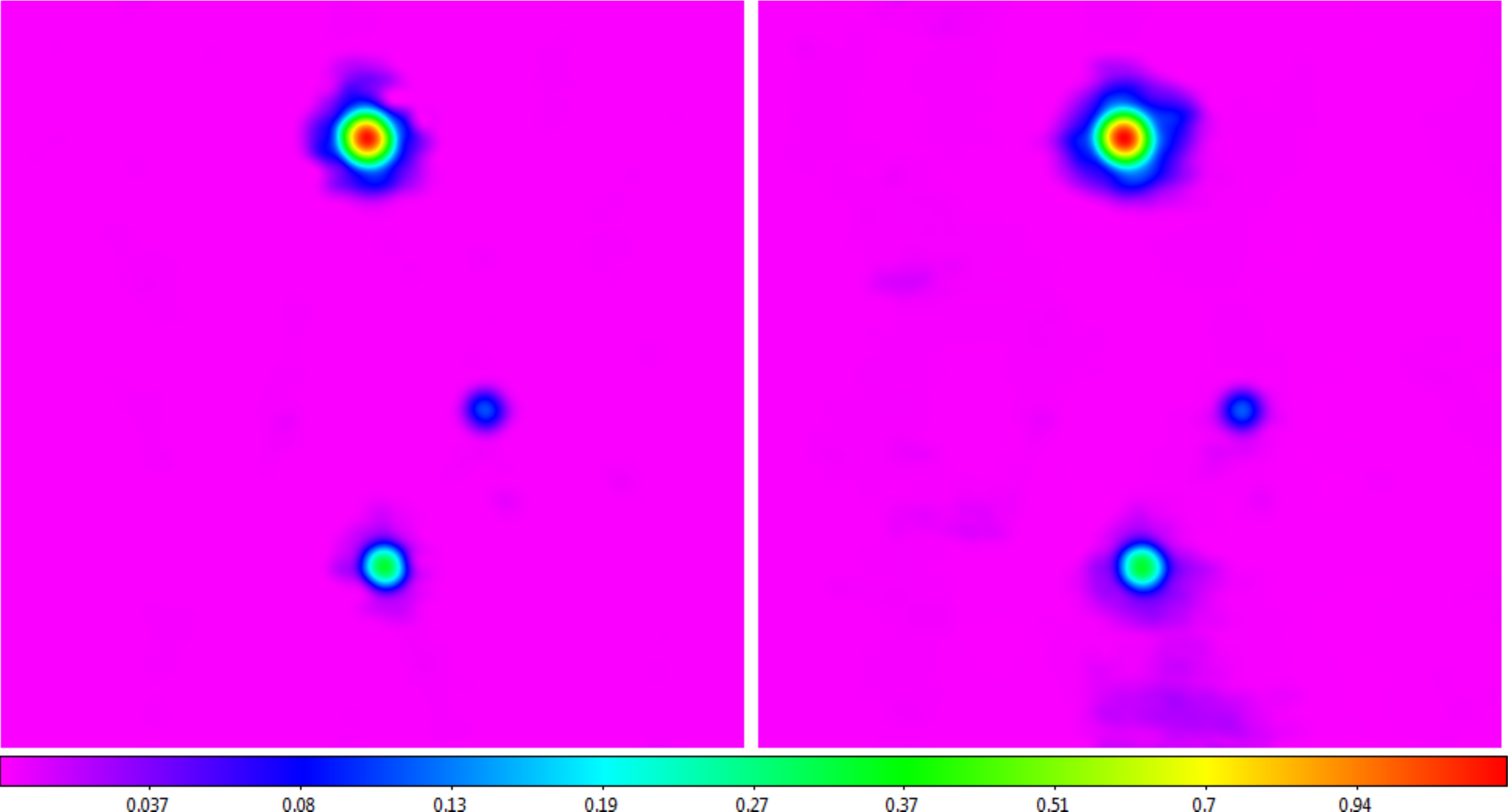}
\caption{20-meter L-band raster from Figure~5 background-subtracted, with 7- (left; Table~1) and 24- (right) beamwidth scales (Figure~23), and RFI-subtracted, with a 0.9-beamwidth scale (Table~2).  RFI, both long-duration intersecting 3C 273 and short-duration near Virgo~A, as well as en-route drift across the entire image, are successfully eliminated.  Locally modeled surfaces have been applied for visualization, with a minimum weighting scale of 2/3 beamwidths (\textsection1.2.1, see \textsection3.7).  Hyperbolic-arcsine scaling is used to emphasize fainter structures.}
\end{figure*}

\begin{figure*}
\epsscale{0.95} 
\plotone{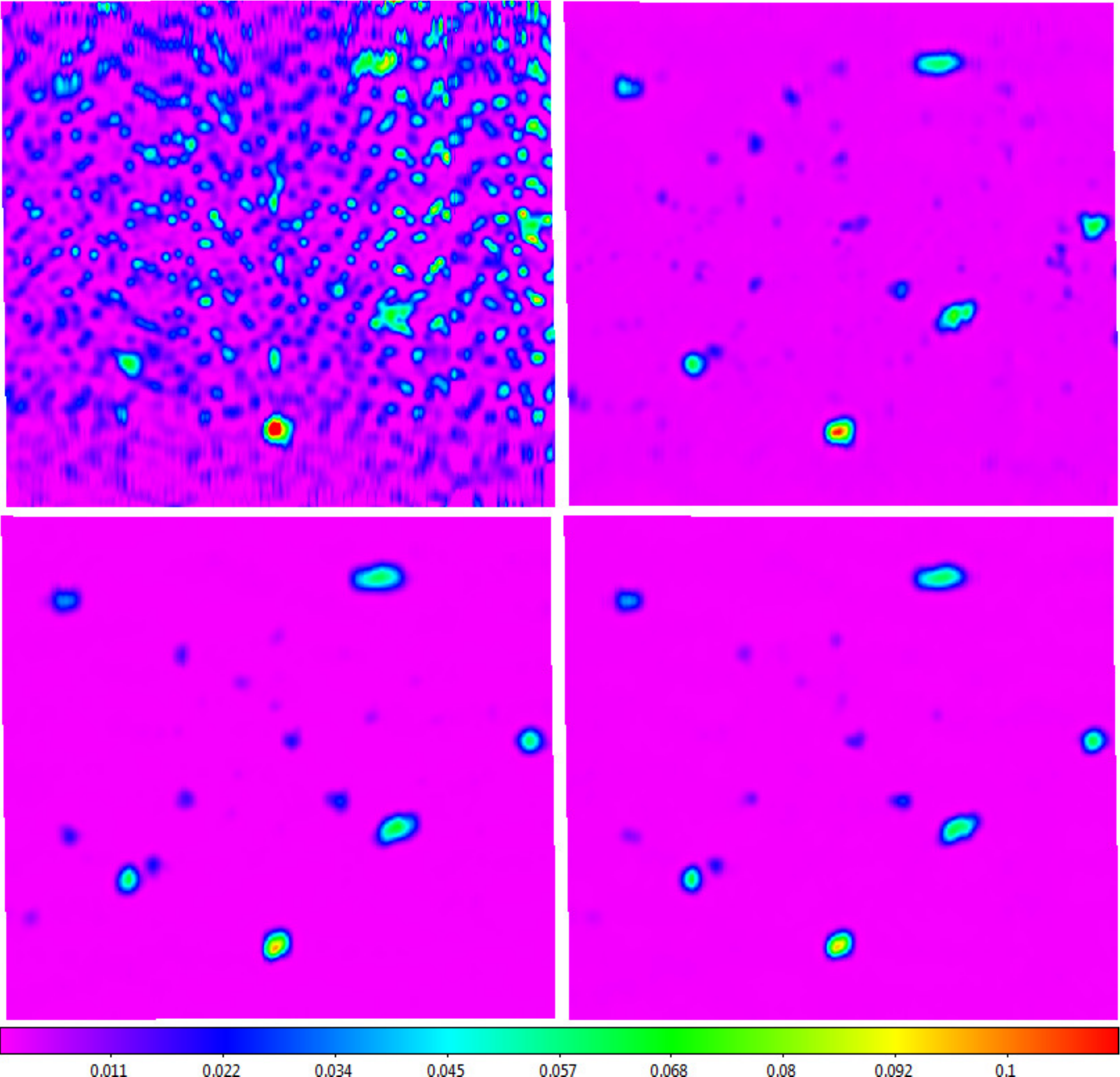}
\caption{\textbf{Top Left:}  Background-subtracted mapping of, from right to left, 3C~84, NRAO~1560/1650, 3C~111, 3C~123, 3C~139.1, and 3C~147, as well as fainter sources, acquired with the 40-foot in L band, using a maximum slew speed nodding pattern.  The data are heavily contaminated by linearly polarized, broadband RFI, affecting only one of the receiver's two polarization channels.  \textbf{Top Right:}  Data from the top-left panel time-delay corrected and RFI-subtracted, with a 0.7-beamwidth scale (Table~2).  \textbf{Bottom Left:}  Identically processed data from the receiver's other, relatively uncontaminated polarization channel, for comparison.  The RFI-subtraction algorithm is not perfect, but performs very well given the original, extreme level of contamination.  \textbf{Bottom Right:}  Background-subtracted and time-delay corrected data from both, equally calibrated polarization channels first appended and then jointly RFI-subtracted.  Locally modeled surfaces have been applied for visualization (\textsection1.2.1, see \textsection3.7).}
\end{figure*}

\begin{figure*}
\plotone{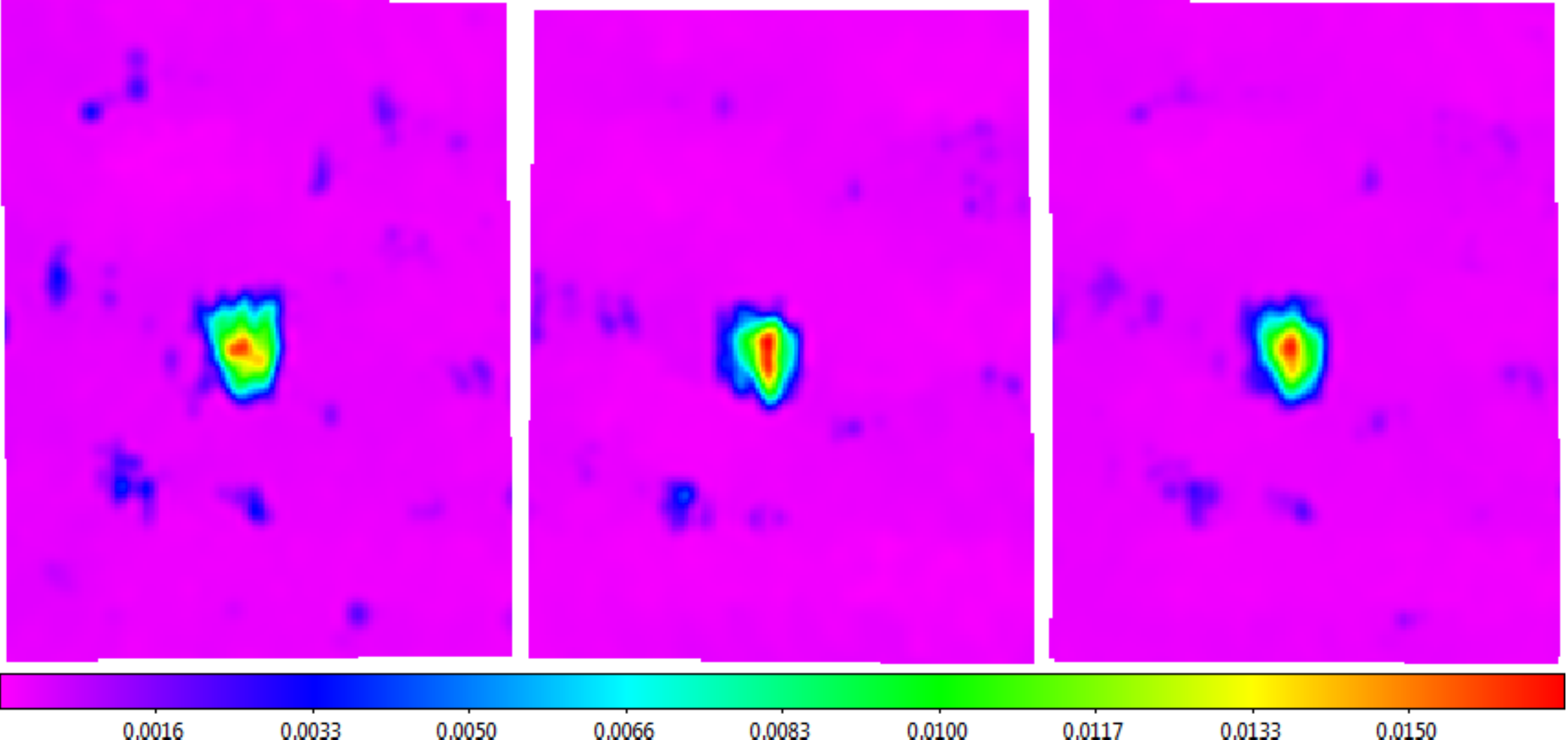}
\caption{Background-subtracted and time-delay corrected 40-foot noddings of Andromeda from Figure~24 (1)~separately RFI-subtracted (left and middle), and (2)~appended and then jointly RFI-subtracted (right), with a 0.7-beamwidth scale (Table~2).  These maps are relatively free of RFI; as such, the appended map is nearly identical to what one gets from averaging the first two maps, but is nearly twice as efficient to produce, computationally (see below).  Locally modeled surfaces have been applied for visualization, with a minimum weighting scale of 1/3 beamwidths (\textsection1.2.1, see \textsection3.7).}
\end{figure*}

\begin{figure*}
\plotone{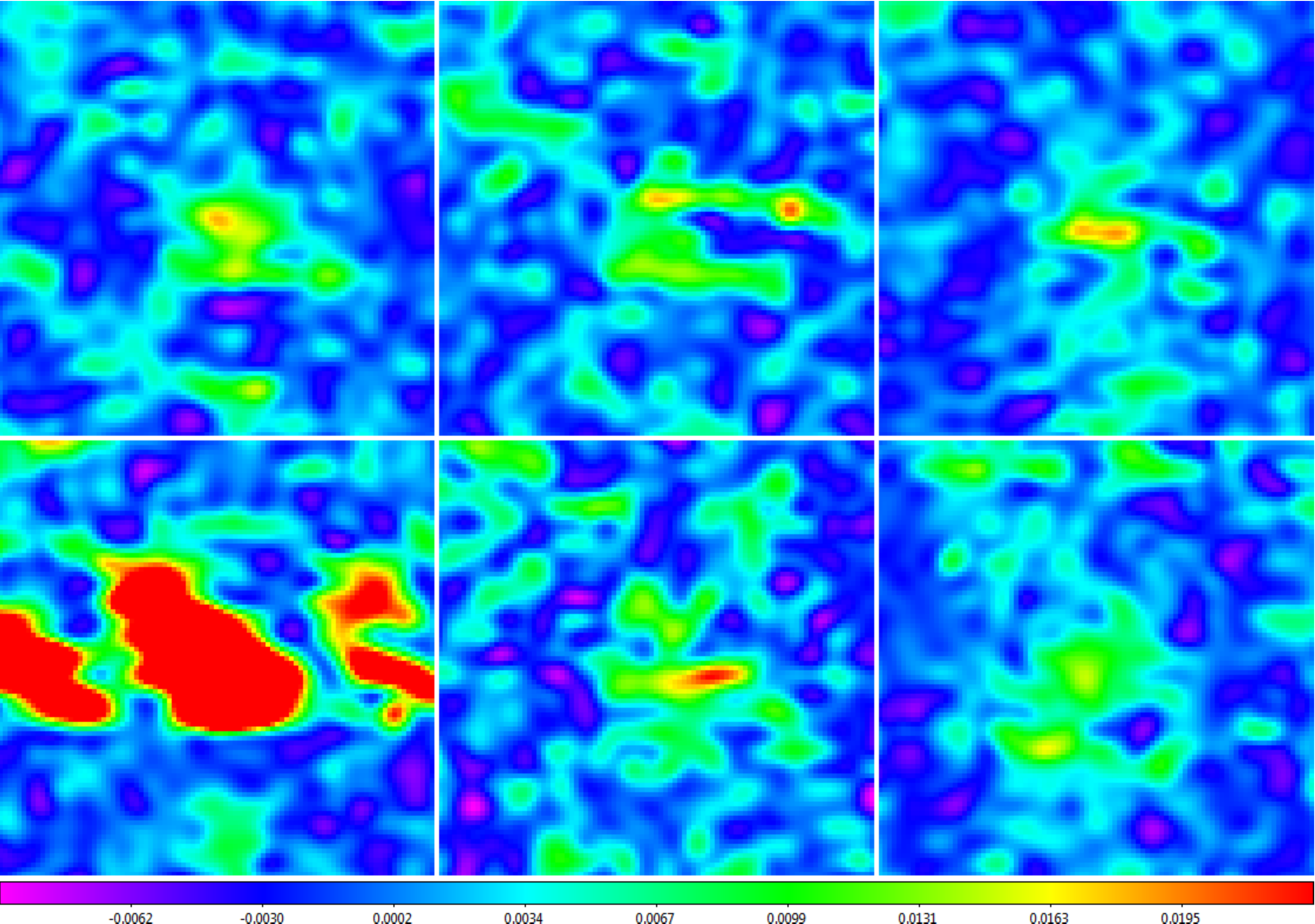}
\caption{Six background-subtracted mappings of Jupiter, acquired with the 20-meter in L band, using 1/5-beamwidth rasters.  Jupiter is only marginally detected in each, except for the fourth, which is significantly contaminated by RFI.  Locally modeled surfaces (\textsection1.2.1, see \textsection3.7) have been applied for visualization only.  }
\end{figure*}

\begin{figure*}
\plotone{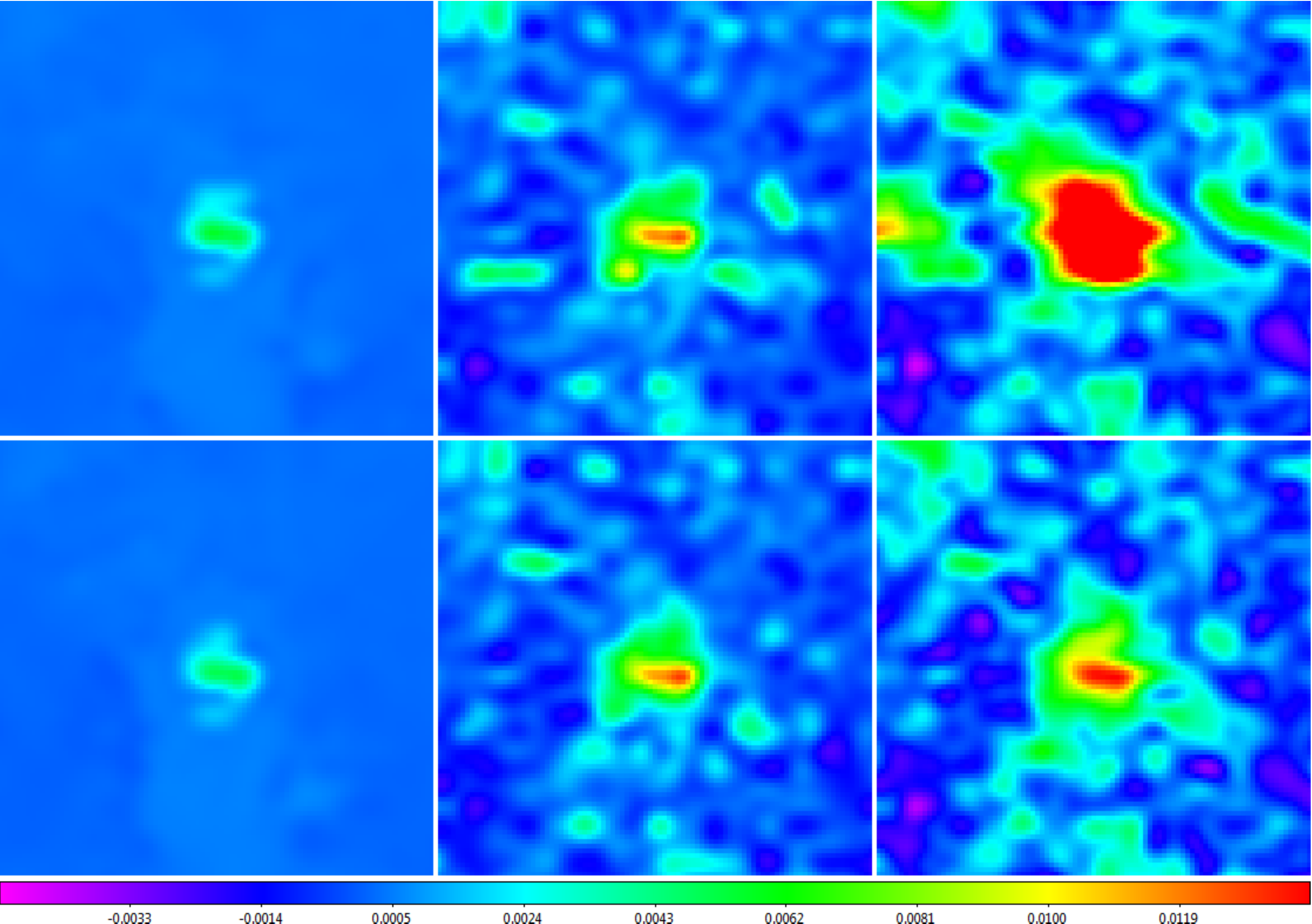}
\caption{\textbf{Top Row:}  The six background-subtracted mappings of Jupiter from Figure~41 appended and jointly RFI-subtracted, with 0.9- (left), 0.1- (middle), and $\approx$0- (right) beamwidth RFI-subtraction scales.  \textbf{Bottom Row:}  The same, but excluding the fourth, significantly RFI-contaminated mapping from Figure~41.  Smaller RFI-subtraction scales recover more near noise-level signal.  Because multiple 0.2-beamwidth mappings are used, the RFI-subtraction scale can be as low as 0.1-beamwidths and still be completely effective at eliminating RFI.  Locally modeled surfaces have been applied for visualization, with a minimum weighting scale of 1/3 beamwidths (\textsection1.2.1, see \textsection3.7).}
\end{figure*}

\begin{figure*}
\plotone{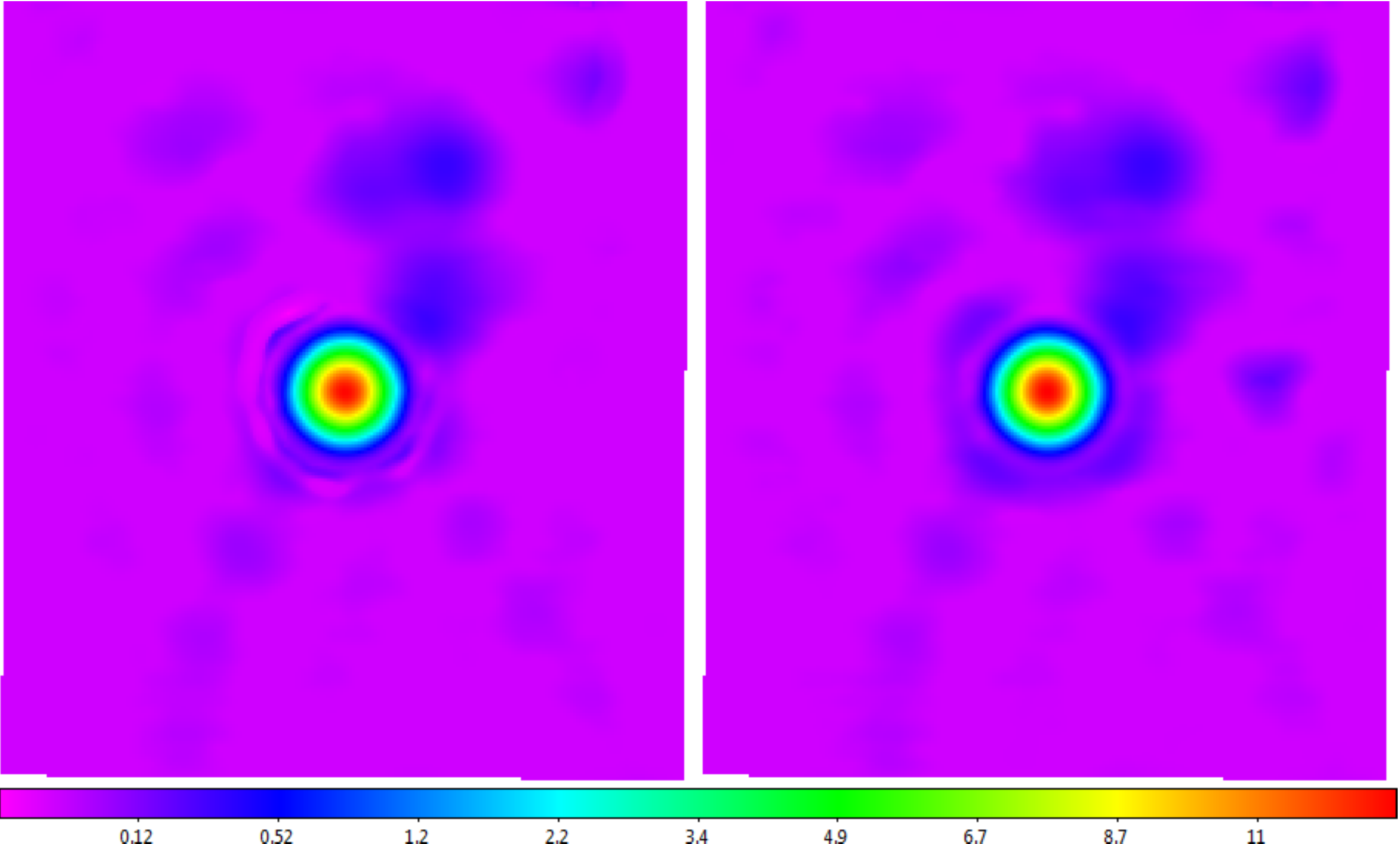}
\caption{Background-subtracted 20-meter L-band raster of Cassiopeia~A from Figure~27 RFI-subtracted, (1)~with the maximum recommended RFI-subtraction scale from Table~2 (0.8 beamwidths), which partially eliminates the first Airy ring (left), and (2)~with half of this scale, which retains this structure (right).  Locally modeled surfaces have been applied for visualization, with a minimum weighting scale of 1/3 beamwidths (\textsection1.2.1, see \textsection3.7).  Square-root scaling is used to emphasize fainter structures.}
\end{figure*}

\begin{figure}
\plotone{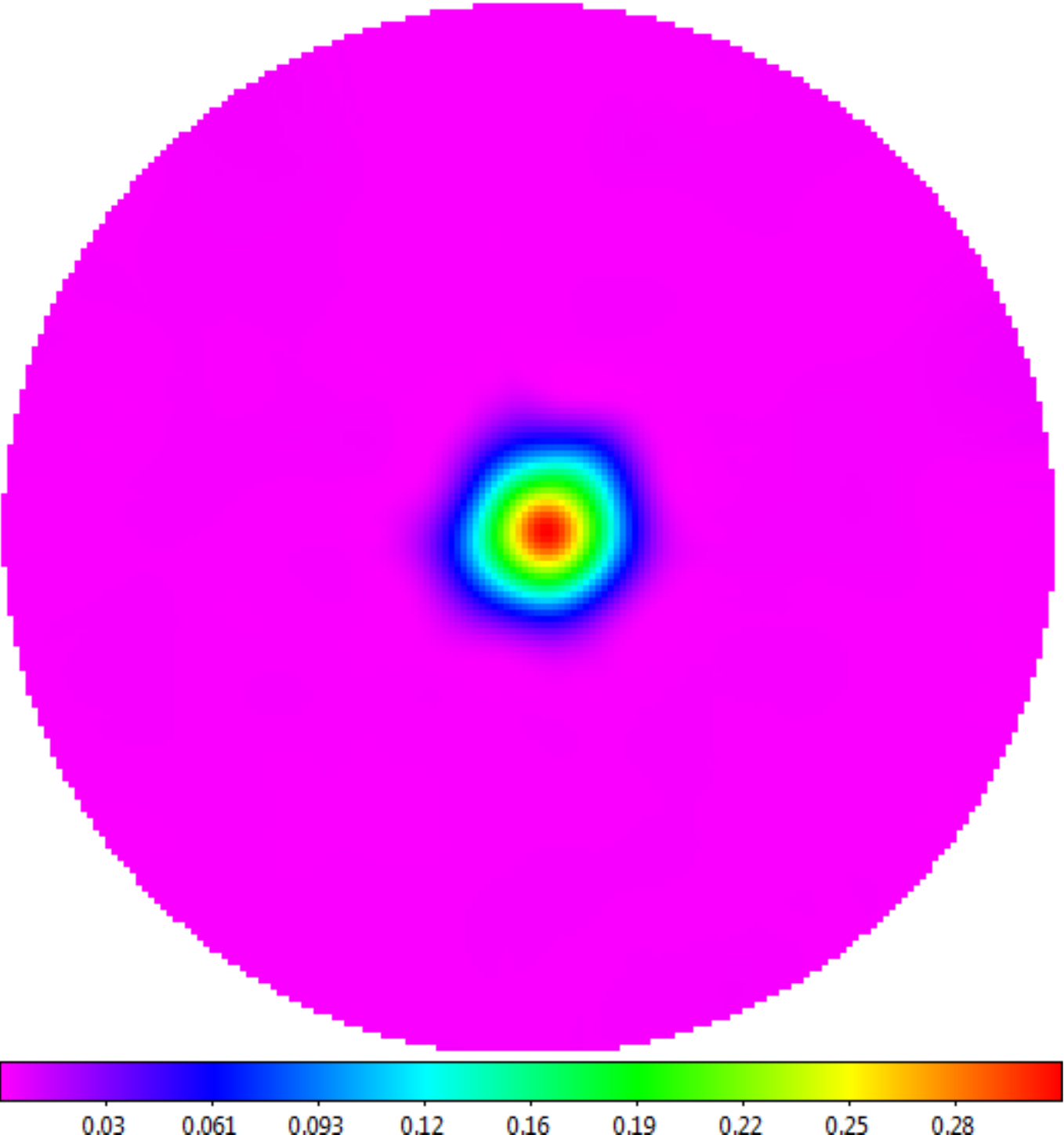}
\caption{Background-subtracted 20-meter X-band daisy of 3C 84 from Figure~25 RFI-subtracted, with a 0.8-beamwidth scale (Table~2).  Locally modeled surfaces have been applied for visualization, with a minimum weighting scale of 2/3 beamwidths (\textsection1.2.1, see \textsection3.7).}
\end{figure}

\begin{figure*}
\plotone{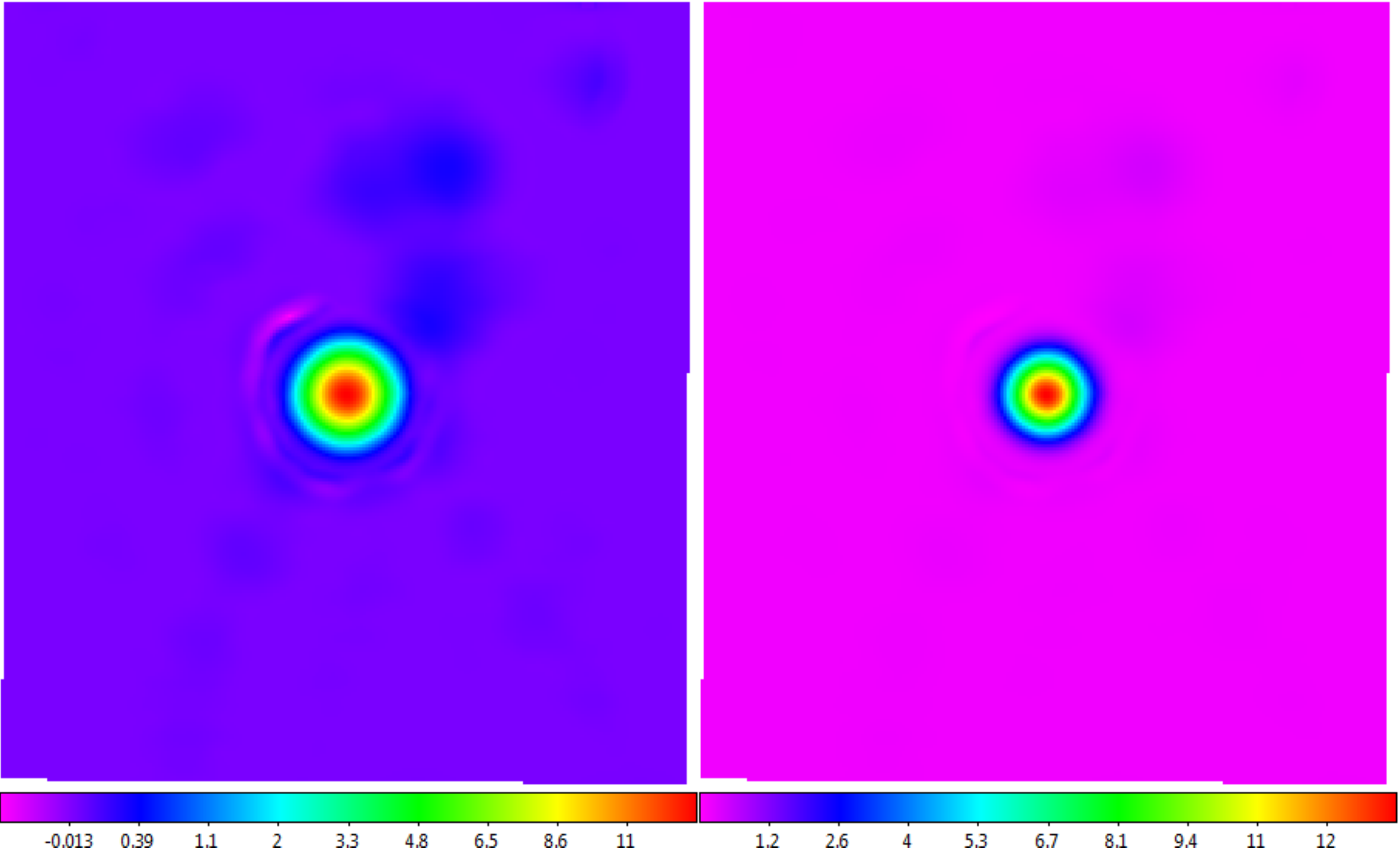}
\caption{The left panel of Figure~43, instead processed without the noise-level prior (see Footnote~27), and visualized with square-root scaling on the left, and with regular, linear scaling on the right.  The surface model undershoots at the base of this high-S/N, well-focused, point source, especially where the first Airy ring has been partially eliminated by the RFI-subtraction algorithm.  Consequently, the noise-level prior is normally included (Figure~43).  Note, however, that even without the noise-level prior, this is a small effect, and is only barely noticeable when visualized on regular, linear scaling (right).}
\end{figure*}

In theory, the RFI-subtraction scale, $\theta_{RFI}$, need only be marginally less than the true FWHM of the beam pattern.  However, in practice, the beam pattern might be more peaked than Equation~9 with $\theta_{RFI} = 1$, or might be asymmetric, being more compact along one axis.  In such cases, one must use a smaller RFI-subtraction scale, or astronomical signal will be mistaken for RFI and removed.  Given this, we have determined maximum recommended values for the telescopes and receivers of \textsection2, empirically, by decreasing this scale until the measured brightness (see \textsection4) of bright sources, observed at multiple parallactic angles, plateaued.  We list these values in Table~2.

\subsubsection{Simulation:  Gaussian Random Noise}

As in §3.3, we test this algorithm by applying it to simulated data of increasing complexity.  We begin by applying it to the background-subtracted Gaussian random noise from the top row of Figure~14, to evaluate its performance in the absence of small- or large-scale structures.  We RFI-subtract these data on a near-full beamwidth scale (0.95 beamwidths; see Figure~34, top row), as well as on a smaller, half-beamwidth scale (Figure~34, bottom row), for comparison.  We find that:  (1)~The RFI-subtracted data are not biased high nor low; and (2)~The noise level of the RFI-subtracted data is indeed significantly less than the noise level of the original, background-subtracted data, having incorporated all of the information within 0.95- and 0.5-beamwidth radius regions, respectively, about each point (Figure~34).  

\subsubsection{Simulation:  Small-Scale Structures}

Next, we apply our RFI-subtraction algorithm, adopting the 0.95-beamwidth scale, to the background-subtracted data from the top row of Figure~17, which includes simulated point sources and short-duration RFI (see Figure~35, top row).  We present residuals in the middle row of Figure~35, but here we have additionally subtracted off the Gaussian random noise residuals from the top row of Figure~34, to help distinguish residuals that are due to the small-scale structures from those that are due to the Gaussian random noise.

We find that the short-duration RFI, which was reduced only marginally by background subtraction, is now reduced by a factor of $\approx$19,000, to $\approx$3\% of the noise level, when background-subtracted on the scale of the map (24 beamwidths), and is reduced to immeasurable levels when background-subtracted on the 6-beamwidth scale, as well as on smaller scales (Figure~20).  

We also find that the source residuals, compared to their post-background subtraction/pre-RFI subtraction values from Figure~17, (1)~are reduced from sub-noise to completely negligible levels beyond where the sources intersect the noise level, (2)~are increased slightly at these boundaries, but (3)~are otherwise fairly consistent with the three categories of post-background subtraction residuals that we identified in \textsection3.3.2, within these boundaries.  As such, this bias is again smaller by a factor of $\approx$2 -- 3 in the more realistic case of a less-winged beam function (Figure~35, bottom row).  

\subsubsection{Simulation:  1D Large-Scale Structures}

Next, we apply our RFI-subtraction algorithm, again adopting a 0.95-beamwidth scale, to the background-subtracted data from the top row of Figure~19, which includes residual en-route drift and long-duration RFI (see Figure~36, top row).  We present post-RFI subtraction residuals in the bottom row of Figure~36, but here we have additionally subtracted off the residuals from the top and middle rows of Figures 34 and 35, to help distinguish residuals that are due to the 1D large-scale structures from those that are due to the Gaussian random noise and the small-scale structures, respectively.  

Beyond where the sources intersect the noise level, we find that the long-duration RFI is now reduced by a combined (background and RFI subtraction) factor of $\approx$19,000, to $\approx$1\% of the noise level, when background-subtracted on the scale of the map (24 beamwidths), and is reduced to immeasurable levels when background-subtracted on the 6-beamwidth scale, as well as on smaller scales (Figure~20).  

Also beyond where the sources intersect the noise level, we find that the en-route drift is now reduced by a combined factor of $\approx$20, to $\approx$16\% of the noise level, when background-subtracted on the scale of the map; by a factor of $\approx$730, to $\approx$0.4\% of the noise level, when background-subtracted on the 6-beamwidth scale; and by even greater factors when background-subtracted on even smaller scales (Figure~20).  For reference, the basket-weaving technique of Winkel, Floer \& Krauss (2012) (\textsection1.2.3) reduces en-route drift by a factor of $\approx$10, and then only under idealized circumstances.

Within these boundaries, the residuals are fairly consistent with the post-background subtraction residuals of Figure~19.  These are biased neither high nor low, and are noise-level.  

\subsubsection{Simulation:  2D Large-Scale Structures}

Next, we apply our RFI-subtraction algorithm, again adopting a 0.95-beamwidth scale, to the background-subtracted data from the top row of Figure~22, which includes residual large-scale astronomical and elevation-dependent signal (see Figure~37, top row).  We present post-RFI subtraction residuals in the bottom row of Figure~37, but here we have additionally subtracted off the residuals from the top, middle, and bottom rows of Figures 34, 35, and 36, to help distinguish residuals that are due to the 2D large-scale structures from those that are due to the Gaussian random noise, the small-scale structures, and the 1D large-scale structures, respectively.  

Away from the sources, we find that the elevation-dependent signal is now reduced by a combined factor of $\approx$2800, to $\approx$4\% of the noise level, when background-subtracted on the scale of the map (24 beamwidths), and is reduced to immeasurable levels when background-subtracted on the 6-beamwidth scale, as well as on smaller scales (Figure~20).

Also away from the sources, we find that the large-scale astronomical signal is now reduced by a combined factor of $\approx$800, to $\approx$11\% of the noise level, when background-subtracted on the scale of the map; by a factor of $\approx$5900, to $\approx$2\% of the noise level, when background-subtracted on the 6-beamwidth scale; and by even greater factors when background-subtracted on even smaller scales (Figure~20).  

Within the vicinity of the sources, the residuals are fairly consistent with the post-background subtraction residuals of Figure~22.  On the smaller background-subtraction scales, these are as likely to be biased high as low, and are noise level.  However, on the 24-beamwidth scale, these remain non-negligible.  

\subsubsection{20-Meter and 40-Foot Data}

We now apply this algorithm to real data.  First, we apply it to the 20-meter L-band raster from Figures 5 (raw) and 23 (background-subtracted on small and large scales; see Figure~38).  RFI and en-route drift are eliminated on both scales, as is elevation-dependent signal.  Large-scale astronomical signal is eliminated on the smaller background-subtraction scale and is significantly reduced on the larger background-subtraction scale.

Next, we test the algorithm in the limit of extreme RFI contamination (see Figure~39, top-left panel).  The source of the RFI is a broadband transmitter at the Roanoke, VA airport, about 100 miles south of Green Bank.  The signal is linearly polarized, affecting only one of the receiver's two polarization channels.  Not only is the RFI almost completely eliminated (Figure~39, top-right panel, compared to the other polarization channel, bottom-left panel), our time-delay correction algorithm (\textsection3.5), which must be applied before RFI subtraction, correctly aligns the scans despite the extreme contamination, and despite having only lower signal-to-noise sources to inform its measurement.  It measures a time-delay of 0.30 seconds, which is nearly the same as the value that it measures from the uncontaminated polarization channel (0.24 seconds).

Note, since this algorithm is independent of the telescope's mapping pattern (\textsection1.2.2), multiple observations can be combined trivially.  After determining each observation's 2D noise model (\textsection3.5), one simply appends all of the observations' background-subtracted and time-delay corrected scans, and each scan's 2D noise model value, as if they were a single observation, and applies the RFI-subtraction algorithm to them jointly.\footnote{For a sequence of observations centered on a moving object, such as a planet (e.g., see Figure~42), we switch to object-centered coordinates before appending.  Failure to do this can cause the object to be partially, or completely, mistaken for RFI, and eliminated.}  We demonstrate this in Figure~40, for the two observations of Andromeda from Figure~24 (as well as in the bottom-right panel of Figure~39, for the two polarization channels of that observation).

A distinct advantage of this approach is that not just narrow, but broad structures that exist in one, or multiple, mappings, but not others, such as an extended period of RFI, spanning multiple scans, or the sun's diffraction pattern, if different between different mappings, are eliminated.  However, structures that exist in all of the mappings, at the same level, at least within each mapping's separately measured 2D noise level, survive.  

Furthermore, by appending multiple mappings, one can eliminate RFI and similar unwanted structures with a substantially smaller RFI-subtraction scale, since each measurement can be compared to other measurements that were taken at nearly the same sky location, but at different times, instead of only to surrounding measurements, some of which were taken at nearly the same time (and hence may be similarly contaminated).  In fact, if enough observations are taken, this scale can be as small as half of the (local, see \textsection3.7) gap in the mapping pattern.  This is particularly advantageous when mapping low-S/N sources, since our RFI algorithm underestimates signal near the noise level (Figure~35).  In other words, there is a trade-off between RFI subtraction and low-S/N photometry, but this can be mitigated by taking multiple observations and using a minimal RFI-subtraction scale.\footnote{Note, this does not apply to two polarization channels acquired simultaneously (e.g., Figure~39), since (unpolarized, or differently polarized) RFI, as well as en-route drift, etc., will likely be correlated between the two channels.}$^,$\footnote{Alternatively, or additionally, underestimated photometry can be corrected using an empirically-determined expression that we introduce in \textsection4.  However, this expression is only an approximation, and consequently performs better the less that is requested of it (e.g., if the RFI-subtraction scale is already small, which appending multiple mappings can facilitate).}

We demonstrate both of these advantages with six low-S/N mappings of Jupiter, one of which is significantly contaminated by an extended period of RFI, spanning multiple scans (see Figure~41).  In the top row of Figure~42, we append all six of these mappings and jointly RFI subtract them (1)~on the maximum recommended RFI-subtraction scale from Table~2 (0.9~beamwidths, left panel), (2)~on the minimum recommended RFI-subtraction scale for appended mappings (0.1~beamwidths, since these are 0.2-beamwidth rasters, middle panel), and (3)~on an RFI-subtraction scale of almost zero (corresponding to almost no RFI subtraction).  While RFI from the fourth mapping ruins the result on the $\approx$0-beamwidth scale, it is completely eliminated on the 0.1-beamwidth scale (as well as on larger scales).  This can be seen by comparing the top row to the bottom row, where we have repeated the processing, but excluding the contaminated fourth mapping.  As expected, the smaller the RFI-subtraction scale, the more near-noise level signal remains (of course, more noise remains as well).

Yet another advantage of this approach is that it is very efficient computationally (especially when smaller RFI-subtraction scales can be employed).  Instead of RFI-subtracting and surface-modeling (\textsection1.2.2; see \textsection3.7) multiple mappings, and then finding a way to combine the resulting images such that structures that differ from image to image are eliminated, but otherwise, the noise level is reduced, these algorithms need be applied only once.  However, each mapping needs to be calibrated to the same level, else data that were calibrated higher will be eliminated in favor of data that were calibrated lower, if separated by more than their respective noise levels.  Since all of the observations in Figures 41 and 42 were taken within the same day, and both of the observations in Figures 24 and 40 were taken within two days of each other, gain calibration (\textsection3.1) is sufficient.  However, if observations are separated by extended periods of time, or if they are taken from different polarization channels (e.g., Figure~39), the receiver's noise diodes should be cross-calibrated between these observations, ideally by mapping and photometering (see {\color{black}\textsection4}) one or more standard sources, such as {\color{black}Cas}~A, {\color{black}Cyg}~A, {\color{black}Tau}~A, or {\color{black}Vir}~A (e.g., Baars et al.\@ 1977, Trotter et al.\@ 2017).\footnote{These calibration sources are appropriate for small, single-dish radio telescopes, like the 20-meter and the 40-foot, but are not necessarily appropriate for larger single-dish telescopes, or for interferometers.  Also note that Cas~A and Tau~A are fading, on timescales of years and decades, respectively (e.g., Trotter et al.\@ 2017).  However, they can still be used to cross-calibrate observations on shorter timescales.}

Otherwise, one need only take care when appending mappings that do not fully coincide.  Background-subtracted signal can be underestimated near the ends of scans (\textsection3.3.2), in which case data that were not taken near the end of a scan in one mapping could be eliminated in favor of underestimated data that were taken near the end of a scan in another mapping.  We recommend clipping data within $\approx$2 beamwidths of the ends of scans before appending such, non-coincident mappings (\textsection3.3.2).  

When mapping high-S/N point sources, Airy rings are sometimes visible, particularly if the telescope is well focused (see Table~1).  However, Airy rings are typically half the width of the point-spread function, and consequently are partially eliminated when the maximum recommended RFI-subtraction scale is used, the first Airy ring in particular (see Figure~43, left panel).  If one wishes to retain these structures, one need only halve the RFI-subtraction scale (Figure~43, right panel), though at greater risk of RFI contamination. 

Finally, we apply the RFI-subtraction algorithm to the 20-meter X-band daisy from Figure~25, to demonstrate its application to a non-rectangular mapping pattern (see Figure~44).  Although these data were not contaminated with RFI, residual en-route drift is eliminated.

\subsection{2D Surface-Modeling}

In this section, we present our algorithm for ``regridding'' data into an image.  Unlike most algorithms, (1)~ours uses weighted modeling instead of weighted averaging, to avoid blurring the image beyond the telescope's native resolution, and (2)~ours can be applied after the data have been contaminant-cleaned, instead of having to replace the data with an approximation of itself before contaminant-cleaning can begin (\textsection1.2.2).  On this note, our algorithm can be applied at any stage of the contaminant-cleaning process, if visualization is desired.  For example, Figures 3, 5, 13, 16, 18, 21, 24 (top row), and 25 (left panel) are visualizations of raw data; Figures 14, 17, 19, 22, 23, 25 (right panel), 26 (left panel), 27 (left panel), and 39 (top-left panel) are visualizations post-background subtraction (\textsection3.3); Figures 24 (bottom row), 26 (right panel), 27 (right panel), 33 (top panel), and 41 are visualizations post-time-delay correction (\textsection3.4), and Figures 33 (middle and bottom panels), 34 -- 38, 39 (all but top-left panel), 40, 42, 43 and 44 are visualizations post-RFI subtraction (\textsection3.6).  

Another advantage of this approach is that any pixel density can be selected for these images.  Most algorithms have to limit the pixel density to streamline computation, since contaminant-cleaning is done on the regridded data, instead of on the original data.  But since we contaminant-clean prior to regridding, this is not an issue.  Our default pixel scale is 1/20th of a beamwidth, but this is user-configurable.

For each pixel, we fit a flexible surface model to all data that are within 1 beamwidth, weighting data that are closest to the pixel the highest (see below).  We evaluate the fitted model only at this pixel, so it need only fit well here (e.g., Figure~4).  We have found that third-order, 2D polynomials (1)~are sufficiently flexible to, pixel by pixel, reproduce most all diffraction-limited structures (e.g., Figure~3),\footnote{The only exception is at the base of sufficiently high S/N point sources, if the telescope is well-focused and the point-spread function falls off steeply (i.e., does not have long wings), and especially if the first Airy ring is partially or completely eliminated by the RFI-subtraction algorithm (\textsection3.6.5).  In this case, Equation~11, which is the product of two third-order polynomials, where we have dropped fourth- and higher-order terms, recovers a bit too slowly, causing the pixel-by-pixel fitted surface to undershoot in this region, albeit only marginally.  Normally, this is barely noticeable, unless the final image is viewed with a non-linear scaling, such as square-root or hyperbolic-arcsine, to emphasize fainter structures (e.g., see Figure~45).  Fourth-order polynomials would be better in this case, but there is seldom enough data to constrain (actually, to over-constrain, see Footnotes 28 and 30) this many parameters.  Consequently, we instead provide the user with two surface-model fitting options.  The first is just regular regression for weighted data, again, evaluated at each pixel in the final image.  The second imposes a noise-level prior when the first option yields $a_{00}<0$, suppressing all negative values:
\begin{equation}
\begin{split}
p(a_{ij}) \propto \prod_n & \exp\left[-\frac{w_n}{2}\left(\frac{a_{00}}{\sigma_z}\right)^2\right] \\
& \times \exp\left\{-\frac{w_n}{2}\left[\frac{z_n-z(\Delta x_n,\Delta y_n|a_{ij})}{\sigma_z}\right]^2\right\}.
\end{split}
\end{equation}
\noindent (This manifests itself only in the first term of the regression matrix:  $\sum w_n \rightarrow 2\sum w_n$.)  We normally employ this latter option (e.g., Figure~43), unless surface-modeling (1)~pre-RFI subtraction data, or (2)~low-S/N sources, post-RFI subtraction (e.g., Figure~42).} but (2)~use few enough parameters (ten) such that these parameters can be well constrained for most all sampling densities and mapping patterns:\footnote{In the event of a very sparse mapping, or a very sparse region of a mapping, Equation~11 can, of course, be underdetermined.  Consequently, before applying it, we confirm that, within this 1-beamwidth radius region, we are fitting to at least ten points, spread over at least five scans, with at least one scan having at least five points.  These last two conditions ensure overdetermination along each axis, preventing the possibility of wild oscillations between scans or points.  If these conditions are not met, we downgrade to second-order, 2D polynomials.  In this case, we first confirm that we are fitting to at least six points, spread over at least four scans with at least one scan having at least four points (for the same reasons as above).  If these conditions are also not met, we again downgrade, to first-order, 2D polynomials (i.e., a plane).  In this case we first confirm that we are fitting to at least three points, spanning at least two scans, with one scan having at least two points (two instead of three, because planes cannot oscillate).  Failing this, we excise the pixel from the final image.  Note:  These downgraded versions of Equation~11 are built in if needed, but are rarely called upon (an exception to this would be in the outskirts of low petal-number, or oversized, daisies).}
\begin{equation}
z(\Delta x, \Delta y) = \sum_{i=0}^3 \sum_{j=0}^{3-i} a_{ij}(\Delta x)^i(\Delta y)^j,
\end{equation}
where $z$ is the locally modeled signal, $\Delta x$ and $\Delta y$ are angular distances from the central pixel along each coordinate, and $a_{ij}$ are the polynomial coefficients.  At this pixel, Equation~11 simplifies to $z(0,0)= a_{00}$, which streamlines the computation.  We repeat this process for all pixels in the image.

It is important that the (radial) weighting function drop to zero at or before this 1-beamwidth hard limit, else there will be discontinuities in the modeled surface as the central pixel's location changes, as individual data points enter and exit the set of data that is being fit to.  In the absence of the 1-beamwidth hard limit, dropping the weight to zero also has the advantage of limiting the amount of data that each local model is being fit to, again streamlining the computation.  For the weighting function, we use a function similar to Equation~9, but raised to a power:
\begin{equation}
w(\Delta\theta) = 
\begin{cases}
\cos^\alpha\left(\frac{\pi\Delta\theta}{2\,\rm{beamwidths}}\right) & \text{if $\Delta\theta < 1$ beamwidth} \\
0 & \text{otherwise}
\end{cases},  
\end{equation}
\noindent where:
\begin{equation}
\alpha = -\frac{\log2}{\log\left[\cos\left(\frac{\pi\theta_w}{4\,\rm{beamwidths}}\right)\right]},
\end{equation}
\noindent and where $\theta_w$ is the user-defined FWHM of the weighting function (see Figure~46).  Equation~12 approximates the beam function of the telescope when $\theta_w=1$ beamwidth (Figure~31).  

Smaller values of $\theta_w$ favor the data that are closer to the central pixel, resulting in a more accurately modeled value at this pixel.  However, smaller values of $\theta_w$ also reduce the weighted number of data points that contribute to the fit, resulting in a less precisely modeled value at this pixel, and hence a noisier image overall.  We have found that Equation~11 is sufficiently flexible to reproduce most all (Footnote~27) diffraction-limited structures, to $>$99\% accuracy, if $\theta_w\la1/3$ beamwidths (see Figure~47, top two rows).  That said:

\begin{figure}
\plotone{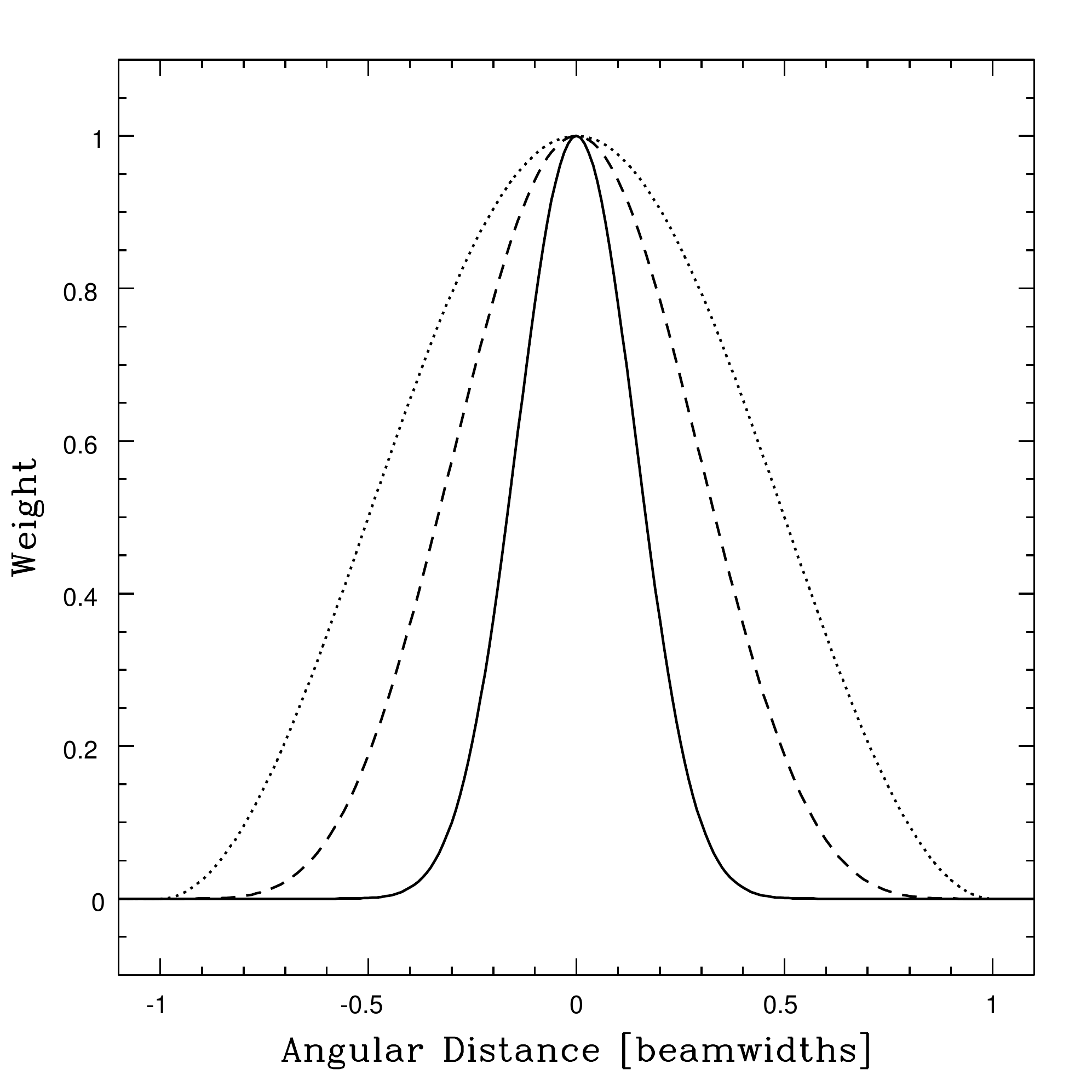}
\caption{Weighting function (Equation~12), with FWHM $\theta_w = 1/3$ (solid), $2/3$ (dashed), and 1 (dotted) beamwidths.}
\end{figure}

\begin{figure*}
\epsscale{0.95}
\plotone{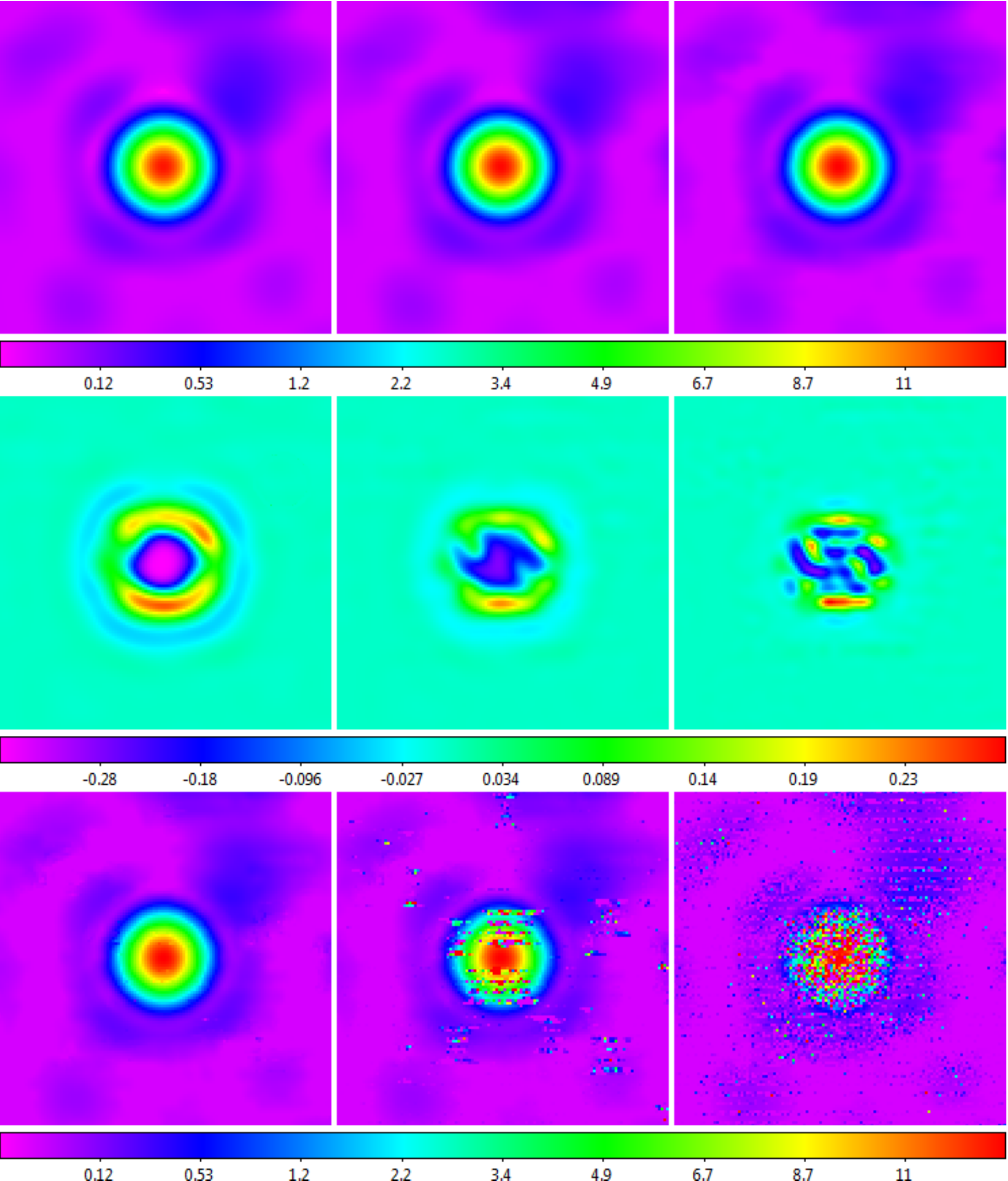}
\caption{\textbf{Top Row:}  Background- and RFI-subtracted data from the right panel of Figure~43 surface-modeled with fixed weighting scales $\theta_w=2/3$ (left), $1/2$ (middle), and $1/3$ (right) beamwidths, and zoomed in on the source.  Square-root scaling is used to emphasize fainter structures.  \textbf{Middle Row:}  Surface models from the top row, at the locations of the background- and RFI-subtracted data (1)~minus these data (residuals), and (2)~re-surface-modeled for visualization.  These weighting scales underestimate the peak of the source, but only by $\approx$4\%, $\approx$2\%, and $<$1\%, respectively, and overestimate the base of the source, corresponding to additional blurring of the source, but only by $\approx$1\%, $\approx$1/2\%, and $<$1/4\%, respectively.  Weighting scales $\theta_w<1/3$ beamwidths serve only to better visualize sub-Nyquist scale structures, such as contaminants.  \textbf{Bottom Row:}  Same as the top row, but with fixed weighting scales $\theta_w=1/6$, $1/9$, and $1/12$ beamwidths.  These weighting scales are too small (i.e., $\theta_w$ is sufficiently smaller than $4/3 \times \theta_{gap} \approx 4/3 \times 1/5$ beamwidths $=0.27$ beamwidths) that the surface model is not always well constrained between data points.}
\end{figure*}

\begin{figure*}
\plotone{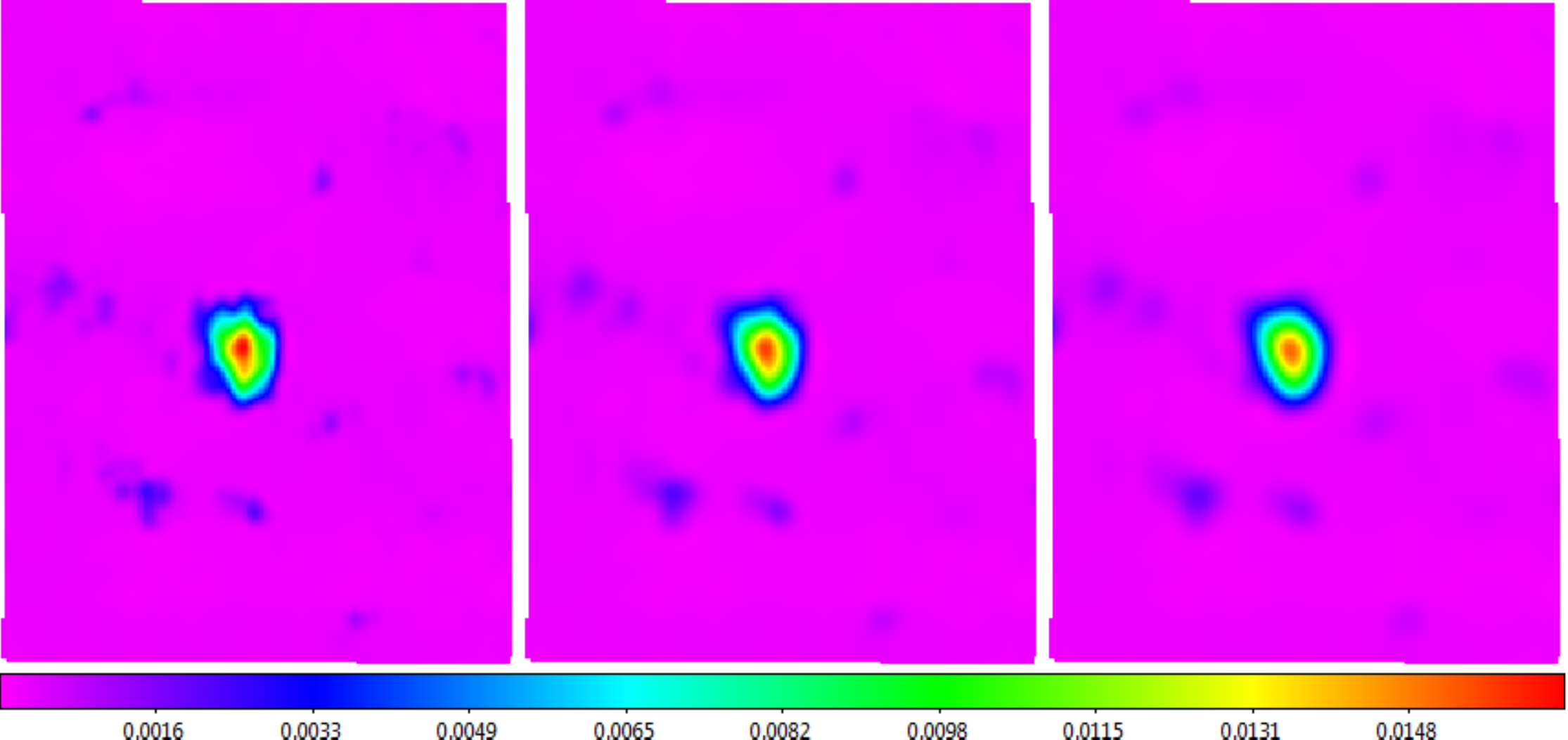}
\caption{Background- and RFI-subtracted data from Figure~40 surface-modeled with fixed weighting scales $\theta_w=1/3$ (left), $1/2$ (middle), and $2/3$ (right) beamwidths.  The higher weighting scales result in significantly smoother images, but without additionally blurring the source by more than $\approx$1/2\% and $\approx$1\%, respectively (Figure~47).}
\end{figure*}

\begin{figure*}
\plotone{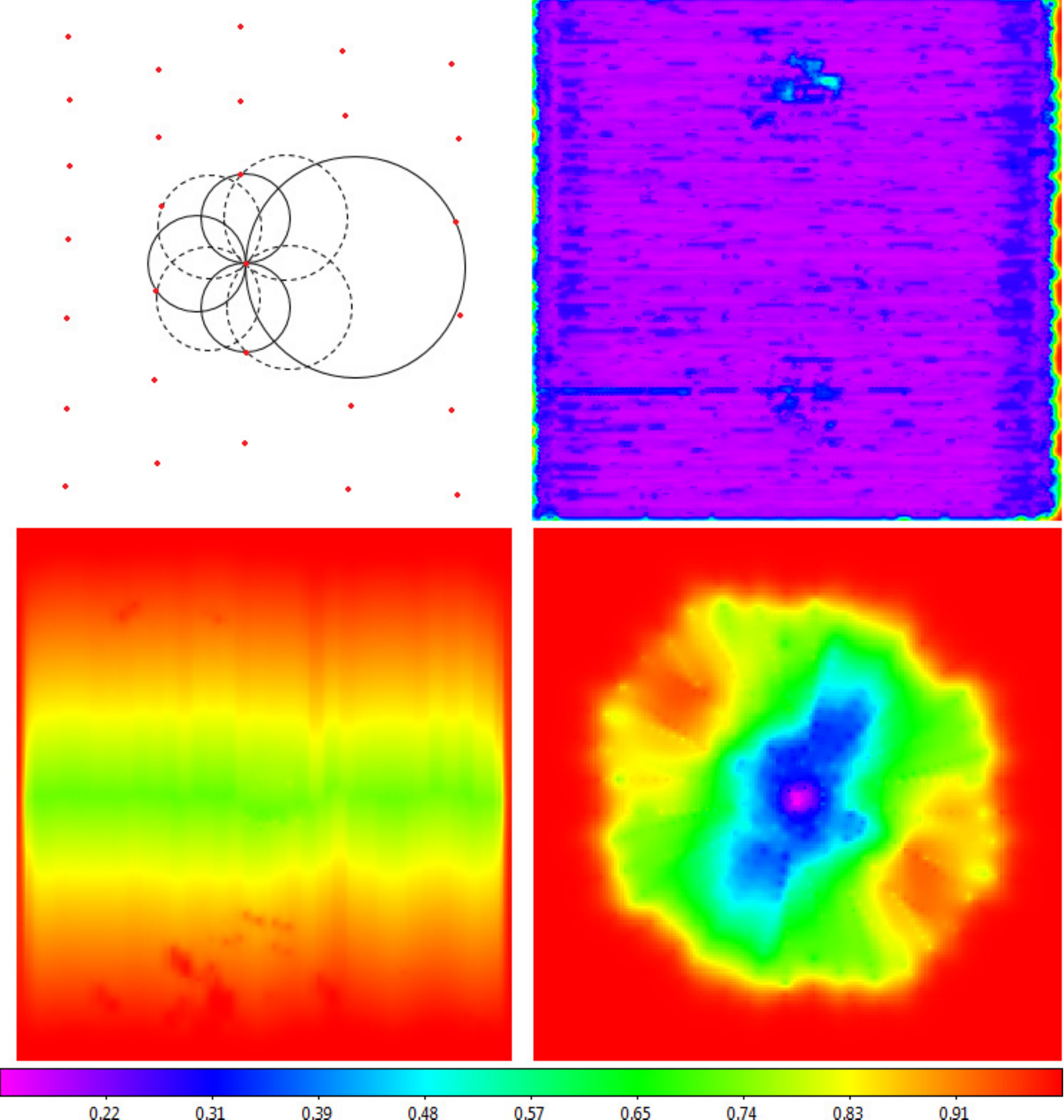}
\caption{\textbf{Top Left:}  From each data point, we ``blow a bubble'' in each of the coordinate system's eight cardinal directions, until it intersects another data point that is within $45\degr$ of the blow direction, and that is more than a minimum angular distance away.  We determine this minimum angular distance by taking all angular distances between consecutive measurements in the observation and performing outlier rejection on them, determining the minimum angular distance that is not rejected.\footnote{For daisies, the telescope's slew speed is variable, with the smallest angular distances between consecutive measurements occurring at the ends/beginnings of scans, which is also where this is most difficult to measure accurately:  The telescope's rapid transition between deceleration and acceleration at the ends/beginnings of scans is often messy, resulting in data-point clustering.  Consequently, we instead measure these angular distances at the center of each scan, where the telescope is moving fastest, with minimum acceleration/deceleration, and then divide by half of the number of scans, which gives what these angular distances should have been at the ends/beginnings of scans.}$^,$\footnote{We reject outliers as described in {\color{black}\textsection4 -- \textsection6} of Maples et al.\@ {\color{black}2018}, using iterative bulk rejection followed by iterative individual rejection (using the mode $+$ broken-line deviation technique, followed by the median $+$ 68.3\%-value deviation technique, followed by the mean $+$ standard deviation technique), using the smaller of the low and high one-sided deviation measurements.  Data are weighted equally.}$^,$\footnote{In the case of appended mappings (\textsection3.6.5), we adopt the minimum of each mapping's minimum non-rejected angular distance, in each overlap region.}  Employing this minimum angular distance decreases computation time, and prevents undersized bubbles from being blown, due to data-point clustering, usually at the ends/beginnings of scans, where the telescope changes direction.  We take the largest bubble's diameter as our local measure of $\theta_{gap}$, from it calculate $\theta_w$ (Equation~14), and then interpolate between these values at each pixel in the final image (Footnote~28).  \textbf{Top Right:}  Raster example:  Resulting values of $\theta_w=min\{\frac{4}{3}\times\theta_{gap},1\}$ for the 20-meter horizontal raster from Figures 5, 23, and 38.  Larger gaps are measured where the telescope's momentum caused it to overshoot when changing directions, near the ends/beginnings of scans, and where the wind pushed the telescope across its direction of motion.  Larger gaps are also measured where the RFI-subtraction algorithm removed data points, both due to RFI (e.g., the scan intersecting 3C 273), and in the outskirts of the beam pattern around Virgo~A.  \textbf{Bottom Left:}  Nodding example:  Resulting values of $\theta_w=min\{\frac{4}{3}\times\theta_{gap},1\}$ for the 40-foot nodding from Figure~26.   This is a sparse mapping, with only $\approx$0.4-beamwidth gaps between scans at its middle declination, which, since this is a nodding pattern, yields gaps that are twice as large at the top and bottom of the mapping (Figure~2, middle panel).  \textbf{Bottom Right:}  Daisy example:  Resulting values of $\theta_w=min\{\frac{4}{3}\times\theta_{gap},1\}$ for the 20-meter daisy from Figures 25 and 44.  The pattern is asymmetric, because gravity's pull on the telescope resulted in narrower petals along the lower-left to upper-right diagonal, and wider petals orthogonally (see Figure~53).}
\end{figure*}

1.  Unless one is trying to visualize narrow contaminants, such as RFI, en-route drift, etc., there is no reason to use values of $\theta_w<1/3$ beamwidths, because there is no additional diffraction-limited information to be gained on sub-Nyquist scales.

2.  Values of $\theta_w>1/3$ beamwidths result in only marginally reduced accuracy, but in significantly greater precision.  For example, $\theta_w=1/2$, $2/3$, and 1 beamwidths result in only $\approx$2\%, $\approx$4\%, and $\approx$6\% underestimates at the source's peak, respectively, and these underestimates are almost perfectly compensated by overestimates at the source's base (Figure~47, top two rows).\footnote{Compare this to $\approx$20\%, $\approx$30\%, and $\approx$40\% underestimates at the source's peak, and consequently significantly blurred reconstructions, in the case of weighted averaging (Figure~3).}  This corresponds to additional blurring of the source, but only by $\approx$1/2\%, $\approx$1\%, and $\approx$2\%, respectively.  However, this does result in significantly smoother images (see Figure~48).

3.  $\theta_w$ must be sufficiently large to encompass enough data to constrain, and preferably to over-constrain, Equation~11.  Otherwise, the fitted 2D polynomials may vary wildly between the data points (Figure~47, bottom row).  To avoid this, we find that $\theta_w$ should be larger than $\approx$$4/3\times\theta_{gap}$, where $\theta_{gap}$ is the largest gap between data points in the vicinity of the central pixel (see below).\footnote{The factor of $4/3$ ensures that at least five scans are encompassed within an $\approx$$\theta_w$-radius circle about the central pixel.  This is the minimum number of scans necessary to over-constrain Equation~11 (see Footnote~28); scans beyond this radius carry too little weight to as meaningfully contribute (Figure~46).}   

Consequently, we recommend that:
\begin{equation}
\theta_w=max\left\{\theta_{min},min\left\{\frac{4}{3}\times\theta_{gap},1\,\rm{beamwidth}\right\}\right\},
\end{equation}
\noindent where $\theta_{min}=0$ beamwidths when trying to visualizing sub-Nyquist scale contaminants, $\theta_{min}\geq1/3$ beamwidths when trying to visualize diffraction-limited structures, and we have capped $\theta_w\leq1$ beamwidth, because larger values make the weighting function (Equation~12) increasingly flat-topped, and only diminishingly wider.  For our post-RFI subtraction images of diffraction-limited structures, we have used:  (1)~$\theta_{min}=1/3$ beamwidths, when evaluating the performance of various aspects of the algorithm (e.g., Figures 34, 35, 36, 37, 39 -- here, $4/3\times\theta_{gap}>\theta_{min}$ everywhere in the image, e.g., see Figure~49, bottom-left panel -- 40, 42, 43 and 45), and (2)~our default value of $\theta_{min}=2/3$ beamwidths, when a more polished, final image is desired (e.g., Figures 38 and 44).

\begin{figure*}
\plotone{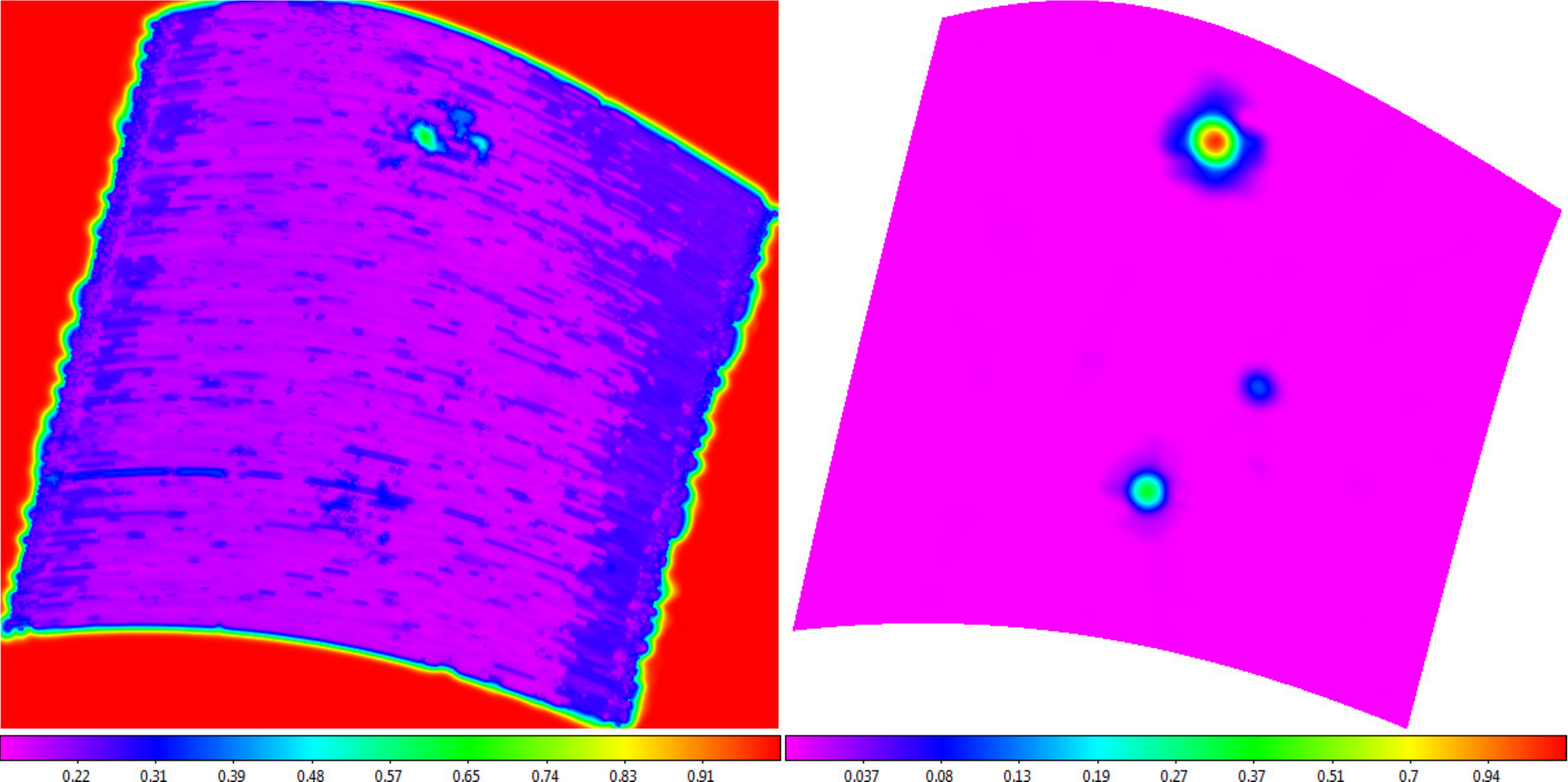}
\caption{\textbf{Left:}  The weighting scale map from the top-right panel of Figure~49, but instead processed after converting to Galactic coordinates.  \textbf{Right:}  The corresponding final image, also processed in Galactic coordinates, after imposing a minimum weighting scale of 2/3 beamwidths, so it can be compared to the left panel of Figure~38.  In Figure~38, we process the same data, but in its original, equatorial coordinate system.  Furthermore, this field is at high Galactic latitude, and consequently serves as a good example of the equal-areas and equal-distances properties of our sinusoidal projection (see below).  Equal areas means that sources cover the same number of pixels, and consequently should yield approximately the same photometry (see \textsection4):  The three sources in this map yields the same photometry as in the left panel of Figure~38 to within 3\%, despite the greater (diagonal) distortion that these sources can experience at high Galactic latitudes.  Equal distances refers to distances along horizontal lines, as well as along the central vertical axis, being distortion-free.}
\end{figure*}

\begin{figure*}
\plotone{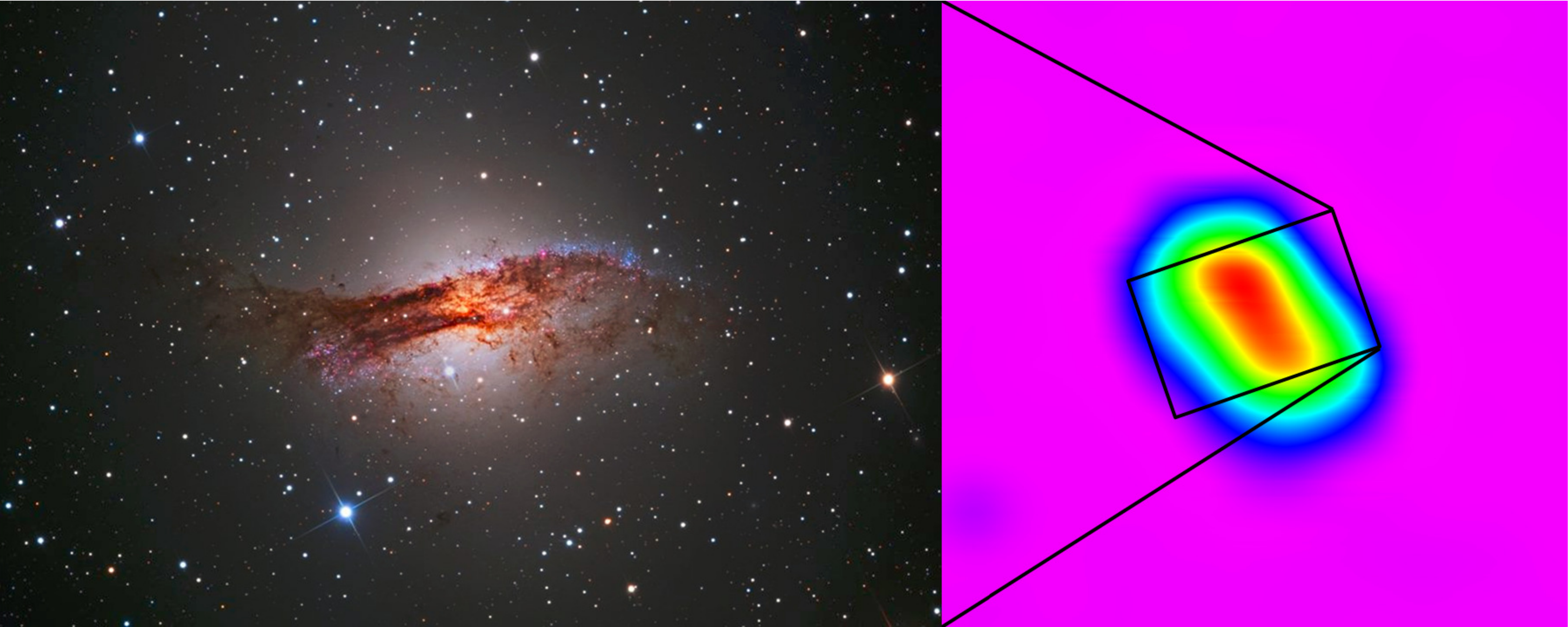}
\caption{20-meter X-band raster of Centaurus~A, background-subtracted, with a 7-beamwidth scale (larger than the minimum recommended scale from Table~1, given the size of the source), time-delay corrected, RFI-subtracted, with a 0.8-beamwidth scale (Table~2), and surface-modeled, with a minimum weighting scale of 2/3 beamwidths (right).  The radio jet is marginally resolved, and oriented correctly with respect to the galaxy (left).  Optical picture of NGC 5128 taken with Skynet's PROMPT-2 telescope at Cerro-Tololo Inter-American Observatory, courtesy of the Star Shadows Remote Observatory astrophotography group.}
\end{figure*}

\begin{figure}
\plotone{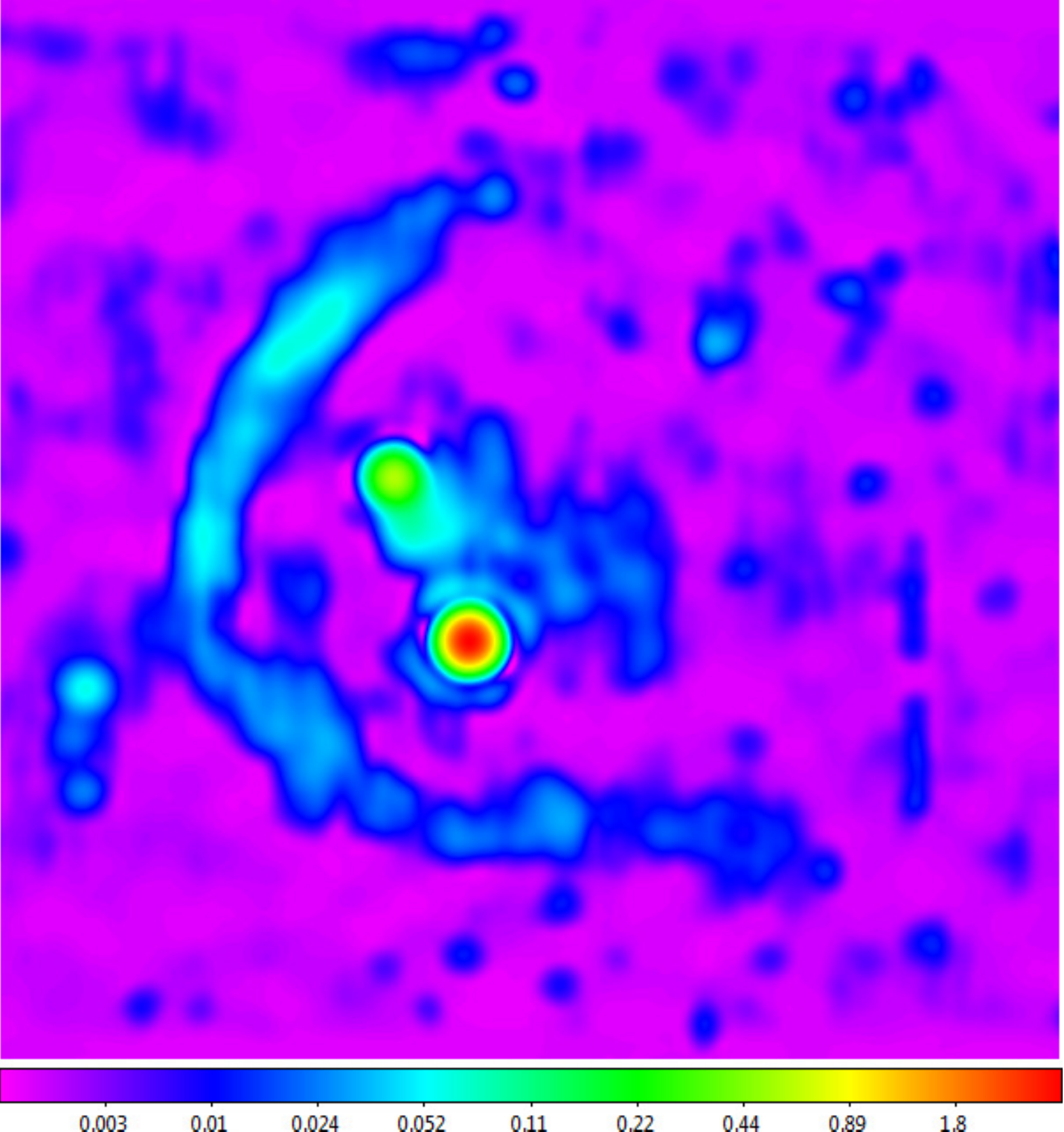}
\caption{Four 20-meter L-band rasters of Orion~A, Orion~B, and Barnard's Loop, separately background-subtracted, with a 20-beamwidth scale (larger than the minimum recommended scale from Table~1, given the size of Barnard's Loop), time-delay corrected, appended, jointly RFI-subtracted, with a 0.4-beamwidth scale (to preserve Airy rings), and surface-modeled, with a minimum weighting scale of 2/3 beamwidths.  Logarithmic scaling is used to emphasize Barnard's Loop and fainter sources.}
\end{figure}

\begin{figure*}
\plotone{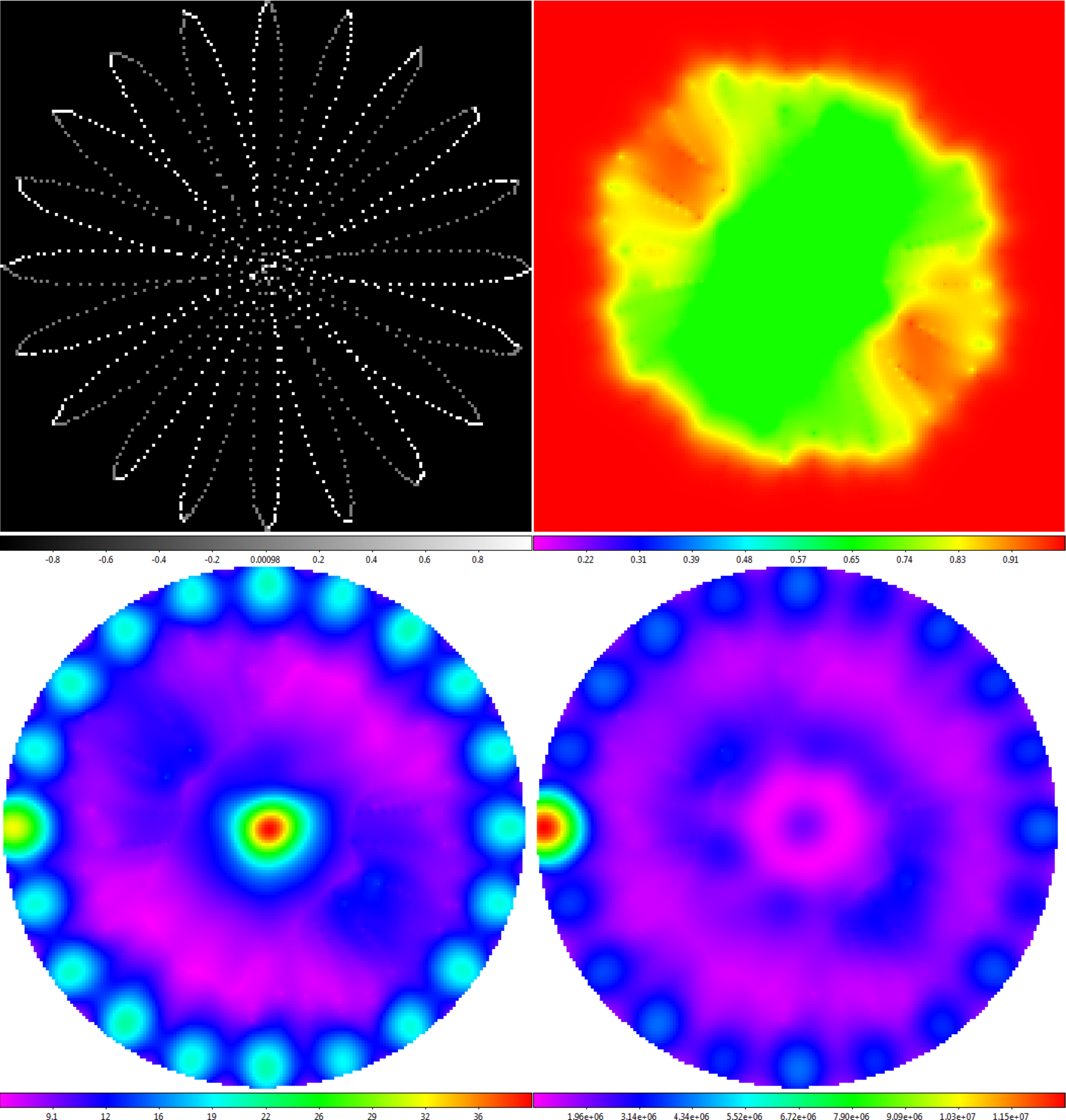}
\caption{\textbf{Top Left:}  Path map of the 20-meter X-band daisy from Figures 25 and 44.  This is a 20-petal daisy, though the user may select as few as four petals (however, see \textsection4).  The path began and ended on the left edge, and some points, particularly near the center, have been removed by the RFI-subtraction algorithm.  Alternating scans alternate between white and gray.  \textbf{Top Right:}  The corresponding scale map for our default minimum weighting scale of $\theta_{min}=2/3$ beamwidths (Equation~14), which can be compared to the top-right panel of Figure~49, where $\theta_{min}=0$ beamwidths.  This is the scale map that was used to surface-model Figure~44.  The exterior yellow, orange, and red regions, corresponding to $\theta_w>3/4$ beamwidths, are not appropriate for photometry, but this is not a problem because the source does not extend this far out (Figure~44).  \textbf{Bottom Left:}  Weighted number of data points that contributed to each pixel when surface-modeling Figure~44, given the path map and the scale map.  Extra data were acquired at the telescope's start/stop position, on the left edge.  \textbf{Bottom Right:}  The full weight map, which includes weights for each of the post-RFI subtraction values that were fitted to when surface-modeling Figure~44.  Less information informs these values in the vicinity of the source.  Weight maps are important (1)~when stacking images, since not all regions are equally well determined, and (2)~when doing photometry, both when determining the background level and when calculating error bars.}
\end{figure*}

\begin{figure}
\plotone{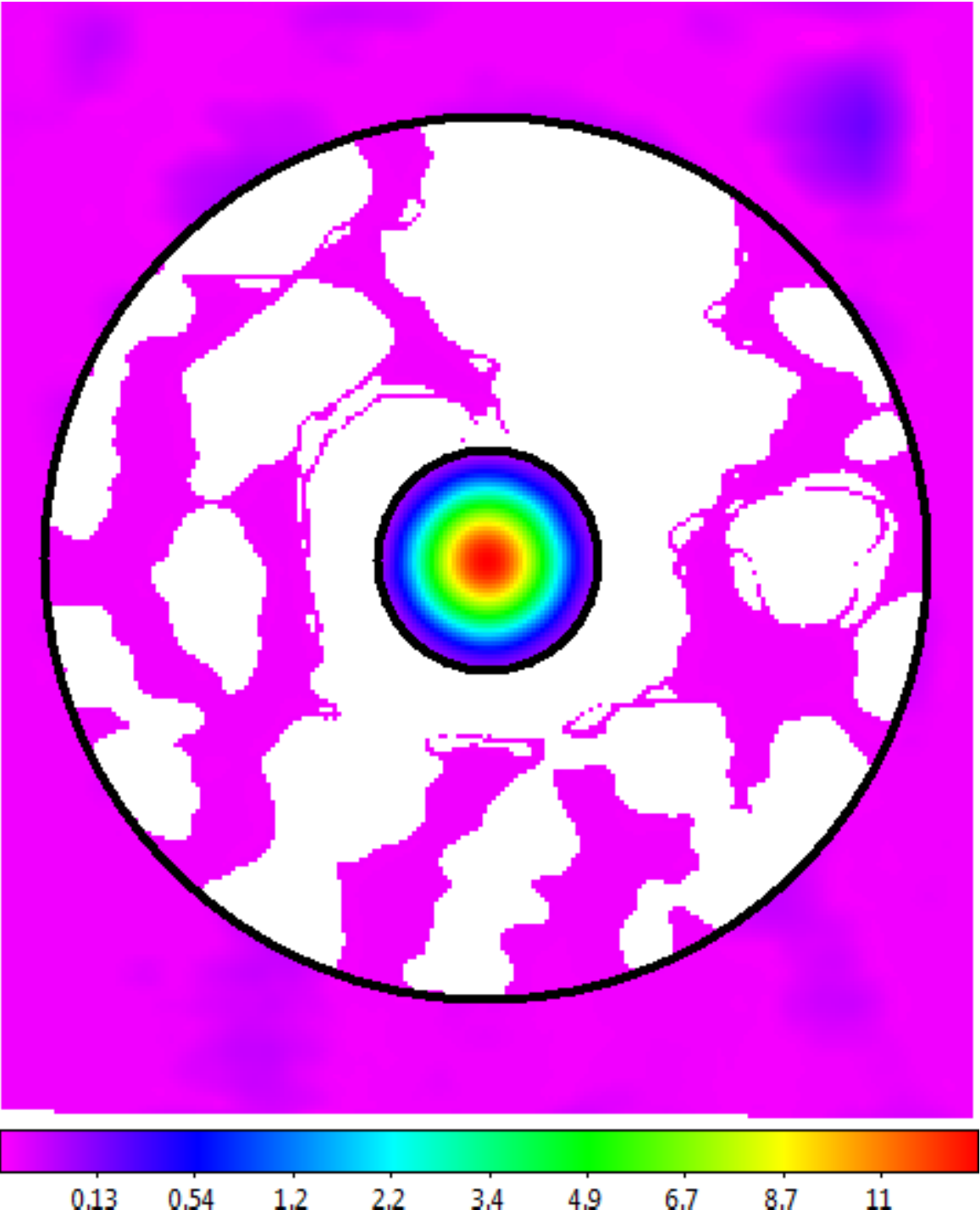}
\caption{2.5-beamwidth diameter aperture and 10-beamwidth diameter annulus that we use in Table~4, centroided on Cassiopeia~A, from the right panel of Figure~43.  Outlying pixels within the annulus, corresponding to Airy rings, other sources, etc., have been eliminated (compare to Figure~43).  Square-root scaling is used to emphasize fainter structures.}
\end{figure}

Although $\theta_{gap}$ can be modeled globally, based on the expected mapping pattern, we find it better to measure $\theta_{gap}$ locally, in case the telescope did not move exactly as expected (e.g., due to acceleration at the beginnings of scans, due to wind, etc.), or if data points are removed by the RFI-subtraction algorithm.  To do this, we employ a custom ``bubble-blowing'' algorithm about each data point (see Figure~49).  Generally, this gives values equal to, or slightly greater than, whichever is larger, the gap between the data point and the closest data point in the preceding scan, or the gap between the data point and the closest data point in the proceeding scan.  If a data point is on the edge of the mapping pattern, its $\theta_{gap}$ is infinite, corresponding to a limiting weighting scale of $\theta_w=1$ beamwidth (Equation~14).  We then determine the weighting scale that should be used at each pixel by interpolating between the weighting scales measured at each data point.\footnote{We interpolate between these values by employing proximity-weighted averaging, where we have selected a weighting function of:
\begin{equation}
w(\Delta\theta)=
\begin{cases}
(-\ln\Delta\theta)^a & \text{if $\Delta\theta < 1$ beamwidth} \\
0 & \text{otherwise}
\end{cases}, 
\end{equation}  
\noindent because (1)~$w(\Delta\theta \rightarrow 0) \rightarrow \infty$ (for $a > 0$) ensures that the weighted average yields the same value as the data point when $\Delta\theta=0$, and (2)~only data points within one beamwidth need be included in the weighted average, keeping it computationally efficient.  The exponent:
\begin{equation}
a = -2.329\ln(\theta_{gap}/2)-0.510
\end{equation}
is chosen such that the area under the weighting function between $-\theta_{gap}/2$ and $\theta_{gap}/2$ approximately matches the area under a Gaussian between $-\theta_{gap}/2$ and $\theta_{gap}/2$.  This is the smallest range that one can use and still get relatively smooth transitions between data-point values; smaller ranges result in more abrupt transitions, and a tessellated, or patchwork pattern, with one patch per data point.}

With rasters, $\theta_{gap}$ is approximately constant across the interior of the mapping (Figure~49, top right).  Consequently, rasters should be designed with gaps, both between and along scans, that are not larger than $\approx$3/4 $\times$ (1/3 -- 2/3 beamwidths) $=$ 1/4 -- 1/2 beamwidths.  Our recommendation is 1/5 -- 1/3 beamwidths, since OTF mappings usually result in marginally larger $\theta_{gap}$ measurements, particularly at the ends/beginnings of scans, where the telescope changes direction, and under windy conditions.

With noddings and especially with daisies, $\theta_{gap}$ increases toward the extremities of the mapping, where the telescope changes direction (Figure~2; Figure~49, bottom row), and consequently this condition cannot always be met.  However, as long as there are enough, and sufficient (Footnote~28), data to fit to, this algorithm will still model the surface.  Although the accuracy of the reconstruction cannot be guaranteed beyond a largest gap of $\approx$0.4 beamwidths, corresponding to Nyquist sampling, undersampled mappings, or undersampled regions of mappings, can still be useful for quick-look, and student, results.

Although we have focused on three mapping patterns in this paper, this surface-modeling algorithm, as well as our RFI-subtraction algorithm, work regardless of the design and details of the mapping pattern.  We demonstrate this in Figure~50, where we have instead processed the top-right panel of Figure~49 and the left panel of Figure~38 in Galactic coordinates, even though these data were acquired on a horizontal raster pattern in equatorial coordinates.\footnote{Our time-delay correction algorithm, as well as our 2D noise measurement algorithm, make use of measurements that were made at, or nearly at, the same position along adjacent scans.  This requires a coordinate along the scans, and consequently, this is most easily done in the coordinate system in which the mapping pattern was carried out.  Consequently, we keep everything in this coordinate system, be it equatorial, Galactic, etc., through this part of the algorithm.  If a different end-product coordinate system is requested by the user, we convert to it before applying our RFI-subtraction and surface-modeling algorithms.}  We currently allow users to configure and execute mapping patterns in either equatorial or Galactic coordinates, and to process their images in either of these coordinate systems, regardless of which was used when acquiring the data.  We make use of this in Paper II, where we demonstrate many of these papers' techniques simultaneously, in the processing of a 100$\degr$ X-band survey of the Galactic plane.

Finally, we set the edge(s) of the map.  Upon entry, we translate each right ascension/Galactic longitude coordinate value such that the map is centered on zero, and then we multiply each translated value by the cosine of its corresponding declination/Galactic latitude coordinate value.  This is known as a sinusoidal, or Sanson-Flamsteed, or Mercator equal-area, projection, which minimizes map distortion not only near the map's center, but also along the horizontal and vertical lines that pass through the map's center (Cossin 1570).  Furthermore, the 20-meter maps in these sinusoidally-corrected coordinates.  Consequently, for rasters, their four edges are given by fitting straight lines to the sinusoidally-corrected coordinate values of the first and last scans, and of the highest and lowest (or leftmost and rightmost) points in each scan, rejecting any outliers.\footnote{We fit this model to the data, simultaneously rejecting outliers, as described in {\color{black}\textsection8} of Maples et al.\@ {\color{black}2018}, using iterative bulk rejection followed by iterative individual rejection (using the generalized mode $+$ 68.3\%-value deviation technique, followed by the generalized median $+$ 68.3\%-value deviation technique, followed by the generalized mean $+$ standard deviation technique), using the smaller of the low and high one-sided deviation measurements.  Data are weighted equally.}  For daisies, its single, circular edge is given by averaging the sinusoidally-corrected angular distances between the daisy's center and the most distant point in each petal, again rejecting outliers.\footnote{We reject outliers as described in {\color{black}\textsection4 -- \textsection6} of Maples et al.\@ {\color{black}2018}, using iterative bulk rejection followed by iterative individual rejection (using the mode $+$ broken-line deviation technique, followed by the median $+$ 68.3\%-value deviation technique, followed by the mean $+$ standard deviation technique), using the smaller of the low and high one-sided deviation measurements.  Data are weighted equally.}  The 40-foot, on the other hand, is a meridian-transit telescope, and consequently maps in equatorial coordinates without being able to pre-correct for map distortion.  Hence, for noddings, their four edges are determined as in the raster case, but using the original, pre-corrected coordinate values.  All calculated edges are then carried forward, into whichever final coordinate system the user selects (changing their shape if this is a different coordinate system).  We also give the user the option of excising data within, typically two beamwidths, of edges that were fitted to the beginnings and ends of scans (\textsection3.3.2), which can be important when appending mappings that do not fully coincide (\textsection3.6.5).

{\color{black}\subsubsection{\color{black}Application to Asymmetric Structures}}

We have applied this algorithm to simulated and real data throughout this paper.  Most of these applications have been to symmetric structures, such as point sources.  However, Equation~11, due to its cross terms, imposes no artificial symmetries, allowing it to equally well model, e.g., asymmetric beam patterns (Figure~3, third row), asymmetric sources (see Figure~51), and asymmetric regions (see Figure~52).\\

{\color{black}\subsubsection{\color{black}Default Data Products}}

Upon completion of each 20-meter mapping, Skynet automatically produces the following data products:

\textbf{Raw Maps:}  We gain-calibrate the data, but otherwise do not process it, except for the application of a surface model so the user can visualize the pre-processed, raw data.  For this, we use a minimum weighting scale of $\theta_{min}=0$ beamwidths (Equation~14), to better visualize sub-beamwidth contaminants (\textsection3.7).  We do this for each polarization channel, and for the {\color{black}average} of both channels (after multiplying each by {\color{black}appropriate flux-density conversion factors}, if this information is available; Footnote~9).

\textbf{Contaminant-Cleaned Maps:}  We fully process the data using our default settings:  (1)~We gain-calibrate the data (\textsection3.1); (2)~We background-subtract the gain-calibrated data, using the minimum recommended scale from Table~1 (\textsection3.3); (3)~We time-delay correct the background-subtracted data (\textsection3.4); (4)~We apply our RFI-subtraction algorithm to the time-delay corrected data, using the maximum recommended scale from Table~2, or, if the mapping contains a sufficiently high S/N source, using half of this value, to preserve visible Airy rings (\textsection3.6); and (5)~We apply our surface-modeling algorithm to the RFI-subtracted data, using a minimum weighting scale of $\theta_{min}=2/3$ beamwidths, as well as the noise-level prior from Footnote~27 (\textsection3.7).  Again, we do this for each polarization channel, and for the sum of both channels.  (These settings can be changed by the user afterward, and the data re-processed.) 

\textbf{Path Maps:}  Path maps plot one point at the coordinates of each signal integration in the OTF mapping, alternating between white and gray for alternating scans (e.g., see Figure~53, top-left panel).  We generate two path maps, one with the raw maps, using pre-time delay corrected coordinates, and including all of the measurements, and one with the contaminant-cleaned maps, using post-time delay corrected coordinates, and including all of the measurements save those that are removed by the RFI-subtraction algorithm.

Path maps are useful for determining if the telescope had difficulty maintaining the desired spacing between scans, due to momentum-driven overshooting when changing directions (in which case a slower, or possibly denser, or simply differently patterned, mapping might be considered), due to wind, or due to motor or encoder error.  They are also useful for seeing which measurements, if any, were removed by the RFI-subtraction algorithm.  Since our algorithms are fairly insensitive to the design and details of the mapping pattern, or path, these effects should not significantly affect the results, unless they leave too large of a gap in the mapping pattern for the surface-modeling algorithm to be effective (\textsection3.7).

\textbf{Scale Maps:}  Scale maps are calculated from the information in the path maps, and visualize the weighting scale that is used to calculate the surface model, at each pixel in the final image (\textsection3.7).  Examples are given in Figure~49 for $\theta_{min}=0$ beamwidths (Equation~14), but often $\theta_{min}>min\{4/3\times\theta_{gap},1\}$, in which case the scale map, at least in such regions, is simply single-valued (Figure~53, top-right panel).

Scale maps are also important when doing photometry.  For most maps, Nyquist sampling ($\approx$0.4 beamwidths) results in largest-gap values of $\theta_{gap}\approx0.5$ -- 0.55 beamwidths, which in turn results in weighting-scale values of $\theta_w=4/3\times\theta_{gap}\approx2/3$ -- 3/4 beamwidths (for $\theta_{min}\leq2/3$ beamwidths).  Hence, if $\theta_w\ga3/4$ beamwidths within a photometric aperture, the result should be taken with a grain of salt (\textsection3.7).  (This however is not the case within a photometric annulus, as the background need not be sampled on Nyquist scales; see \textsection4).

\textbf{Weight Maps:}  Weight maps record the weighted number of data points to which Equation~11 was fitted, for each pixel in the final image.  Each data point's weight is given by (1)~the product of (a) the proximity-dependent weight that was used in the fit, given by Equations 12 -- 14, and (b) the weighted number of dumps that contributed to the data point's post-RFI subtraction value (Footnote~22), (2)~divided by a number that, at least approximately, corrects for the non-independence of this value over a scale that is related to the RFI-subtraction scale (see Appendix~D).  We demonstrate the effect of just (1)~in the bottom-left panel of Figure~53, where we see larger values (a) where there's a higher density of data points, and (b) where the weighting scale is larger, resulting in larger weights per data point.  We demonstrate the effect of both (1)~and (2)~in the bottom-right panel of Figure~53.  Since (far) fewer data points contribute to each RFI-subtraction local model in the vicinity of sources (most are rejected; e.g., Figure~33, top panel), less information informs the post-RFI subtraction values in the vicinity of sources.

Weight maps are important if stacking (averaging) images, since different maps, and different regions of the same map, can be more constraining than others, and consequently should be weighted appropriately.  (That said, it is still preferable to append all of the mappings' scans before RFI subtraction, and surface-model only once; \textsection3.6.)  Weight maps are also important when doing photometry, both when determining the background level within an annulus, and when determining what the error bar should be, since (again, far) less information informs the surface model in the vicinity of sources, and hence in the aperture, than in the annulus (see {\color{black}\textsection4}).

\textbf{Correlation Maps:}  Correlations maps record the scale over which pixel data are correlated, which is important when calculating photometric error bars from a single image.  Calculational details are provided in Appendix~D.

The calculation of photometric error bars makes use of the image, the image's weight map, and the image's correlation map.  If calculating photometric error bars from a weighted average of multiple images, we recommend (1) producing a new weight map by summing their individual weight maps, and (2) producing a new correlation map by taking the weighted average of their individual correlation maps.  (But again, the better approach would be to append all of the mappings' scans before RFI subtraction, and surface-model only once, producing a single image, a single weight map, and a single correlation map.) 

\section{Aperture Photometry}

Given an accurately modeled, contaminant-cleaned surface, spanning a rectangular grid of pixels, one can perform operations on it, much as one would on a reduced optical image.  In this section, we present an aperture photometry algorithm, which we use to test the accuracy of our small-scale structure maps.  

However, unlike in optical images, where each pixel is measured directly, in these images, each pixel is a reconstruction, based on the data points around it.  And since we cannot guarantee the accuracy of this reconstruction in regions where the weighting scale $\theta_w\geq3/4$ beamwidths (\textsection3.7.2), we do not recommend photometering sources in such regions.  This is easy enough to avoid by designing a sufficiently dense mapping, though additional care must be taken with noddings and daisies, for which $\theta_w$ increases toward the extremities of the mapping (\textsection3.7).\footnote{It is not difficult to show that this condition is met within $\approx$$N/8$ beamwidths of a daisy's center, where $N$ is the number of petals in the daisy.  Our users can chose between $N=4$, 8, 12, 16, and 20.  We recommend $N\geq12$ for higher-quality maps, and especially for photometry.}   However, this condition need only be met within the aperture (which for daisies, will lie, with the source, at the daisy's lower-$\theta_w$ center).  It is not necessary to meet this condition throughout the annulus (\textsection3.7.2).

Assuming that this condition is met, we begin by centroiding the aperture and annulus on the source.  We do this by fitting a simpler, second-order version of Equation~11 (see Footnote~28), with a fixed weighting scale of $\theta_w=1/3$ beamwidth, to every pixel within a 1-beamwidth radius of a first-guess pixel.  The centroid is given by the extremum of this function.  We then iterate until convergence.

\begin{deluxetable*}{ccccccccccccc}[!t]
\tablewidth{0pt}
\tablecaption{Aperture Photometry of Simulated Sources from \textsection3.3 and \textsection3.6\tablenotemark{a}}
\tablehead{
\colhead{Source} & & & & & & & & & & & \\
\colhead{Ratio} & & \multicolumn{3}{c}{Brightness} & & \multicolumn{2}{c}{Percent Error} & & \multicolumn{4}{c}{Error Bars (\%)}\\
\cline{1-1} \cline{3-5} \cline{7-8} \cline {10-13}
\noalign{\smallskip}
& & \colhead{Measured} & \colhead{Corrected} & \colhead{True} & & \colhead{Measured} & \colhead{After} & & \multicolumn{2}{c}{Internal} & \colhead{External} & \colhead{Total}\\
& & & & & & & \colhead{Correction} & & \colhead{Calculated} & \colhead{Simulated} & \colhead{(from} & \colhead{(from calculation}\\
& & & & & & & & & & & \colhead{correction)} & \colhead{and correction)}}
\startdata
2 / 1 & & 0.252 & 0.252 & 0.255 & & -1.3 & -1.0 & & 0.2 & 0.1 & 0.2 & 0.2 \\
3 / 1 & & 0.110 & 0.111 & 0.112 & & -1.7 & -0.2 & & 0.2 & 0.4 & 0.6 & 0.7 \\
4 / 1 & & 0.061 & 0.064 & 0.062 & & -2.5 & 2.7 & & 0.4 & 0.8 & 2.0 & 2.0 \\
5 / 1 & & 0.038 & 0.040 & 0.040 & & -5.2 & -0.1 & & 1.0 & 1.0 & 2.0 & 2.2 \\
6 / 1 & & 0.028 & 0.032 & 0.031 & & -12.4 & 1.6 & & 1.7 & 0.7 & 5.3 & 5.6 \\
7 / 1 & & 0.018 & 0.026 & 0.022 & & -18.8 & 16.4 & & 1.9 & 3.3 & 10.6 & 10.8 \\
8 / 1 & & 0.016 & 0.019 & 0.020 & & -23.1 & -4.6 & & 2.1 & 1.3 & 7.3 & 7.6 \\
9 / 1 & & 0.009 & 0.015 & 0.012 & & -26.8 & 22.1 & & 2.7 & 2.3 & 13.1 & 13.4 \\
10 / 1 & & 0.007 & 0.012 & 0.010 & & -35.1 & 14.8 & & 2.3 & 2.5 & 13.8 & 14.0
\enddata
\tablenotetext{a}{Relative to the brightest source.  This particular data set was generated using the less-winged beam function of Equation~9 and Figure~31, and included the exact same contaminants that we use in Figure~21.  The data were then processed using a 6-beamwidth background-subtraction scale, a 0.5-beamwidth RFI-subtraction scale, a 0.5-beamwidth surface-modeling (minimum weighting) scale, and the noise-level prior.  Aperture photometry was carried out using a 2-beamwidth diameter aperture and an annulus of 2- and 10-beamwidth inner and outer diameters, respectively.  (We have repeated these measurements for more-winged beam functions, for a wide range of background-subtraction, RFI-subtraction, and minimum-weighting scales, both with and without the noise-level prior, and for a wide range of aperture and annulus diameters, {\color{black}and mapping densities,} with similar results.)  Measurements are presented both before and after correcting for dimming caused by the RFI-subtraction algorithm, and to a lesser extent, by the surface-modeling algorithm (see Equation~21 below), which affects the lower-S/N sources in particular.  We calculate internal error bars using Equation~17 above, and these are consistent with those measured by re-simulating, re-processing, and re-photometering this data set 100 times, where we have randomized the noise and en-route drift in each simulation.  We calculate external error bars as part of the low-S/N dimming correction (see Equation~23 below).  Total error bars are given by adding our internal and external error bars in quadrature, and these are consistent with the measurement errors in our post-correction photometry (except for the highest-S/N source, which suggests that our uncertainty is probably never less than $\sim$1\%, regardless of Equations 17 and 20).  Note, in this example, the internal error bars are small -- often negligibly small -- compared to the external error bars.  However, this is because these (simulated) sources are very densely sampled by this mapping; with lower-density mappings, the internal error bars can be much larger.  Also note that when measuring a source's signal-to-noise, only its internal error bar matters; the rest is calibration uncertainty.}
\end{deluxetable*}

Next, we measure the background level, $\mu$, the standard deviation about the background level, $\sigma$, and the uncertainty in our measurement of the background level, $\mu_\sigma$, from the pixels in the annulus (the background level should be approximately zero in our small-scale structure maps, since these have already been background-subtracted).\footnote{Given the availability of our small-scale structure maps, which have been more carefully background-subtracted than one can achieve with just an annulus, we recommend using these over their accompanying large-scale structure maps (see \textsection2 of Paper II) when photometering sources (except perhaps within $\approx$1 -- 2 beamwidths of the beginnings and ends of scans, where the background-subtraction algorithm can underestimate the brightness of sources in our small-scale structure maps; \textsection3.3.2).}   We measure these by calculating the weighted mean, the weighted standard deviation, and the weighted uncertainty in the mean, respectively, of the pixels in the annulus, making use of (1) the weight map (\textsection3.7.2), and (2) robust Chauvenet rejection (\textsection1.3), to eliminate pixels that are contaminated by Airy rings, other sources, etc.\@ (see Figure~54).\footnote{We reject outliers as described in {\color{black}\textsection4 -- \textsection6} of Maples et al.\@ {\color{black}2018}, using iterative bulk rejection followed by iterative individual rejection (using the mode + broken-line deviation technique, followed by the median + 68.3\%-value deviation technique, followed by the mean $+$ standard deviation technique), using the low one-sided deviation measurement to reject low outliers and the high one-sided deviation measurement to reject high outliers.  This is because the low deviations can be artificially suppressed, by the (default) noise-level prior (Footnote~27; however, see Appendix~D and Footnote~41).  (Pixel) data are weighted by the weight map.}    We then sum the pixel values in the aperture, subtracting the weighted-mean background level from each.  (Aperture and annulus sizes are user-configurable.)

The total photometric error bar is then given by:
\begin{equation}
\sigma_{phot}=f_{\sigma_{phot}}\sqrt{\sigma^2 \sum_{i=1}^{N_{ap}} \frac{\langle w\rangle_{an}}{w_i/N_i}+\left(N_{ap}\sigma_\mu\right)^2},    
\end{equation}
\noindent where the sum is over the number of pixels in the aperture, $N_{ap}$, $w_i$ is each pixel's weight-map value, $\langle w\rangle_{an}$ is the average weight-map value of the pixels in the annulus that were used to measure $\sigma$, $N_i$ is a number that, at least approximately, corrects for the non-independence of the $i$th-pixel's value over both the RFI-subtraction and surface-modeling (weighting) scales, and $f_{\sigma_{phot}}$ is an empirically determined correction factor.  Calculational details are provided in Appendix~D.

We now test the accuracy of our small-scale structure mapping algorithm (as well as the accuracy of Equation~17), by photometering the simulated sources from \textsection3.3 and \textsection3.6.  Results for a middle-of-the-road set of processing parameters are presented in Table~3, and the measured values underestimate the true values, only marginally for the highest-S/N sources, but increasingly so for lower-S/N sources.  Furthermore, these underestimates are significant relative to the expected level of uncertainty, measured both by Equation~17, and from 100 re-simulations of the data, where we have randomized the noise and en-route drift in each.  

However, this is expected:  In \textsection3.6.3, we demonstrated that the RFI-subtraction algorithm can eliminate noise-level signal from sources, dimming low-S/N sources in particular, especially when $\theta_{RFI}$ is large.  To a lesser degree, this is also the case with the surface-modeling algorithm, when $\theta_w$ is large.  We have measured these effects for a wide range of $\theta_{RFI}$ and $\theta_{min}$ values (for cases where $\theta_w$ is dominated by $\theta_{min}$; Equation~14), as well as for different aperture sizes, {\color{black}beam functions, and mapping densities}, and have assembled the following empirical correction factor:
\begin{equation}
\begin{split}
f_{phot}={} & {\color{black}1+\left[f_{lim}^{-0.82}+0.052\left(\frac{\theta_{ap}}{2.5}\right)^{-0.11}\right.}\\
& {\color{black}\times \left.\Theta^{-0.29}\left(\theta_{RFI},\theta_{min},\theta_{ap}\right)\right]^{-1.22}},
\end{split}
\end{equation}
\noindent {\color{black}where:
\begin{equation}
\begin{split}
f_{lim}={} & 0.22\left(\frac{\theta_{ap}}{2.5}\right)^{0.52}\left(\frac{z_{peak}}{1000\sigma}\right)^{-1.20\left(\frac{\theta_{ap}}{2.5}\right)^{-0.39}}\\
& \times \Theta\left(\theta_{RFI},\theta_{min},\theta_{ap}\right),
\end{split}
\end{equation}
\noindent and} where $z_{peak}$ is the signal level at the peak of the source,\footnote{This is given by summing the $\mu$-subtracted pixel values in the aperture out to a user-selected radius, and dividing this by the sum of the peak-normalized beam function, evaluated at these same locations.  For the cosine-squared beam function in Figure~31, the latter sum is approximately given by:
\begin{equation}
\frac{1}{\theta_{pix}^2}\left(\frac{\pi\theta^2}{2}+\frac{\cos{\pi\theta}}{\pi}+\theta\sin{\pi\theta}-\frac{1}{\pi}\right),    
\end{equation}
\noindent where $\theta<1$ is the user-selected radius in beamwidths, and $\theta_{pix}$ is the number of beamwidths per pixel (our default value is 0.05; \textsection3.7).  For the Gaussian beam function in Figure~31, this is instead given by:
\begin{equation}
\frac{1.13309-1.13309e^{-2.77259\theta^2}}{\theta_{pix}^2}.
\end{equation}
\noindent These two expressions are nearly identical for $\theta<0.7$ beamwidths, and differ by only $\approx$13\% at $\theta=1$ beamwidth.  If the point-spread function is not known, we recommend using either of these functions, but with $\theta=min\{0.7,\theta_{ap}/2\}$.} $\sigma$ is the standard deviation of the pixel values in the annulus from above, $\theta_{RFI}$ and $\theta_{min}$ are measured in beamwidths, $\theta_{ap}=min\{\text{aperture diameter in beamwidths}, 2.5\}$, and:
\begin{equation}
\begin{split}
\Theta\left(\theta_{RFI},\theta_{min},\theta_{ap}\right)={} & \theta_{RFI}^{2.64\left(\frac{\theta_{ap}}{2.5}\right)^{-0.54}}\\
& +0.34\left(\frac{\theta_{ap}}{2.5}\right)^{0.73}\theta_{min}^{1.66\left(\frac{\theta_{ap}}{2.5}\right)^{-0.18}}.
\end{split}
\end{equation}
\noindent The uncertainty in $f_{phot}$ is approximately given by:
\begin{equation}
\begin{split}
\sigma_{f_{phot}}={} & {\color{black}\left[\sigma_{f_{lim}}^{-0.82}+0.065\left(\frac{\theta_{ap}}{2.5}\right)^{-0.02}\right.}\\
& {\color{black}\times \left.\Theta^{-0.29}\left(\theta_{RFI},\theta_{min},\theta_{ap}\right)\right]^{-1.22}},
\end{split}
\end{equation}
\noindent {\color{black}where:
\begin{equation}
\begin{split}
\sigma_{f_{lim}}={} & 0.082\left(\frac{\theta_{ap}}{2.5}\right)^{0.32}\left(\frac{z_{peak}}{1000\sigma}\right)^{-1.35\left(\frac{\theta_{ap}}{2.5}\right)^{-0.08}}\\
& \times \Theta\left(\theta_{RFI},\theta_{min},\theta_{ap}\right).
\end{split}
\end{equation}
\noindent We} find that these expressions hold relatively independently of whether the beam function is narrow- or broad-winged{\color{black}, and of the density of the mapping}.  We apply this correction in Table~3, where it raises the measured values of the sources, and the lower-S/N sources in particular, to within, on average, one total error bar (equal to the internal error bar, given by Equation~17, and the external error bar, given by Equation~23 {\color{black}times the measured value}, added in quadrature) of their true values.  (Note, we do not apply this correction to the internal error bars themselves, as they are not significant functions of source brightness; see Appendix~D.  Also, these expressions are only for values of $\theta_{min}$ between $\approx$1/3 beamwidths and $\approx$2/3 beamwidths; \textsection3.7)

\begin{deluxetable*}{cccc}[!t]
\tablewidth{0pt}
\tablecaption{Flux-Density Ratios of Primary Calibration Sources\tablenotemark{a}}
\tablehead{
\colhead{Source Ratio} & \colhead{Measured Flux-Density Ratio} & \colhead{Modeled Flux-Density Ratio} & \colhead{Modeled Flux-Density Ratio} \\
& \colhead{(Average of 24)} & \colhead{(Baars et al.\@ 1977)} & \colhead{(Trotter et al.\@ 2017)}}
\startdata
Cas~A / Cyg~A & $1.1046\pm0.0071$ & $0.951\pm0.091$\tablenotemark{b} & $1.104\pm0.018$ \\
Tau~A / Cyg~A & $0.5348\pm0.0043$ & $0.592\pm0.065$\tablenotemark{b} & $0.5287\pm0.0078$ \\
Vir~A / Cyg~A & $0.1411\pm0.0071$ & $0.135\pm0.016$ & $0.1370\pm0.0021$ 
\enddata
\tablenotetext{a}{Measured from 96 (24 for each source) 20-meter L-band maps, each processed separately with recommended/default settings, and expected values from temporal and spectral models that have been fitted to nearly 60 years of measurements.  For each measurement of Cas~A, Tau~A, and Vir~A, we took the measurement of Cyg~A that was closest in time, usually within hours, and vice versa, to minimize differences in calibration, took their ratio, and averaged these ratios, rejecting outliers using Maples et al.\@ (2017), as described in Trotter et al.\@ (2017).  Our averaged ratios agree with the expected ratios to within the uncertainties.}
\tablenotetext{b}{Baars et al.\@ (1977) overestimated the fading of Cas~A, and did not model (and hence underestimated) the fading of Tau~A.  These have been corrected, and the spectral models of Baars et al.\@ (1977) also improved upon, in Trotter et al.\@ (2017; see also Reichart \& Stephens 2000).}
\end{deluxetable*}

Finally, we test the accuracy of our small-scale structure mapping algorithm by photometering real sources, specifically the primary calibration sources {\color{black}Cas}~A, {\color{black}Cyg}~A, {\color{black}Tau}~A, and {\color{black}Vir}~A, and by comparing their flux-density ratios to previously modeled expectations.  Using the 20-meter in L band, we collected 24 rasters of each source, all within a few days.  In this case, we background-subtracted each with a 6-beamwidth scale (Table~1), time-delay corrected each, RFI-subtracted each with a 0.8-beamwidth scale (Table~2), and surface-modeled each with our default minimum weighting scale of 2/3 beamwidths (and our, also default, noise-level prior).  We photometered each with a 2.5-beamwidth diameter aperture, corresponding to the apparent minimum between the source and the first Airy ring (e.g., Figure~54), and took ratios of these values, and averages of these ratios, as we have done previously in Trotter et al.\@ (2017).  We list these average ratios in Table~4, and they match their expected values (Baars et al.\@ 1974; Trotter et al.\@ 2017) within the uncertainties.

\section{Conclusion}

In this paper, we have presented a single-dish mapping algorithm with the following features:

1.  We use robust Chauvenet rejection (RCR; \textsection1.3) to improve gain calibration, making this procedure insensitive to RFI contamination (as long as it is not total, or nearly so), to catching the noise diode in transition, and to the background level ramping up or down (linearly), for whatever reason, during the calibration (\textsection3.1).

2.  We again use RCR to measure the noise level of the data, in this case from point to point along the scans, also allowing this level to ramp up or down (again, linearly) over the course of the observation (\textsection3.2).  We then use this noise model to background-subtract the data along each scan, without significantly biasing these data high or low (\textsection3.3).  We do this by modeling the background locally, within a user-defined scale, instead of globally and hence less flexibly (as, e.g., basket-weaving approaches do).  This significantly reduces, if not outright eliminates, most signal contaminants:  en-route drift, long-duration (but not short-duration) RFI, astronomical signal on larger scales, and elevation-dependent signal (e.g., Figure~20).  Furthermore, this procedure requires only a single mapping (also unlike basket-weaving approaches).

3.  We use RCR to correct for any time delay between our signal measurements and our coordinate measurements (\textsection3.4).  This method is robust against contamination by short-duration RFI and residual long-duration RFI.  (In general, this procedure requires that the telescope's slew speed remain nearly constant throughout the mapping, or at least during its scans if not between them, though we do offer a modification such that it can also be applied to variable-speed, daisy mapping patterns, centered on a source.)

4.  We again measure the noise level of the data, but this time from point to point across the scans, again allowing this level to ramp up or down (again, linearly) over the course of the observation (\textsection3.5).  We then use this noise model to RFI-subtract the data, again without significantly biasing these data high or low (\textsection3.6).  We do this by modeling the RFI-subtracted signal locally, over a user-defined scale; structures that are smaller than this scale, either along or across scans, are eliminated, including short-duration RFI, residual long-duration RFI, residual en-route drift, etc.\@ (e.g., Figure~20).  This scale can be set to preserve only diffraction-limited point sources and larger structures, or it can be halved to additionally preserve Airy rings, which are visible around the brightest sources.  Furthermore, this procedure can be applied to multiple observations simultaneously, in which case even smaller scales can be used (better preserving noise-level signal, and hence faint, low-S/N sources).

5.  To interpolate between signal measurements, we introduce an algorithm for \textit{modeling} the data, over a user-defined weighting scale (though the algorithm can increase this scale, from place to place in the image, if more data are required for a stable, local solution; \textsection3.7).  Advantages of this approach are:  (1) It does not blur the image beyond its native, diffraction-limited resolution; (2) It may be applied at any stage in our contaminant-cleaning algorithm, for visualization of each step, if desired; and (3) Any pixel density may be selected.  This stands in contrast to existing algorithms, which use weighted \textit{averaging} to regrid the data:  (1) This does blur the image beyond its native resolution, often significantly; (2) It is usually done before contaminant cleaning takes place, because existing contaminant-cleaning algorithms -- unlike ours -- require gridded data; and (3) The pixel density is then necessarily limited to what these contaminant-cleaning algorithms can handle, computationally (\textsection1.2.3).  

Furthermore, since our surface-modeling algorithm does not require gridded data, images can be produced in any coordinate system, regardless of how the mapping pattern was designed.  And since our surface-modeling algorithm does not assume any coordinate system-based symmetries, it works equally well with asymmetric structures.  In addition to the final image, we produce a path map, a scale map, a weight map, and a correlation map, the latter three of which are important when performing photometry on the final image.

6.  Lastly, we introduce an aperture-photometry algorithm for use with these images (\textsection4).  In particular, we introduce a semi-empirical method for estimating photometric error bars from a single image, which is non-trivial given the non-independence of pixel values in these reconstructed images (unlike in, e.g., CCD images, where each pixel value is independent, and consequently the statistics are simpler).  We also provide an empirical correction for low-S/N photometry, which can be underestimated in these reconstructed images.

Additionally, we have provided a technical description of Green Bank Observatory's 20-meter diameter telescope, which we refurbished as part of this effort, and is now being used by thousands of students and researchers each year, remotely, through our Skynet interface (\textsection2.1).

{\color{black}And while developed because of this telescope, the single-dish mapping and aperture-photometry algorithms presented in this paper are not specifically for this telescope.  We demonstrated this by applying these algorithms to a smaller, and less capable, telescope, but there is no reason why they may not also be applied to larger, and more capable, telescopes (e.g., we recently applied them, successfully, to Green Bank Telescope mapping observations).  These algorithms could also be applied to similarly collected data, with similar noise and beam characteristics, at other wavelengths, as well as from non-astronomical experiments.\footnote{\color{black}Furthermore, components of these algorithms could be used to improve interferometric reductions:  E.g., RCR could be used to eliminate outlying interferometric measurements, and our surface-modeling algorithm could be modified to fill gaps in the uv plane.}}

In the second paper in this series (Dutton et al.\@ 2018), we expand on the small-scale structure mapping algorithm that we have presented in this paper, in two ways:  (1) We introduce an algorithm to additionally contaminant-clean and map large-scale structures, and (2) We introduce an algorithm to contaminant-clean and map spectral data, as opposed to just continuum data, as in this paper.  Finally, we present an X-band survey of the Galactic plane, from $-5\degr < l < 95\degr$, to further demonstrate, and further test, the techniques that we present in both of these papers.



\acknowledgments

We gratefully acknowledge the support of the National Science Foundation, through the following programs and awards:  ESP 0943305, MRI-R$^2$ 0959447, AAG 1009052, 1211782, and 1517030, ISE 1223235, HBCU-UP 1238809, TUES 1245383, and STEM$+$C 1640131.  We are also appreciative to have been supported by the Mt.\@ Cuba Astronomical Foundation, the Robert Martin Ayers Sciences Fund, and the North Carolina Space Grant Consortium.  Finally, we thank the approximately 400 students and educators who have participated in \textit{Educational Research in Radio Astronomy (ERIRA)} at Green Bank Observatory since 1992, many of whom helped to collect the 40-foot data that we present in this paper.  We also thank Green Bank Observatory for hosting our group for all of these years. 





\appendix

\section{A. Local Model Weights for Construction of Global Background Model}

The local model for background subtraction is given by:
\begin{equation}
z(x) = a + b(x - \mu)+ \delta c(x - \mu)^2,
\end{equation}
\noindent where $x$ is angular distance along a scan, $\mu$ is the dump-weighted mean angular distance of all of the non-rejected points to which the local model is fitted, $a$, $b$, and $c$ are the fitted parameters, and $\delta$ is zero for linear models and one for quadratic models (\textsection3.3).

Let $\sigma_a$, $\sigma_b$, and $\sigma_c$ be the uncertainties in the fitted values of $a$, $b$, and $c$, respectively.  Given the construction of the local model, these uncertainties will be largely uncorrelated ({\color{black}\textsection8} of Maples et al.\@ {\color{black}2018}).  Consequently, we approximate the variance of the fitted model, as a function of position in the model, by:
\begin{equation}
\sigma^2(x) \approx \sigma_a^2 + \sigma_b^2(x - \mu)^2 + \delta\sigma_c^2(x - \mu)^4.
\end{equation}
\noindent Furthermore, each term will contribute to the variance approximately equally, yielding:
\begin{equation}
\sigma_a^2 \approx \sigma_b^2 \overline{(x - \mu)^2}
\end{equation}
\noindent for linear models and
\begin{equation}
\sigma_a^2 \approx \sigma_b^2 \overline{(x - \mu)^2} \approx \sigma_c^2 \overline{(x - \mu)^4}
\end{equation}
\noindent for quadratic models.  Consequently:
\begin{equation}
\sigma^2(x) \propto 1 + \left(\frac{x - \mu}{\sigma}\right)^2 + \delta\left(\frac{x - \mu}{\kappa}\right)^4,
\end{equation}

\pagebreak

\noindent where $\sigma^2 = \overline{(x-\mu)^2}$ and $\kappa^4 = \overline{(x - \mu)^4}$ (Equation~4).  

Finally, the variance scales inversely with the number of non-rejected dumps to which the local model is fitted, and weight scales inversely with variance, yielding Equation~3.

\section{B. Analytic Fitting of Local RFI Model}

The local model for RFI subtraction is given by Equation~9, and is parameterized by $f$ and $z_0$.  This model, $z(\Delta\theta|f,z_0)$, can be fitted to data analytically.  Consider the following posterior probability distribution:
\begin{equation}
p(f,z_0) \propto \prod_{mn} \exp\left[-\frac{N_{mn}}{2}\left(\frac{z_0}{\sigma_m}\right)^2\right]\exp\left\{-\frac{N_{mn}}{2}\left[\frac{z_{mn}-z(\Delta\theta_{mn}|f,z_0)}{\sigma_m}\right]^2\right\},
\end{equation}
\noindent where $z_{mn}$ is the $n$th background-subtracted signal measurement in the $m$th scan, $N_{mn}$ is its dump weight, $\Delta\theta_{mn}$ is its angular distance from the model's fixed center, and $\sigma_m$ is the 2D noise model of the $m$th scan (\textsection3.5).  The first exponential is a prior probability distribution on $z_0$, constraining it to be noise level.  The second exponential is the likelihood function.  The best fit is given by maximizing $p(f,z_0)$:
\begin{equation}
\frac{\partial p(f,z_0)}{\partial f} = \frac{\partial p(f,z_0)}{\partial z_0} = 0.
\end{equation}
\noindent Solving for $f$ and $z_0$ yields:
\begin{equation}
f = \frac{2\left(\sum_{mn}\frac{N_{mn}}{\sigma_m^2}\right)\left[\sum_{mn}\frac{N_{mn}z_{mn}}{\sigma_m^2}\cos^2\left(\frac{\pi\Delta\theta_{mn}}{2\theta_{RFI}}\right)\right]-\left(\sum_{mn}\frac{N_{mn}z_{mn}}{\sigma_m^2}\right)\left[\sum_{mn}\frac{N_{mn}}{\sigma_m^2}\cos^2\left(\frac{\pi\Delta\theta_{mn}}{2\theta_{RFI}}\right)\right]}{2\left(\sum_{mn}\frac{N_{mn}}{\sigma_m^2}\right)\left[\sum_{mn}\frac{N_{mn}}{\sigma_m^2}\cos^4\left(\frac{\pi\Delta\theta_{mn}}{2\theta_{RFI}}\right)\right]-\left[\sum_{mn}\frac{N_{mn}}{\sigma_m^2}\cos^2\left(\frac{\pi\Delta\theta_{mn}}{2\theta_{RFI}}\right)\right]^2}
\end{equation}
\noindent and
\begin{equation}
z_0 = \frac{\sum_{mn}\frac{N_{mn}z_{mn}}{\sigma_m^2}-f\sum_{mn}\frac{N_{mn}}{\sigma_m^2}\cos^2\left(\frac{\pi\Delta\theta_{mn}}{2\theta_{RFI}}\right)}{2\sum_{mn}\frac{N_{mn}}{\sigma_m^2}}.
\end{equation}

\section{C. Local Model Weights for Construction of Global RFI Model}

The local model for RFI subtraction is given by Equation~9.  Let $\sigma_f$ and $\sigma_{z_0}$ be the uncertainties in the fitted values of $f$ and $z_0$, respectively.  As in Appendix~A, we approximate these uncertainties to be uncorrelated, in which case the variance of the fitted model, as a function of position in the model, is given by:
\begin{equation}
\sigma^2(\Delta\theta) \approx \sigma_f^2\left[\cos^2\left(\frac{\pi \Delta\theta}{2\theta_{RFI}}\right)-\overline{\cos^2\left(\frac{\pi \Delta\theta}{2\theta_{RFI}}\right)}\right]^2+\sigma_{z_0}^2.
\end{equation}
\noindent However, unlike in Appendix~A, each term does not contribute to the variance equally, because $z_0$ is additionally constrained by a prior probability distribution to be noise level (Appendix~B).  Approximating the variance of $z_0$ (from zero) from the prior probability distribution to be comparable to, and almost fully correlated with, the variance of $z_0$ (from its, typically near-zero, best-fit value) from the likelihood function, instead yields:
\begin{equation}
\sigma_f^2\overline{\left[\cos^2\left(\frac{\pi \Delta\theta}{2\theta_{RFI}}\right)-\overline{\cos^2\left(\frac{\pi \Delta\theta}{2\theta_{RFI}}\right)}\right]^2} \sim 2\sigma_{z_0}^2.
\end{equation}
\noindent Consequently:
\begin{equation}
\sigma^2(\Delta\theta) \propto 1 + 2\frac{\left[\cos^2\left(\frac{\pi \Delta\theta}{2\theta_{RFI}}\right)-\overline{\cos^2\left(\frac{\pi \Delta\theta}{2\theta_{RFI}}\right)}\right]^2}{\overline{\left[\cos^2\left(\frac{\pi \Delta\theta}{2\theta_{RFI}}\right)-\overline{\cos^2\left(\frac{\pi \Delta\theta}{2\theta_{RFI}}\right)}\right]^2}}.
\end{equation}

Finally, as in Appendix~A, the variance scales inversely with the number of non-rejected dumps to which the local model is fitted, and weight scales inversely with variance, yielding:
\begin{equation}
w_{ij} \sim \frac{\sum_j N_{ij}}{1 + 2\frac{\left[\cos^2\left(\frac{\pi \Delta\theta_{ij}}{2\theta_{RFI}}\right)-\overline{\cos^2\left(\frac{\pi \Delta\theta_{ik}}{2\theta_{RFI}}\right)}\right]^2}{\overline{\left[\cos^2\left(\frac{\pi \Delta\theta_{ij}}{2\theta_{RFI}}\right)-\overline{\cos^2\left(\frac{\pi \Delta\theta_{ik}}{2\theta_{RFI}}\right)}\right]^2}}},
\end{equation}
\noindent where $\sum_j N_{ij}$ is the number of non-rejected dumps that contributed to the $i$th local model, $w_{ij}$ is the weight of the $j$th point from the $i$th local model, $\Delta\theta_{ij}$ is the 2D angular distance of this point from the model's center, and $\theta_{RFI}$ is the user-defined RFI-subtraction scale (\textsection3.6).  Admittedly, the factor of two in this equation is crudely determined, but its inclusion is fairer than using, e.g., $w_{ij}\approx\sum_j N_{ij}$, or $w_{ij}\approx1$.  Regardless, the resulting global model is not very sensitive to how the local models are weighted.

\section{D:  Averaging and Summing Errors Correlated Over a Scale}

Consider $N$ measurements, each with uncertainty $\sigma_i$.  If these measurements are independent of each other, the uncertainty of their average is given by:
\begin{equation}
\frac{1}{\sigma_{ave}^2} = \sum_i^N\frac{1}{\sigma_i^2},    
\end{equation}
\noindent and the uncertainty of their sum is given by:
\begin{equation}
\sigma_{sum}^2 = \sum_i^N\sigma_i^2.    
\end{equation}
\noindent If all of the $\sigma_i$ are equal, $\sigma_{ave}=\sigma_i/\sqrt{N}$ and $\sigma_{sum}=\sqrt{N}\sigma_i$, which are the expected results.

However, these equations must be modified if some of the measurements are correlated.  First, consider the extreme case where all of the measurements are correlated.  Then, $\sigma_{ave}=\sigma_i$ and $\sigma_{sum}=N\sigma_i$, or:
\begin{equation}
\frac{1}{\sigma_{ave}^2} = \sum_j^1\frac{1}{\sigma_j^2} = \sum_i^N\frac{1}{N\sigma_i^2},    
\end{equation}
\noindent and:
\begin{equation}
\sigma_{sum}^2 = \sum_j^1(N\sigma_j)^2 = \sum_i^N\frac{(N\sigma_i)^2}{N} = \sum_i^NN\sigma_i^2.    
\end{equation}

Now, consider the intermediate, but still simple, case where the $N$ measurements consist of $M$ independent subgroups, each with $N_j$ correlated measurements.  Then:
\begin{equation}
\frac{1}{\sigma_{ave}^2} = \sum_j^M\frac{1}{\sigma_j^2} = \sum_i^N\frac{1}{N_i\sigma_i^2},    
\end{equation}
\noindent and:
\begin{equation}
\sigma_{sum}^2 = \sum_j^M(N_j\sigma_j)^2 = \sum_i^N\frac{(N_i\sigma_i)^2}{N_i} = \sum_i^NN_i\sigma_i^2,    
\end{equation}
\noindent where $N_i$ is shorthand for $N_j$ of the $i$th measurement.

Finally, consider the case where each measurement is correlated with either a definable subset, or all, of its surrounding measurements, over a scale $\theta$.  In this case, the above equations apply, with $N_i$ instead being the number of measurements (1) within a $\theta/2$ radius of the $i$th measurement (2) that are also part of this subset, and (3) that are also part of the sum.

We first make use of this when averaging RFI-subtraction local-model values, to produce a global-model value (\textsection3.6).  Each local-model value is correlated with a definable subset of its surrounding local-model values, over the $2\theta_{RFI}$-diameter scale of the local model (Equation~9).  If the local model were a one-parameter model, each local-model value would be correlated with all of its surrounding local-model values over this scale.  However, it is instead a two-parameter model, with a correlated $f$-dominated inner region, and a separately correlated (and perhaps even partially anti-correlated) $z_0$-dominated outer region.  If a local-model value is above its corresponding 2D noise-model, we take it to be part of the $f$-dominated region; else we take it to be part of the $z_0$-dominated region.  

Consequently, when calculating the weight of the global-model value (Footnote~22), we use Equation~D5:  We sum the weights ($\propto$ $1/\sigma_i^2$) of the non-rejected local-model values (Appendix~C), but not before dividing each of these weights by the number of non-rejected local-model values (1) within a $\theta_{RFI}$ radius of that weight's local-model location (2) that are above their 2D noise-model values if the global-model value is above its 2D noise-model value, or that are below their 2D noise-model values if the global-model value is below its 2D noise-model value, and (3) that are also part of the sum (i.e., that are also within a $\theta_{RFI}$ radius of the global-model location).

We again make use of this when surface-modeling the post-RFI subtraction data (i.e., when surface-modeling the above global-model values; \textsection3.7).  These data are correlated over a scale, $\theta_{corr}$, which we measure directly from the non-rejected local-model data that contributed to each global-model value.\footnote{For a given global-model, or post-RFI subtraction, value, consider the non-rejected local-model data that contributed to this value, with each data point weighted (1) by the number of times that it contributed, and (2) by its number of dumps.  We then take $\theta_{corr}$ to be proportional to twice the standard deviation of these points' locations about this value's location, where we set the factor of proportionality, $f_{corr}$, empirically below.  Consequently, each post-RFI subtraction data point has its own $\theta_{corr}$ value, just as each has its own $\theta_w$ value (\textsection3.7).  Away from sources, where few local-model data points are rejected, $\theta_{corr} \approx 1.3f_{corr}\theta_{RFI}$.  But near sources, this scale can be much smaller.}   Consequently, when calculating the weight map (\textsection3.7.2), which is again proportional to Equation~D5, we divide each term in the sum by the number of data points (1) within a $\theta_{corr}/2$ radius of that term's data point (2) that are also within the 1-beamwidth radius region over which the surface model is being fitted.

The weight map is then used in photometry ({\color{black}\textsection4}).  We use it to measure or model (1) the background level, (2) the standard deviation of the pixel values about the background level, (3) the uncertainty in any pixel value in the image, (4) the total uncertainty from summing all of the pixel values in the aperture, (5) the uncertainty in the background level, and (6) the total photometric error:

1.  The background level is simply given by:
\begin{equation}
\mu = \frac{\sum_i^{an}w_iz_i}{\sum_i^{an}w_i},
\end{equation}
\noindent where $z_i$ is the value of the $i$th pixel, $w_i \propto 1/\sigma_i^2$ is its corresponding weight-map value, and the sums are over all non-rejected pixels in the annulus (Footnote~37; e.g., Figure~54).  

2.  The standard deviation of the pixel values about the background level is then given by:
\begin{equation}
\sigma \approx \sqrt{\frac{\sum_i^{an}w_i(z_i-\mu)^2}{\sum_i^{an}w_i}},    
\end{equation}
\noindent where these sums are over only the $z_i>\mu$ subset of the non-rejected pixels in the annulus, because the $z_i<\mu$ pixels can be artificially suppressed, by the (default) noise-level prior (Footnotes 27 and 37).\footnote{Even so, use of the noise-level prior results in artificially smaller values of $\sigma$ as the measured value of $\mu$ approaches zero, or becomes negative.  We have measured this effect for a wide range of $\theta_{RFI}$ and $\theta_{min}$ values (for cases where $\theta_w$ is dominated by $\theta_{min}$; Equation 14), and have assembled the following empirical correction factor:
\begin{equation}
f_\sigma = \ln\left(4.0e^{-4.2\frac{\mu}{\sigma}}+e\right).
\end{equation}
\noindent This correction factor becomes unimportant as $\mu \rightarrow \sigma$.  But in most cases, $\mu < \sigma$, so this factor should not be neglected when photometering images that were surface-modeled using the noise-level prior.}

3.  Equation~D8 is the average uncertainty per pixel, but only in the annulus.  The uncertainty per pixel varies across the image, because the weighted number of dumps that contributed to each pixel (i.e., the weight map) varies across the image.  Hence, we take the uncertainty per pixel, at any pixel in the image, to instead be given by:
\begin{equation}
\sigma_i \approx \sigma\sqrt{\frac{\langle w\rangle_{an}}{w_i}},
\end{equation}
\noindent where $\langle w\rangle_{an}$ is the value of the weight map averaged over the same pixels that were summed over in Equation~D8 (i.e., non-rejected pixels in the annulus with $z_i>\mu$).  (Note, unlike in an optical image, there is no additional uncertainty from signal above the background level, as the signal is background-dominated.)

4.  The total uncertainty from summing all of the pixel values in the aperture is then given by Equation~D6:
\begin{equation}
\sigma_{ap} \approx \sqrt{\sum_i^{ap}N_i\sigma_i^2} \approx \sigma\sqrt{\langle w\rangle_{an}}\sqrt{{\color{black}\sum_i^{ap}}\frac{N_i}{w_i}}.
\end{equation}
\noindent If the pixel values were independent of each other, $N_i$ would be 1, but this is not the case:  These values have been correlated first on the above $\theta_{corr}$-diameter scale, by the RFI-subtraction algorithm, and then on a scale that is proportional to the weighting scale, by the surface-modeling algorithm, where we set the factor of proportionality, $f_w$, empirically below.  Consequently, we take $N_i$ to be the number of pixels (1) within a $[\theta_{corr}^2+(f_w\theta_w)^2]^{1/2}/2$ radius of that term's pixel (2) that are also part of the sum (i.e., that are within the aperture).\footnote{$\theta_{corr}$ is measured at each data point (Footnote~40).  The value of $[\theta_{corr}^2+(f_w\theta_w)^2]^{1/2}$ at each pixel is set by the same proximity-weighted averaging algorithm that we use to set the value of $\theta_w$ at each pixel (Footnote~31).  This is then output as a separate, ``correlation'' map, similar to the scale map and the weight map, for later use in photometry (\textsection3.7.2).}

5.  The uncertainty in the background level is given by:  
\begin{equation}
\sigma_\mu \approx \frac{\color{black}\sqrt{\color{black}\sum_i^{an}N_iw_i^2(z_i-\mu)^2}}{\sum_i^{an}w_i},
\end{equation}
\noindent where the sums are as in Equation~D8, and we again take $N_i$ to be the number of pixels (1) within a $[\theta_{corr}^2+(f_w\theta_w)^2]^{1/2}/2$ radius of that term's pixel (2) that are also part of the sum (in this case, non-rejected pixels in the annulus with $z_i>\mu$).

6.  The background level, per pixel, is then $\mu \pm \sigma_\mu$, and the background level summed over the aperture is $N_{ap}\mu \pm N_{ap}\sigma_\mu$, where $N_{ap}$ is the number of pixels in the aperture.  Hence, the total photometric error is then $\sigma_{phot} = \sqrt{\sigma_{ap}^2+(N_{ap}\sigma_\mu)^2}$ (Equation 17).

As stated above, we set the factors $f_{corr}$ and $f_w$ empirically, to get the best results for a wide range of RFI-subtraction scales, minimum weighting scales, aperture sizes, and point-spread functions.  For each of a representative set of combinations of these, we re-simulated, re-processed, and re-photometered the ten-source mapping from \textsection3.3.2 and \textsection3.6.2 50 times, to determine what their photometric error bars should be.  {\color{black}We find that $f_{corr} = f_w = 3$ yields the best result, consistently getting to within $\approx$30\% of approximately half of the desired result.  Consequently, we adopt these values, and then multiply $\sigma_{phot}$ by this empirically determined correction factor of two.  Although not perfect}, we felt that it was important to give the user a way to least approximate their photometric error bars from a single observation, instead of having to re-observe each target a large-enough number of times to be able to calculate these error bars statistically.  

Finally, it should be noted that the above $N_i$ calculations, both in determining the weights of the post-RFI subtraction data for surface-modeling, and in determining the weights of the post-surface modeling pixel values for photometry (or for image stacking; \textsection3.7.2), are computationally intensive.  Consequently, we give the user the option of forgoing these calculations and instead treating each RFI-subtraction local-model value, and global-model value, as independent of its surrounding values.  This results in significantly quicker processing, but is recommended only if the user does not need to calculate photometric error bars, or stack images.

\end{document}